\documentclass[12pt]{article}\usepackage[hyperfootnotes=false]{hyperref}
\usepackage{epsfig}
\usepackage{float}
\usepackage{empheq}
\usepackage{bbold}
\usepackage{xspace}

\makeatletter

\@addtoreset{equation}{section}
\makeatother
\usepackage[utf8]{inputenc}
\usepackage{amsmath}

\usepackage{caption}

\usepackage{amsmath}
\usepackage{amssymb}
\usepackage{graphicx}
\newlength{\abstractwidth}
\renewcommand{\thefootnote}{\fnsymbol{footnote}}
\renewcommand{\thanks}[1]{\footnote{#1}} % Use this for footnotes
\newcommand{\starttext}{
\setcounter{footnote}{0}
\renewcommand{\thefootnote}{\arabic{footnote}}}

\newcommand{\be}{\begin{equation}}
\newcommand{\bea}{\begin{eqnarray}}
\newcommand{\eea}{\end{eqnarray}}
\newcommand{\beq}{\begin{equation}}
\newcommand{\ee}{\end{equation}}

\newcommand*\widefbox[1]{\fbox{\hspace{2em}#1\hspace{2em}}}
\def\eq{&=&}

\def\la{\langle}
\def\ra{\rangle}
\def\simleq{\; \raise0.3ex\hbox{$<$\kern-0.75em
\raise-1.1ex\hbox{$\sim$}}\; }
\def\simgeq{\; \raise0.3ex\hbox{$>$\kern-0.75em
\raise-1.1ex\hbox{$\sim$}}\; }

\def\bi{\begin{itemize}}
\def\ei{\end{itemize}}

\def\CA{{\cal{A}}}

\def\CF{{\cal{F}}}

\def\CJ{{\cal{J}}}

\def\CL{{\cal{L}}}

\def\CO{{\cal{O}}}

\def\t{\tau}

\def\bsub{ \begin{subequations}
\begin{empheq}[box=\widefbox]{align}  }
\def\esub{ \end{empheq}
\end{subequations}}

\def\1{\(  \mathbb{1} \)}

   \def\l{l_{ads}}

 \def\lf{\left(}
    \def\rg{\right)}

  \def\bn{\bigskip \noindent}

 \def\bm{\begin{bmatrix}}
 \def\em{\end{bmatrix}}

\def\dk{$\text{DSSYK}_{\infty}$\xspace}
\def\dkp{$\text{DSSYK}_{\infty}$\xspace.}
       \def\l{\ell}
            
            \def\f{\frac }
            \def\g{g_{\rm ym}}
             \def\gs{g_{\rm string}}
             \def\N{N_{\rm ym}}
                 \def\t{t_{\rm cosmic}}

\makeatletter
\g@addto@macro\normalsize{%
  \setlength\abovedisplayskip{10pt}
  \setlength\belowdisplayskip{20pt}
  \setlength\abovedisplayshortskip{10pt}
  \setlength\belowdisplayshortskip{20pt}
}
\makeatother

\usepackage{color}

\begin{document}
  
\begin{titlepage}

 \rightline{}
\bigskip
\bigskip\bigskip\bigskip\bigskip
\bigskip

\centerline{\Large \bf {Double-Scaled SYK, QCD, and the Flat}}
\bn
\centerline{\Large \bf {Space Limit of de Sitter Space}}
\bn

\bigskip
\begin{center}
\bf Yasuhiro Sekino$^{3,1}$ and  Leonard Susskind$^{1,2}$   \rm

\bigskip

$^1$SITP, Stanford University, Stanford, CA 94305, USA \vskip0.4em
$^2$Google, Mountain View, CA 94043, USA\vskip0.4em
$^3$Department of Liberal Arts and Sciences,
Faculty of Engineering, \\ Takushoku University, 
%815-1 Tatemachi, 
Hachioji, Tokyo 193-0985, Japan

%\vspace{1cm}
\end{center}

\bn

\Large
\begin{abstract} 

A  surprising connection exists between  double-scaled SYK  at infinite temperature,   and large $N$ QCD. The large $N$ expansions of the two theories have the same form; the 't~Hooft limit of QCD parallels the fixed $p$ limit of SYK (for a theory with $p$-fermion interactions), and the limit of fixed gauge coupling $\g$---the flat space limit in AdS/CFT---parallels the double-scaled limit of SYK. From the holographic perspective  fixed $\g$ is the   far more  interesting  limit of gauge theory, but very little is known about  it. DSSYK allows us to explore it in a more tractable example.  
The connection   is illustrated  by  perturbative and non-perturbative  DSSYK calculations, and comparing the results with known properties of Yang Mills theory.

The correspondence is largely independent of the conjectured duality between DSSYK and de Sitter space, but may have a good deal to tell us about  it.

\end{abstract}

\end{titlepage}

\starttext \baselineskip=17.63pt \setcounter{footnote}{0}

\Large

\tableofcontents

\section{Introduction}
\subsection{A Surprising  QCD-DSSYK$_{\infty}$ Parallel}\label{S: parallel}

   A surprising parallel exits \cite{susskind:confined} between Double-Scaled SYK \cite{Cotler:2016fpe}\cite{Berkooz:2018jqr} at  infinite  temperature (\dk) and large $\N$ QCD. It is an ``empirical" fact whose deeper meaning is as yet unclear.  Although this relation is largely independent of the conjectured holographic duality between \dk and de Sitter space~\cite{susskind:confined, Susskind:2022bia, Susskind:2021esx, Susskind:2022dfz,   Rahman:2023pgt,Rahman:2024vyg, Rahman:2024iiu}, it does cast light on that correspondence\footnote{%
The duality proposed in \cite{susskind:confined, Susskind:2022bia, Susskind:2021esx, Susskind:2022dfz,  Rahman:2023pgt,Rahman:2024vyg, Rahman:2024iiu} is between \dk and semiclassical de Sitter space with a de Sitter radius that tends to
 infinity as $N$ tends to infinity. 
H.~Verlinde~\cite{Narovlansky:2023lfz, Verlinde:2024znh} has proposed a
 different duality between \dk and a Planck scale de Sitter space. 
In Verlinde's version of the duality  the de Sitter
radius is  given by
$\ell_{\rm ds}= 8\pi/\lambda \sim 1.$ Such a duality
 would not have a flat space limit and most of the
 considerations of this paper would not apply.}. 
\dk, Jackiw-Teitelboim(JT)-de Sitter space and large 
$\N$ QCD  seem inextricably related. 
The  QCD-\dk \ parallel includes correspondences in: 
't~Hooft type  $1/N$ expansions and the associated perturbative expansions at each order of $1/N$; non-perturbative confinement mechanisms; and ``flat-space limits" which we will explain. There are also hints of string-like behavior in \dk \ that parallel the confining behavior of QCD\footnote{%
By QCD, we do not intend to mean the theory which describes quarks and gluons in the real world. We have nonabelian gauge theory in mind, but what we say about the structure of perturbative expansions should be valid for any theory which have matrix degrees of freedom described by a single-trace action. On the other hand, the existence of the flat space limit is guaranteed in a limited class of theories, and we do not have reason to expect it for the real-world QCD.}.

\subsection{Temperatures}\label{S: Large}

There are a number of notions of temperature  \cite{Rahman:2024vyg} that apply to \dk:
the Boltzmann temperature which is infinite; the tomperature \cite{Lin:2022nss}, which corresponds to the Hawking temperature experienced by an observer at the pode of de Sitter space; and the Unruh temperature at the stretched horizon denoted by $T_{\rm cord}$ in \cite{Rahman:2024vyg}. $T_{\rm cord}$ describes the environment felt by the overwhelming number of degrees of freedom comprising the de Sitter entropy\footnote{The temperature $T_{\rm cord}$ was identified in  \cite{Rahman:2024vyg}: as the temperature associated with the periodicity of the so-called fake disc of \cite{Lin:2023trc}.}. 
The QCD analog of $T_{\rm cord}$ is the temperature at the 
confinement---de-confinement   transition. Just as confinement---de-confinement is a non-perturbative emergent phenomena in QCD, $T_{\rm cord}$ is similarly non-perturbative and emergent. In both cases perturbation theory contains hints of the emergent phenomena, but demonstrating them is much easier in \dk \ than in QCD. 

\subsection{Flat Space Limits}
The flat space limit of AdS is defined as the limit in which the AdS radius of curvature $\l_{\rm ads}$ goes to infinity while all microscopic length scales stay fixed, one scale being the string scale. It is the limit of ``sub-AdS locality.'' In terms of CFT parameters it is the non-'t~Hooft ultra-strongly coupled limit, $\N \to \infty$ with $\g$ held fixed ('t~Hooft coupling going to infinity). 

The same issue of locality arises in holographic descriptions of de Sitter space. As we will explain, the flat space limit is defined as the limit in which the horizon area goes to infinity, microscopic scales remaining fixed. In terms of \dk parameters it is the double-scaled limit, $N \to \infty$ with $p^2/N$ held fixed (for a theory with $p$-fermion interactions and $N$ fermion species).

\section{QCD at Large $\N$} 
\label{S: Deftions}

 We  begin with a brief review of  the perturbative and large $\N$ expansions of QCD.
 
\begin{enumerate}
      \item The gauge group is $SU(\N)$ where $\N$ is large, allowing an expansion in inverse powers of $\N$. Note that the number of degrees of freedom carried by the gauge fields is $\sim \N^2.$ This, for example, means that the entropy of a hot QCD plasma is of order $\N^2.$
 \item The gauge coupling is $\g.$ In the string description of QCD the closed-string coupling is $\gs = \g^2.$
 \item The 't~Hooft coupling\footnote{We use the notation $\alpha$ instead of the more common $\lambda$ to avoid confusion with the parameter $\lambda$ in double-scaled SYK.}  $\alpha$ is defined by,
 \be 
 \alpha = \g^2 \  \N.
 \label{tooft}
 \ee
The 't~Hooft limit is defined by $\N\to \infty$ with 
$\alpha$ held fixed. The perturbation expansion can be rearranged into an expansion in inverse powers of $\N.$ In QCD without quarks\footnote{%
%%%%%%%%%%%%%%%footnote%%%%%%%%%%%%
In the present paper, we will focus on the case without quarks or with quarks in the adjoint representation. In a subsequent paper~\cite{Miyashita:2025rpt} with S.~Miyashita, we will consider quarks in the fundamental representation, in which case the expansion is in powers of $\N^{-1}$.}
%%%%%%%%%%%%%%%%%%%%%%%%%%%%%%%%%%%
the expansion is in powers of $\N^{-2}.$

\item 
Perturbation theory at each order of $\N$ is given as a power series expansion in $\alpha.$ It should be pointed out that individual orders of the perturbation expansion are infrared divergent.

\item 
The ``flat-space limit" is defined by $\N\to \infty,$ but unlike the 't~Hooft limit, in the flat-space limit the gauge coupling $\g$ is held fixed. This implies that the 't~Hooft coupling goes to infinity;  the flat space limit is therefore ultra-strongly coupled and  highly non-perturbative.  
 
\end{enumerate}
 
 The name ``flat-space limit" is taken from AdS/CFT  where in the fixed $\g$ limit bulk space-time becomes flat, and the flat-space S-matrix can be recovered from suitable limits of AdS/CFT boundary correlation functions~\cite{Polchinski:1999ry}, \cite{Susskind:1998vk}, \cite{Polchinski:1999yd}.
It is also the limit in which the theory exhibits sub-AdS locality.

\subsection {The 't~Hooft Limit}\label{S: tooftlim}

For definiteness we will consider an amplitude with two external gauge boson lines denoted by $\CA$. As long as perturbation theory applies, there is nothing special about the perturbative expansion for this amplitude; the pattern we will describe is very general.

\begin{figure}[H]
\begin{center}
\includegraphics[scale=.4]{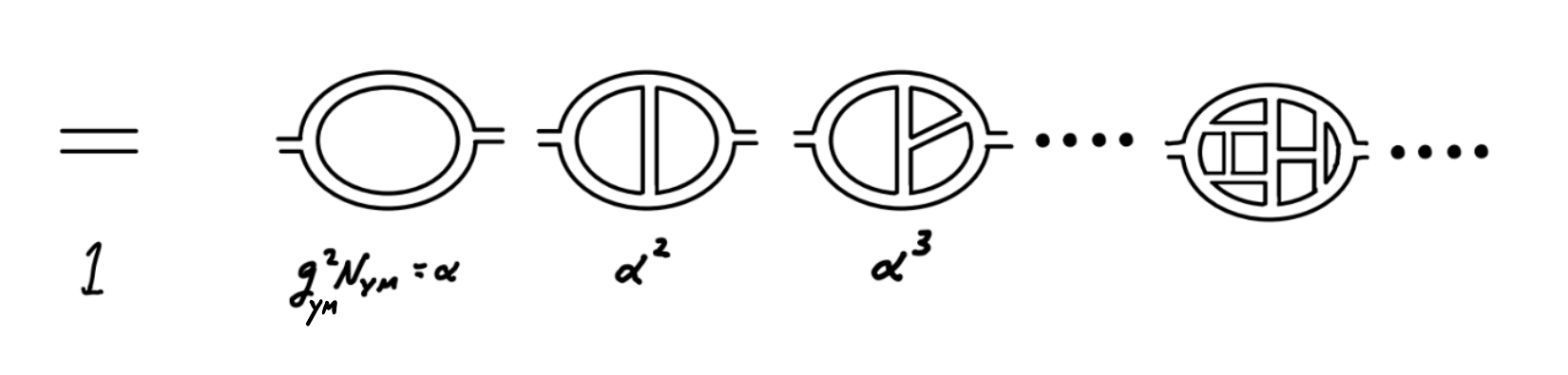}
\caption{Genus zero ($h=0$) diagrams.}
\label{AAh=0}
\end{center}
\end{figure}

By use of 't~Hooft's double-line (ribbon) notation for Feynman diagrams, every diagram can be assigned a genus; namely the smallest genus surface upon which the diagram can be drawn without crossing of lines. The genus of the diagram determines the power of $(\N)^{-2}.$ A genus $h$ diagram with $v$ vertices and two external lines gives a  contribution  to $\CA$ which scales like,
\be  
  \frac{\alpha^v}{(\N)^{2h}}.
\label{scales}
\ee
In Figure~\ref{AAh=0} some genus zero diagrams are shown scaling like 
$1,$   $\alpha,$ $\alpha^2,$  and $\alpha^3.$ These are the beginning of an infinite set of  genus zero diagrams.

For each genus there is a unique lowest order ribbon diagram which we will call a ``primitive" diagram\footnote{A single ribbon diagram represents several conventional Feynman diagrams. Unlike individual Feynman diagrams ribbon diagrams are gauge invariant.}.
For genus zero the primitive diagram is the simple diagram on the far left of Figure~\ref{AAh=0}. The remaining diagrams (an infinite set) are obtained from the primitive diagram by adding vertices and propagators but in such a way as to preserve the genus. We will refer to this process as ``decorating" the primitive diagram. 

\begin{figure}[H]
\begin{center}
\includegraphics[scale=.6]{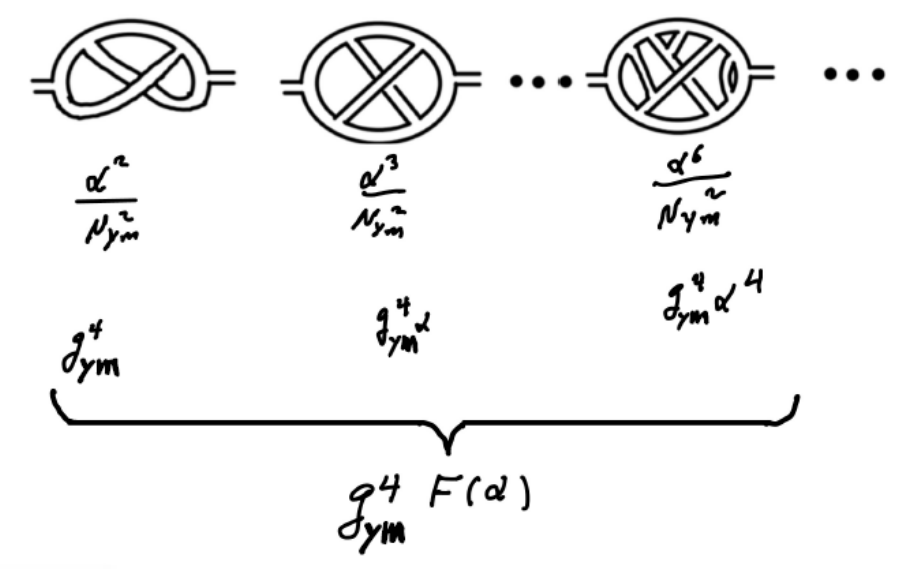}
\caption{Genus one ($h=1$) diagrams. The coefficient for each diagram is expressed by $\N$ and $\alpha$ in the first line below the diagrams, and by $\g$ and $\alpha$ in the second line.}
  \label{AAh=1}
\end{center}
\end{figure}

In Figure~\ref{AAh=1} some of the genus one diagrams are shown. For genus one the primitive diagram is the ``pretzel'' diagram on the far left.

\begin{figure}[H]
\begin{center}
\includegraphics[scale=.3]{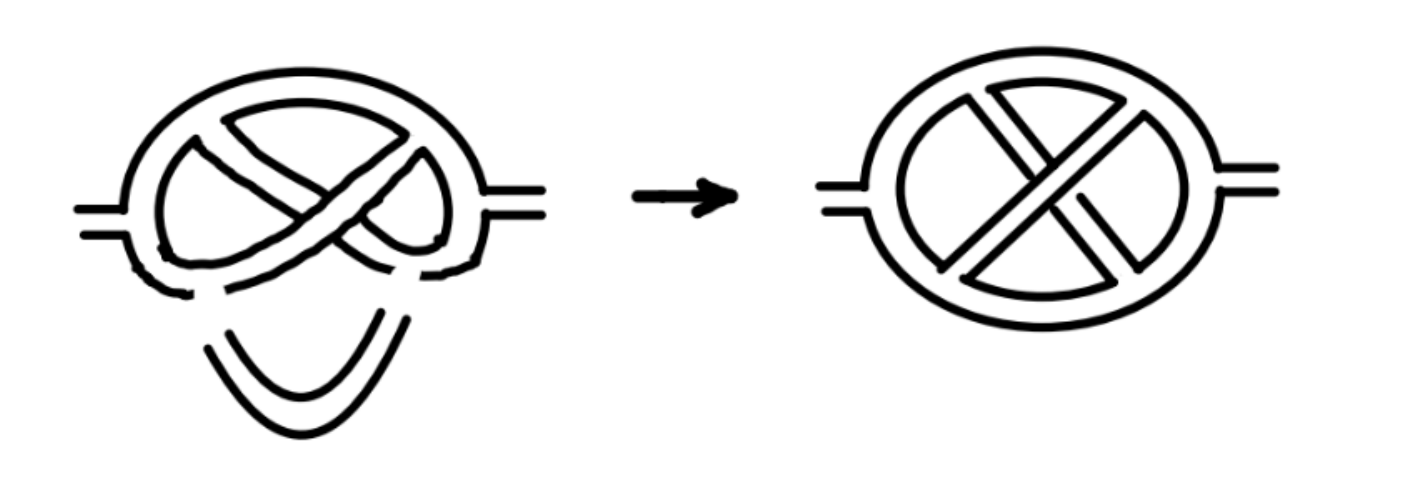}
\caption{Decorating the primitive diagram for $h=1$.}
  \label{pretz+}
\end{center}
\end{figure}

Figure \ref{pretz+} shows how the second  diagram in Figure~\ref{AAh=1} can be obtained by decorating the pretzel diagram with an additional propagator and two vertices.
Although it is standard to express the factors associated with each diagram by $\N$ and $\alpha$, one can also express them by other choices of two independent parameters, such as $\g$ and $\alpha$. 
By inspection of the $h=0$ and $h=1$ cases, we see that 
the diagrams at each genus 
have an overall factor of $\g^{4h}$ multiplied by some factors of 
$\alpha$, where the power of $\alpha$ does not become negative. 

The general pattern of the expansion in $\N$ and $\alpha$ is well known, 
\be 
\boxed{
\CA =  \sum_{h=0}^{\infty} \f{\tilde{\CF}^{(h)}( \alpha)}{(\N)^{2h}}
}
\label{ymNexp1}
\ee
where each $\tilde{\CF}^{(h)}(\alpha)$ is an infinite polynomial in $\alpha.$
Explicitly $\CA$ is a double sum,
\be 
\CA =  \sum_{h=0}^{\infty} \f{1}{(\N)^{2h}} \sum_{n=2h}^{\infty} F^{(h)}_n \alpha^n. 
\label{ymNexp}
\ee
Note that $\tilde{\CF}^{(h)}$ and $F^{(h)}_n$  do \it not \rm mean $\tilde{\CF}$ or $F_n$ raised to the power $h$.
The $F_n^{(h)}$ coefficients are typically infrared divergent and depend on an infrared cutoff which we will discuss later.

There are two types of terms in \eqref{ymNexp}. The first have\footnote{%
It is clear that $n=2h$ is the lowest $n$ for a given $h$, since the power of $\N$ cannot be negative. We can construct a diagram with $n=2h$ by connecting $h$ pretzel diagrams (the far left diagram in Figure~\ref{AAh=1}) in series, by attaching an end of a pretzel to an end of another. This has $2h$ vertices, and 
has no index loop so it scales as $\N^0$, meaning that its genus $h$ satisfies
$n=2h$.}
$n=2h.$ 
They are
the primitive diagrams which provide the starting point for an infinite set of diagrams with a given genus $h$. 
The second have $n> 2h$, and can be obtained
by decorating the primitive diagrams. 
For example the first diagrams in Figure~\ref{AAh=0} ($h=0$) and 
Figure~\ref{AAh=1} ($h=1$) can be decorated with additional loops 
without increasing the genus.

Equation \eqref{ymNexp} is the genus expansion for the 't~Hooft limit in which $\alpha = \g^2 \N$ is held fixed as $\N \to \infty.$ In the following, we will consider the meaning of this expansion when we regard $\g$ and $\alpha$ as independent parameters.

\bn

\subsection{Fixed $\g$ Limit}  \label{S: fixedg} %\label{S: fixedg}
There is another much less studied limit---the fixed $\g$ limit---which in AdS/CFT is called the flat-space limit 
\cite{Polchinski:1999ry}\cite{Susskind:1998vk}\cite{Polchinski:1999yd}. In the flat-space limit $\N \to \infty$ while $\g$ is held fixed. 

The fixed $\g$ limit is  generally avoided, but it is central for some very important purposes which we list here:
\begin{enumerate}
\item In AdS/CFT it is usual to study the 't~Hooft limit for which the bulk string scale is finite in units of the AdS curvature. On the other hand,
in the fixed $\g$ flat-space  limit
  \cite{Polchinski:1999ry}\cite{Susskind:1998vk}\cite{Polchinski:1999yd}
the ratio of the  string length scale to the  AdS radius of curvature tends to zero, and the theory is said to exhibit sub-AdS locality. 

\item  In BFSS matrix theory \cite{Banks:1996vh} the 't~Hooft limit is the  limit of D0-brane black holes in 10-dimensions. The fixed $\g$ limit is the far more interesting but far more difficult limit of flat 11-dimensional M-theory. 

\item
A behavior very similar to AdS/CFT is seen 
in \dk \ \cite{susskind:confined}.
If we accept the \dk-de Sitter correspondence, we may call  the fixed $\lambda$ limit (studied below) the limit of  sub-dS (or sub-cosmic) locality \cite{susskind:confined}.

\end{enumerate}

From a holographic perspective the  fixed $\g$ limits are the  more interesting limits, but  because they are ultra-strongly coupled they are far more difficult,  and are relatively unexplored.  \dk \ gives us an opportunity to explore them in a more tractable form than in ordinary gauge  or matrix theory.

\subsection{Perturbative Expansion in QCD}\label{S: perturbative}

To understand the fixed $\g$ limit we return to \eqref{ymNexp}. The genus zero  contribution $\CA^{(0)}$ is shown in Figure~\ref{AAh=0}, the first term of which is\footnote{Strictly speaking it is a function of the time and distance between the arguments of the correlation function, but it is of order unity with no dependence on $\g$ or $N.$ } 
\be 
F_0^{(0)} = 1.
\label{F1(0)}
\ee
This trivially has a good fixed $\g$ limit.

The remaining terms can be expressed in terms of  a power series in $\alpha,$ multiplying $F_0^{(0)}$,
\bea 
\CA^{(0)} &=& F_0^{(0)}\times(1 +c_1\alpha + c_2\alpha^2 +\cdots) \cr \cr
\eq  \CF^{(0)}(\alpha).
\label{pwrsrs}
\eea
(The coefficients $c_n$ depend on the genus $h$ but in the interests of an uncluttered notation we will leave the $h$ dependence implicit.)
In general the function $\CF^{(0)}(\alpha)$ may depend on the time interval between the initial and final ends of the propagator, but not on $\N$.

In the fixed $\g$ limit, $\alpha\to \infty$ as $\N \to \infty.$ Therefore the condition for a good fixed $\g$ limit requires that the limit
\be  
\lim_{\alpha \to \infty} \CF^{(0)}(\alpha) = \CF^{(0)}
\label{lim00}
\ee
exists.

Let's consider the genus one contribution $\CA^{(1)}$ shown in Figure~\ref{AAh=1}. It may be written in the form\footnote{%
The function $\CF^{(h)}(\alpha)$ is defined from 
$\tilde{\CF}^{(h)}(\alpha) $ that appeared
in \eqref{ymNexp1} by dividing by the power of $\alpha$ 
for the primitive diagram, 
$\CF^{(h)}(\alpha)=\tilde{\CF}^{(h)}(\alpha)/\alpha^{2h}$.},
\bea 
\CA^{(1)} \eq 
 \g^4\times (1 +c_1\alpha + c_2\alpha^2 +\cdots) \cr \cr
\eq  \g^4\CF^{(1)}(\alpha).
\eea
Again  a good fixed $\g$ limit requires the limit
$$ 
\lim_{\alpha \to \infty}\CF^{(1)}(\alpha)
$$
to exist.  In that case the genus one contribution is proportional to $\g^4.$

More generally, going back to \eqref{ymNexp}  and using $\alpha = \g^2 \N$ we rewrite it as,
\be 
\CA= \sum_{h=0}^{\infty}  \g^{4h} \CF^{(h)}(\alpha),
\label{fgsum3}
\ee
where the functions $\CF^{(h)}(\alpha) $ are the sum over the primitive diagram plus  those with added decorations.

The desired fixed $\g$ limit is given by replacing the functions $\CF^{(h)}(\alpha) $ by their large $\alpha$ limits (assuming the limits exist). Thus define,
\be  
\lim_{\alpha \to \infty} \CF^{(h)}(\alpha) = \CF^{(h)}.
\label{lims}
\ee
The fixed $\g$ limit is,
\be 
\CA= \sum_{h=0}^{\infty}  \CF^{(h)} \g^{4h},
\label{fgsumm}
\ee
where each $ \CF^{(h)}$ contains infinitely many diagrams of the same genus.

The factor $\g^{4h}$ accompanying the genus $h$ term has a meaning in string theory where the genus $h$ world-sheet amplitude is proportional to  $\gs^{2h}.$ Thus we  make the identification,
\be 
 \gs =\g^2,
\label{gs=g2}
\ee
leading to the classic formula for the contribution of a genus $h$ diagram in string theory,
\be 
\boxed{
\CA= \sum_{h=0}^{\infty}   \CF^{(h)} \gs^{2h}.}
\label{strhnt}
\ee

It is an assumption that the limits in \eqref{lims} exist.  
For supersymmetric CFT's with a bulk gravitational dual, the existence of the limit is guaranteed by the flat space limit, but in general unlike the 't~Hooft limit, the fixed $\g$ limit may not exist.
One of the main points of this paper is that the corresponding limits for \dk \ do exist at least for small but finite $\lambda,$   despite the absence of supersymmetry.

\section{\dk: Preview of Results} \label{S: preview}

In this section we will preview the results of calculations of \dk \ correlation functions, performed in 
later sections and Appendix~A. For definiteness we will focus on single fermion correlation functions
 \be 
A(T) = \text{tr}  \big( \psi(0) \psi(T)\big) =  \text{tr}\left( \psi e^{iHT} \psi e^{-iHT}\right)
\label{CT1}
 \ee
 (Note that in this context $T$ is a time in cosmic units, not a temperature.)

\subsection{Large $N$ Expansion for DSSYK$_{\infty}$} \label{sub:2.1}

The correlation functions can be formally expanded in a power series in the strength $\CJ$ of the coupling constants. The series is badly infrared divergent; each term being proportional to a positive power of $T$ which grows faster than the previous term, but let's ignore that for now. We will see that the series can be re-organized\footnote{%
We use the symbol $h$ to denote the power of $1/N$ as in QCD. 
We do not mean that the coefficients $F_n^{(h)}$
or the functions $\tilde{\CF}^{(h)}(p)$ are 
the same as $F_n^{(h)}$ or $\tilde{\CF}^{(h)}(\alpha)$ in QCD,
but we use the same symbols to emphasize the similarity in the 
structure of perturbative expansions. As in QCD, $F_n^{(\alpha)}$ 
will depend on the infrared cutoff.} into an expansion in powers of $1/N.$ 
\be
\boxed{ 
A(T)=\sum_{h=0}^{\infty}\f{1}{N^h}
\sum_{n} F_{n}^{(h)}p^n
=\sum_{h=0}^{\infty} \f{\tilde{\CF}^{(h)}(p)}{N^h}.   }
\label{C(T)}
\ee

The functions $\tilde{\CF}^{(h)}(p)$ are defined as infinite polynomials in the parameter $p$ (for a theory with $p$-local Hamiltonian) which we can calculate in perturbation theory and which depend on the time $T.$ The leading term for $h=0$ is the sum of so-called melon diagrams (which play the same role as genus zero diagrams in QCD).

By analogy with QCD we will call the contribution that scales like $1/N^{h}$ the genus $h$ term, recognizing that the attribution of a genus is formal: there is no known   topological meaning to a \dk \ diagram. That of course could change.

\subsection{Correspondence with QCD} %\label{sub:2.1}

Recall  that in the  't~Hooft limit the perturbation expansion of pure QCD (without quarks) can be organized into an expansion in inverse powers of $\N^2$ as in   \eqref{ymNexp1}.
  The similarity between equations \eqref{C(T)} and \eqref{ymNexp1} is obvious and is the basis for much of what follows. 
  
  Comparing \eqref{C(T)} and \eqref{ymNexp1}
   we see that they have the same form with the following correspondences,
\bea
\N^2 	&\leftrightarrow & N, \cr \cr
\alpha 	& \leftrightarrow & p. 
\label{corres}
\eea
That the SYK $N$ corresponds to $\N^2$ is to be expected; both represent the number of degrees of freedom: For example the entropy of \dk \ is of order $N$ while the entropy of a QCD plasma is order $\N^2.$ 
That $\alpha$ and $p$ are corresponding quantities is less obvious but  follows from the forms of the expansions \eqref{C(T)} and \eqref{ymNexp1}. 

Double-scaled SYK is defined by taking the $N\to \infty$ (and $p\to\infty$)
limit with the ratio
\be
\lambda\equiv \frac{2p^2}{N} 
\label{lambda}
\ee
kept fixed. From the definition of $\alpha$ and $\lambda$, it follows 
\be
\g^4   \  \leftrightarrow \ \lambda.
\label{morecorr}
\ee
With this correspondence the  discussion in Section~\ref{S: fixedg} goes through for the fixed $\lambda$ limit with $\g$ replaced by $\lambda^{1/4}$.

\subsection{Two Limits}
\begin{figure}[H]
\begin{center}
\includegraphics[scale=.5]{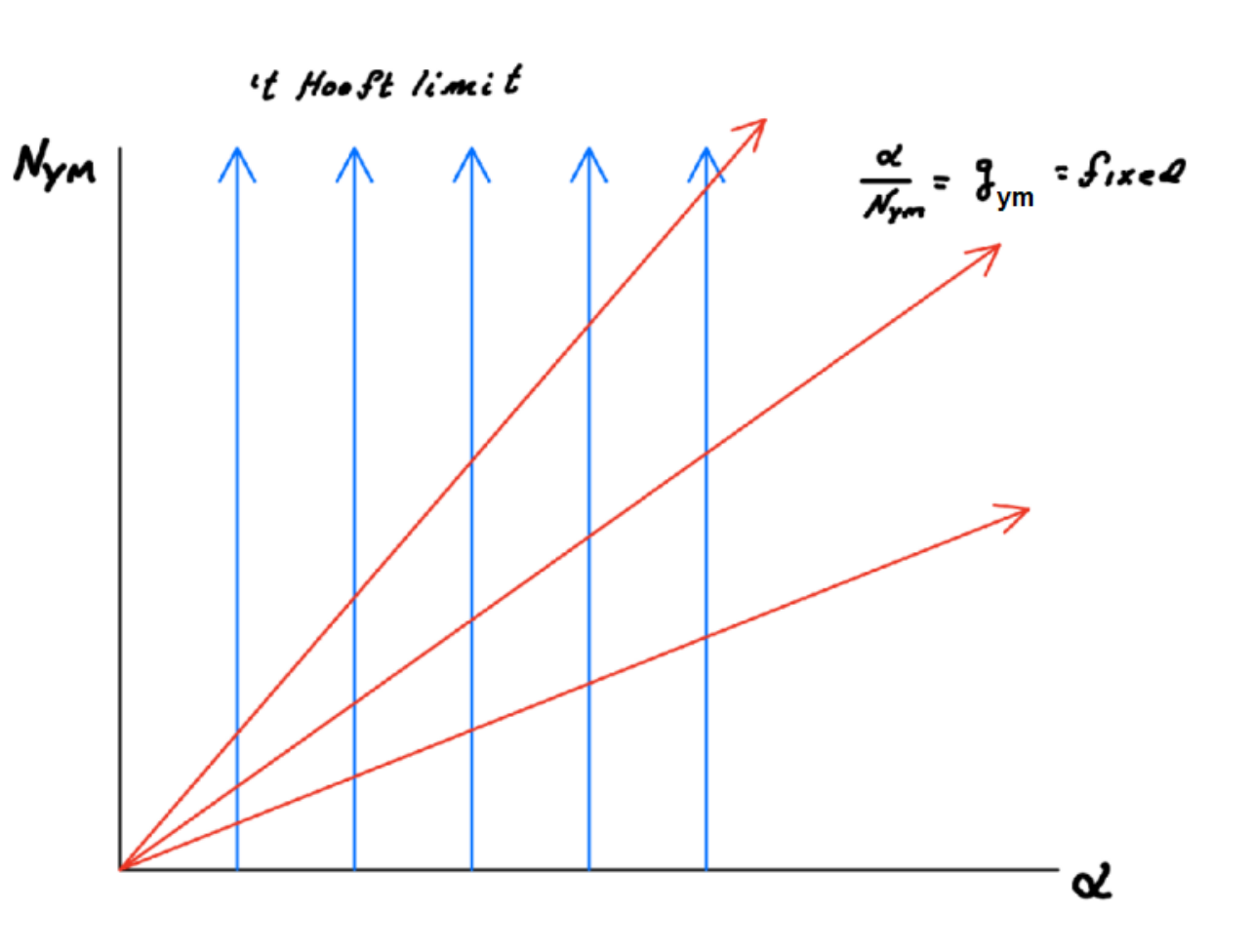}
\caption{The two types of large $\N$ limits in QCD.}
\label{hooffig}
\end{center}
\end{figure}

As we explained earlier there are two types of large $\N$ limits in QCD, the 't~Hooft limit,
\bea
\N &\to& \infty \cr \cr
\alpha &\equiv& \g^2 \N \ \ \ \  \text{fixed}
\label{tooftlim}
\eea
and the fixed $\g^2$ (aka, flat-space) limit,
\bea
\N &\to& \infty \cr \cr
 \g^2  &&   \text{fixed}
\label{fixedg}
\eea
Figure \ref{hooffig} shows a plot of $\N$ versus $\alpha$ and the two types of large $\N$ limits: the 't~Hooft limit of fixed $\alpha$ and the  fixed $\g$ limit. 

\begin{figure}[H]
\begin{center}
\includegraphics[scale=.4]{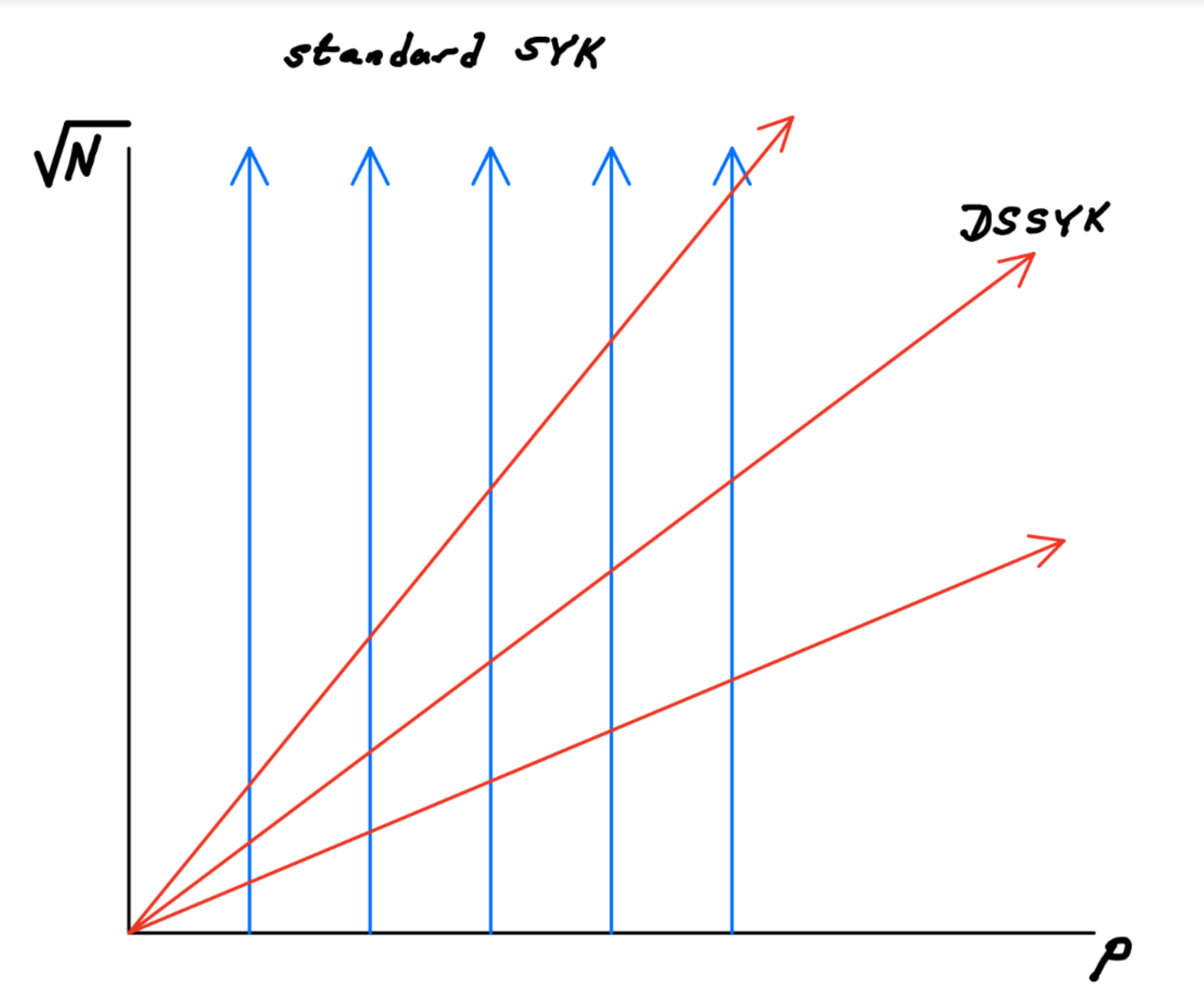}
\caption{The two types of large $N$ limits in SYK.}
\label{dkfig}
\end{center}
\end{figure}

Similarly, Figure~\ref{dkfig} shows a plot of $p$ versus $\sqrt{N}$ for SYK.  
Moving vertically upward in Figure~\ref{dkfig}  defines the conventional SYK limits for various values of $p.$ Moving along the red radial lines defines the double-scaled limit of fixed $\lambda.$
The reason  for this close correspondence  is not at all obvious but as we move ahead we will see more fascinating aspects of it.

\subsection{Sub-dS Locality}

The 't~Hooft limit corresponds to the standard SYK limit $N\to \infty$ with $p$ held fixed. The original SYK theory sets $p=4$ and is analogous to the large  $\N$ limit with small 't~Hooft coupling $\alpha \sim 1.$ The large but fixed $p$ limit corresponds to  the strongly coupled 't~Hooft limit with  $\alpha>>1.$   

For the conjectured de Sitter dual the fixed $p$ theory is a limit in which the string scale is finite in units of  the de Sitter radius, the ratio of the two being~\cite{Susskind:2022bia}
\be
 \f{\ell_{\rm string}}{\ell_{\rm ds}}\sim \f{1}{p}.
 \ee 
We should note that by using the term string scale, we do not imply there is string-like object in the theory. The reason we call it string scale is that it plays the same role as the string scale in QCD.  It provides an emergent IR cutoff in SYK, as string tension does in QCD. 

The analog of the fixed $\g$ limit is the double-scaled limit of fixed $\lambda$ in which the string scale tends to zero in units of the de Sitter radius. It is the limit of  sub-dS locality.  Alternatively the string scale stays fixed as the de Sitter radius goes to infinity.

 \subsection{Does the Fixed $\lambda$ Limit Really Exist?}\label{S: fixlam?}
 
 The fixed $\g$ limit of Yang Mills theory does not necessarily exist. The 't~Hooft limit does exist, at least for small 't~Hooft coupling but in general there is no reason to expect that the limit of large $\alpha,$ let alone the limit in which $\alpha$ grows with $\N,$ exists. Supersymmetry can play a role in insuring analyticity in $\alpha$ but without supersymmetry we lose control at strong coupling.  A phase transition or other singularity can obstruct the continuation to strong coupling.

 What about the analogous large $p,$ or even the double-scaled theory in which $p\to \infty$ with $N$? 
Each term in the expansion in \eqref{C(T)} can be obtained as sum of primitive  graphs and their decorations, analogous to  \eqref{fgsum3}, and is written 
in the form\footnote{%
As in the QCD case, $\CF^{(h)}(p)$ is defined by 
$\CF^{(h)}(p)=\tilde{\CF}^{(h)}(p)/p^{2h}$ from $\tilde{\CF}^{(h)}(p)$
that appeared in \eqref{C(T)}.}
\be 
A= \sum_{h=0}^{\infty} \lambda^{h} \CF^{(h)}(p).
\label{sum}
\ee
Following the arguments in Section~\ref{S: fixedg}, we see
that the requirement for a fixed $\lambda$ limit 
is the same as for the fixed $\g$ limit in Yang Mills theory:
the functions $\CF^{(h)}(p)$ should have limits as $p\to \infty.$ In the \dk \ case the answer is believed to be that the double-scaled  limit does exist for a range of $\lambda$ including $\lambda=0.$  (See for example \cite{Cotler:2016fpe}\cite{Berkooz:2018jqr}.)  Whether or not that range includes arbitrarily large $\lambda$ is not known, at least to the authors.

\section{ Perturbation Theory in DSSYK$_{\infty}$} \label{S:PTDK }

In this section we consider the perturbative and $1/N$ expansions of \dk \ correlation functions.

\subsection{Time Scales}\label{S:tscales}

 As explained in \cite{susskind:confined} there are two well-separated time scales in \dk \ as well as in the semiclassical limit of $(2+1)$-dimensional de Sitter space: the cosmic and  string scales. Time in cosmic units ($\t$)   is measured relative to the de Sitter radius. String scale time runs faster than cosmic time by a factor of $p.$
\be 
t_{\rm string} = p \  \t.
\label{ts=ptc}
\ee

As an example of the importance of these two time scales we can consider the correlation function of two cord (matter chord) operators which consists of $p$ fermions, and compare it with the single fermion correlation function. 

For $p\to \infty$ (to be more precise, in the $N\to \infty$ limit with fixed $p$, followed by $p\to\infty$, which amounts to the double-scaled limit with $\lambda=0$) the matter chord correlation function in string units is~\cite{Maldacena:2016hyu},\cite{Roberts:2018mnp},
\be 
\text{cord correlator} =   \f{1}{\cosh^2({\CJ t_{\rm string}})}.
\label{cordcor1}
\ee
In string units it decays at a finite rate at large time.
By contrast, in cosmic units it is
\be 
\text{cord correlator} = \f{1}{\cosh^2({\CJ p  \ \t})},
\label{cordcor2}
\ee
and decays infinitely rapidly when $p\to \infty.$

The single fermion correlation function for $\lambda=0$ is given by
the $p$-th root of the cord correlation function, since the cord
correlator factorizes in this limit. In cosmic units it is given 
by
\be 
\text{fermion correlator} = 
\lim_{p\to\infty}
\lf  \f{1}{\cosh^2({\CJ p \ \t})} \rg^{1/p}   =     e^{-2\CJ \t}.
\label{fercor1}
\ee
In cosmic units it has a finite decay rate
while in string units it decays infinitely slowly,
\be 
\text{fermion correlator} \sim  e^{-\f{2\CJ}{p}t_{\rm string}}.
\label{fercor2}
\ee
There is no single choice of units in which both correlation functions vary at a finite rate.
 
 For our purposes in this paper cosmic units are most appropriate. If we were discussing  the large $N$ expansion for cord operators string units would be  more useful.
 
From this point forward we will simplify the notation by denoting cosmic time by $t,$
 \be 
 \t \equiv t.
 \ee

\subsection{DSSYK$_{\infty}$}
SYK is a theory of $N$ real anticommuting fermionic degrees of freedom $\psi_i$ coupled through an all-to-all $p$-local Hamiltonian,
\be  
H_{\rm cosmic} = i^{p/2}\sum_{i_1<i_2...<i_p}J^{i_1,i_2,...,i_p} \ \psi_{i_1}\psi_{i_2}...\psi_{i_p}.
\label{Hcosmic}
\ee
The coupling constants $J^{i_1,i_2,...,i_p}$ are drawn randomly and independently from a Gaussian distribution with variance satisfying,
\be  
\la J^2 \ra =\frac{\CJ^2 }{2} \frac{p!}{N^{p-1}}.
\label{Jvariance}
\ee

As explained above we  work in cosmic units
 \cite{susskind:confined}\cite{Rahman:2024vyg}.
The choice of units enters only  into the expression for the variance of the couplings \eqref{Jvariance}.
Readers who are familiar with the SYK literature may find \eqref{Jvariance} unfamiliar. The usual normalization of the variance is \cite{Maldacena:2016hyu}  
\be 
\la J^2 \ra_{\rm usual}= \frac{1}{p^2}{\CJ^2\over 2} \frac{p!}{N^{p-1}},
\label{usualvar}
\ee
corresponding to a Hamiltonian,
$$H_{\rm string} = \frac{1}{p} H_{\rm cosmic}.$$  
$H_{\rm string}$ is
normalized so that energies\footnote{%
What we mean by energy here is really the decay rate that appeared in the previous subsection. It should be also true for energies.} 
of cords remain  finite as $N\to \infty$ and the energy of single fermions vanish like $1/p.$ The cosmic Hamiltonian is normalized so that the energy of single fermions is finite and of order $\CJ$ while cord energies diverge like $p\CJ.$ 
The cosmic and string Hamiltonians are generators of time translation in cosmic and string units,
\bea 
H_{\rm cosmic} &=&i \f{\partial}{\partial \t}, \cr \cr
H_{\rm string} &=&  i  \f{\partial}{\partial t_{\rm string}}.
\label{Hs}
\eea

\bn

Double-scaled SYK is defined by taking the limit $N\to \infty$ with the 
parameter $\lambda$ defined in \eqref{lambda} kept fixed. 
To define \dk there is one more  condition: the temperature in the Boltzmann distribution is taken to be infinite \cite{Rahman:2024vyg}. This means that the density matrix of the static patch is proportional to the identity matrix and is maximally mixed. It follows that expectation values are simply traces. For any operator $\CO$ in \dk 
 \be 
 \la \CO \ra = \text{tr} \ \CO,
 \ee
where $\text{tr} $ denotes the normalized trace such that $\text{tr}  1 = 1.$
 
 \subsection{DSSYK$_{\infty}$ Perturbation Theory}
We turn now to the calculation of correlation functions,
 \be 
A(T) = \text{tr} \big( \psi(0) \psi(T) \big)
=  \text{tr} \left( \psi e^{iHT} \psi e^{-iHT}\right)
\label{CT}
 \ee
 where $T$ is a positive time\footnote{We note again that in this context $T$ does not denote temperature.} measured in cosmic units.   
We use $A(T) $ for definiteness; there is nothing special about it. The conclusions we draw are general and apply to all correlation functions involving a fixed finite number of fermion operators.

There are two ways of constructing the \dk \ perturbation expansion.

The first is based on the Hamiltonian formalism. We expand the exponentials in \eqref{CT} and collect terms of a given order in $\CJ.$ Since $H$ is proportional to $\CJ$ this means collecting terms of a given order in $H.$ It of course is a dangerous thing to do:  we are attempting to expand in the entire Hamiltonian, and doing so will lead to infrared divergences which will get worse with each order. But QCD is also IR-divergent\footnote{The IR divergences in 4-dimensional QCD are logarithmic while in \dk \ they are power law.}. In both cases summing appropriate sets of diagrams will lead to emergent time scales which regulate  the divergences. For now we will just study and compare the formal expansions. 
 
The second method, the Lagrangian path integral method, is used in the following and in Appendix~A. 
The Lagrangian is,
\be 
\CL = -\f{i}{2}\sum_i \psi_i {\dot{\psi_i}} - \  i^{p/2}\sum_{i_1<i_2...i_p}J^{i_1,i_2,...,i_p} \ \psi_{i_1}\psi_{i_2}...\psi_{i_p} 
\ee
and the perturbation expansion is derived by the usual method of Schwinger and Keldysh in which the insertions of the interaction vertex ($H$) occur at definite times between $t=0$ and $t=T.$ The times are then integrated over.
The two methods are of course equivalent.

\begin{figure}[H]
\begin{center}
\includegraphics[scale=.4]{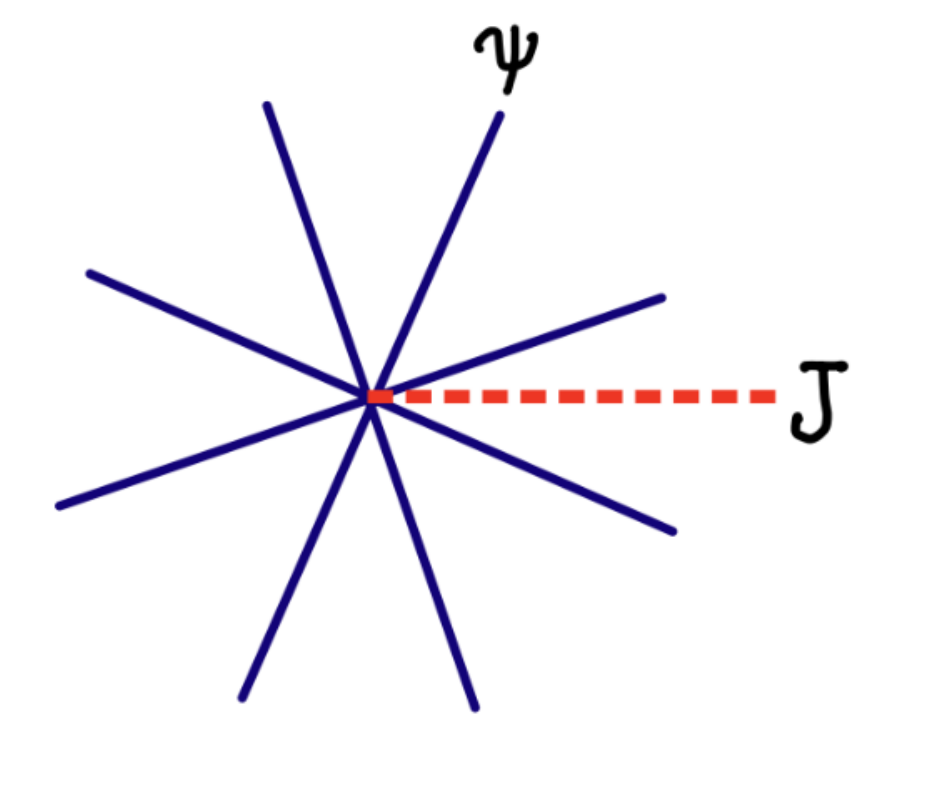}
\caption{Interaction vertex (for $p=8$).}
\label{vertex}
\end{center}
\end{figure}

The vertex   associated with the perturbation involves $p$ fermions as shown in fig \ref{vertex}.
\bn
It has numerical weight $\CJ.$
There is an extra dotted red line coming out of the vertex which represents the coupling constants $J^{i_1,i_2,...,i_p}.$ Ordinarily we would treat the   coupling constants as fixed c-numbers but in SYK we integrate over them in defining the ensemble average. It is convenient to treat them as fields but with no kinetic term.

\begin{figure}[H]
\begin{center}
\includegraphics[scale=.4]{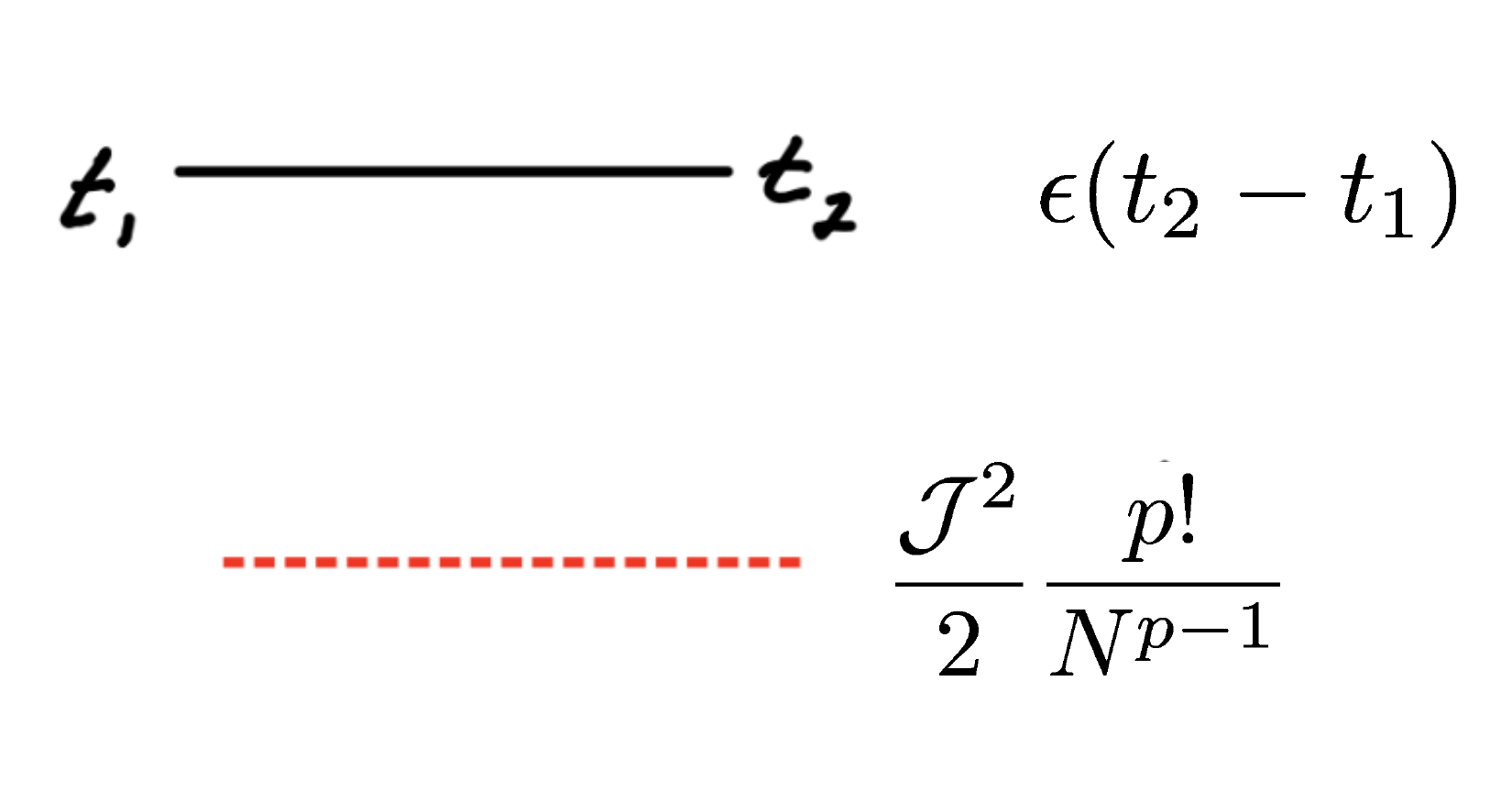}
\caption{Propagators for fermions and for coupling constants.}
\label{propagators}
\end{center}
\end{figure}

The propagators and their values are shown in fig \ref{propagators}. The symbol $\epsilon(t)$ is the sign function, $+1$ for $t>0$ and $-1$ for $t<0.$

\bn
Notice that the $JJ$ propagator (dashed red line) is not labeled with a time coordinate.

We introduce a contour parameterized by a  coordinate $s$ which runs from $0$ to $2T.$ The portion of the contour from $0$ to $T$ represents the propagation by $e^{-iHT}.$ The second half of the contour from $T$ to $2T$ represents the backward propagation by $e^{+iHT}$. At $s=0$ and $s=T$ fermion operators $\psi_i$ are inserted. As usual the ensemble average is carried out.
Odd powers of $H$ vanish after ensemble averaging.

\begin{figure}[H]
\begin{center}
\includegraphics[scale=.5]{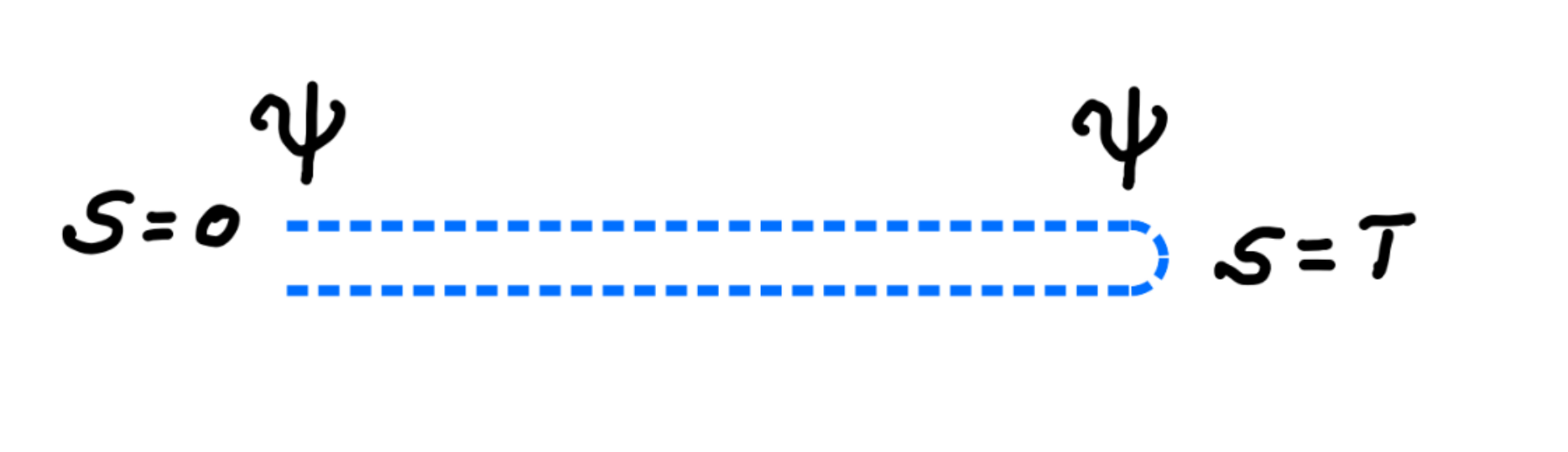}
\caption{Schwinger-Keldysh contour.}
\label{contour1}
\end{center}
\end{figure}

\begin{figure}[H]
\begin{center}
\includegraphics[scale=.5]{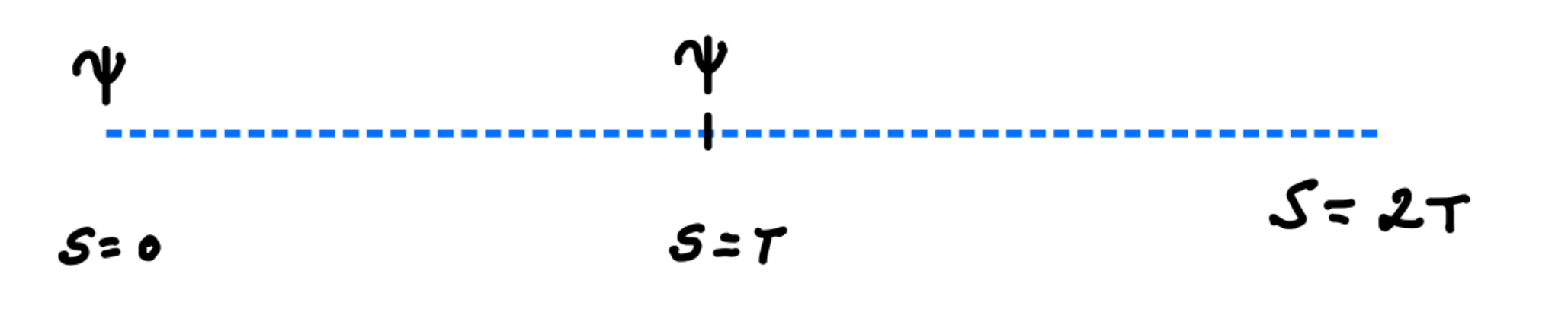}
\caption{Schwinger-Keldysh contour represented as a single straight line.}
\label{contour2}
\end{center}
\end{figure}

The contour and  $\psi$ insertions are shown in Figures \ref{contour1} and \ref{contour2}.
In Figure~\ref{contour1} the contour is shown as returning to the origin. 

In Figure~\ref{contour2} the same contour is represented as going from $s=0$ to $s=2T.$ One accounts for the backward propagation from $s=T$ to $s=2T$ by reversing the sign of the Hamiltonian on the second half of the contour.

The first contribution to the correlation function $A(T)$ is the order zero term with no vertex insertions of $H.$ The diagram representing this contribution is simply the bare fermion
 propagator in Figure~\ref{propagators}. For $T>0$ its value is $1,$ 
\be \boxed{
A_0(T) =1.
}
\label{C0}
\ee 

 The  contributions involving a single vertex insertion, either on the first half of the contour or the second half, as well as all odd orders, vanish upon ensemble averaging, $$A_{\rm odd}(T)=0.$$
 
\subsection*{Order $\CJ^2$}

 The second order contribution  $A_2(T)$ involves two vertex insertions. In the Lagrangian formulation  the vertices occur at  definite values of $s_1$ and $s_2$ and we integrate  $s_1$ and $s_2.$ There are three cases:
 \begin{description}
 \item[Case 1:] 
both vertices inserted in the first half of the contour---$0<s<T.$
 \item[Case 2:] one insertion in the range $0<s<T,$ and another insertion in the range
 $T<s<2T.$
 \item[Case 3:]  both insertions in the range $T<s<2T.$
\end{description}  

The fermions in the vertices and those in the  initial and final $\psi$ insertions must be contracted in all possible ways\footnote{%
There is a disconnected term in which the initial and final fermions are contracted but we can ignore this term.}. 
We will consider all three cases in detail. (In Appendix~A, we will organize the terms in a slightly different manner, by first summing over the contribution from the forward and backward evolving parts of the time contour for a given physical time of the vertex.)

\begin{figure}[H]
\begin{center}
\includegraphics[scale=.5]{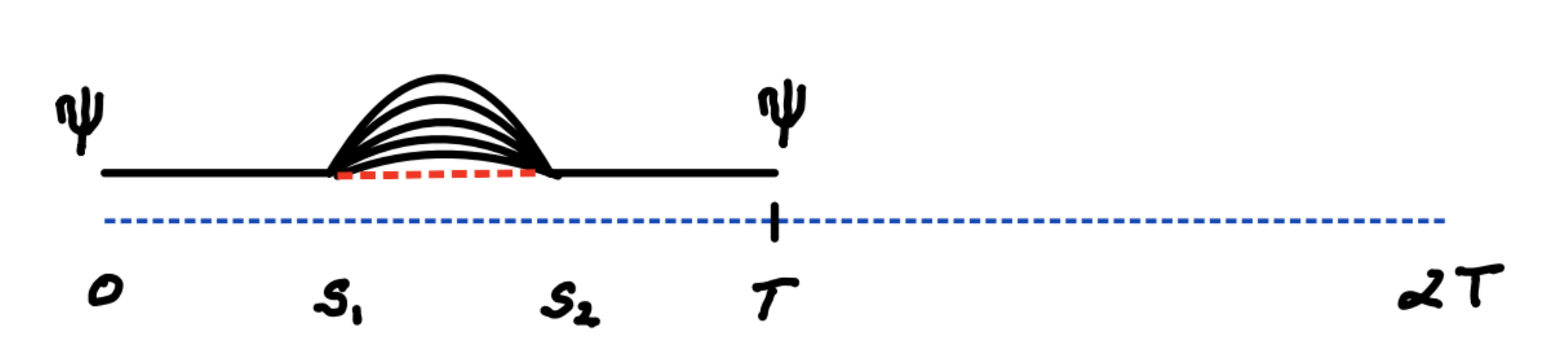}
\caption{A possible contraction for case 1.}
\label{case1a}
\end{center}
\end{figure}

\begin{figure}[H]
\begin{center}
\includegraphics[scale=.4]{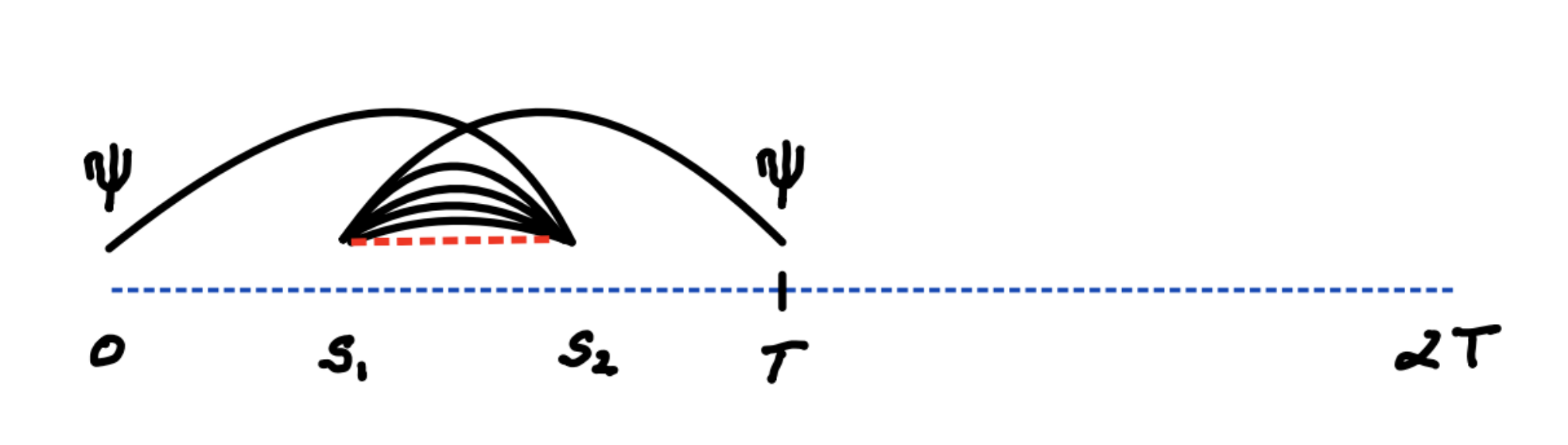}
\caption{Another possible contraction for case 1.}
\label{case1b}
\end{center}
\end{figure}

For case 1, the possible contractions are shown in Figures \ref{case1a}, \ref{case1b}.
%%%
Because of the trivial nature of the propagators the values of diagrams are actually independent of $s_1, \ s_2$, and the integration in the range
$0<s_1<s_2<T$ just gives a factor of $T^2/2$ in each case. (In the Hamiltonian formulation the same $T$-dependence just comes from expanding $e^{iHT}$ to second order.) But the two diagrams in Figures \ref{case1a} and \ref{case1b} exactly cancel due to the minus sign caused by the crossed fermion lines in Figure~\ref{case1b}.

\begin{figure}[H]
\begin{center}
\includegraphics[scale=.4]{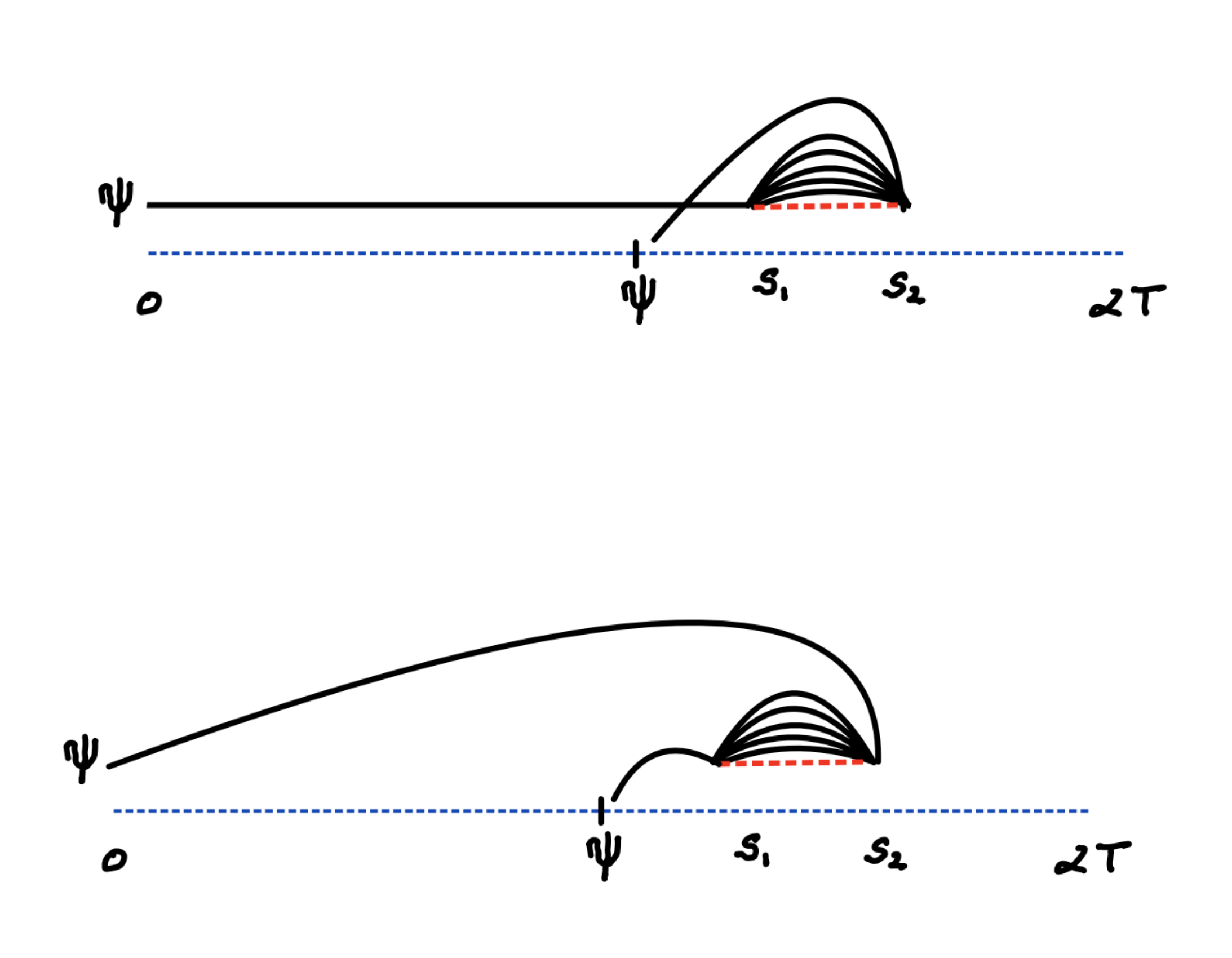}
\caption{Contractions for case 3.}
\label{case3}
\end{center}
\end{figure}

Case 3 is very similar to case 1  and consists of the two canceling diagrams in Figure~\ref{case3}.

\begin{figure}[H]
\begin{center}
\includegraphics[scale=.5]{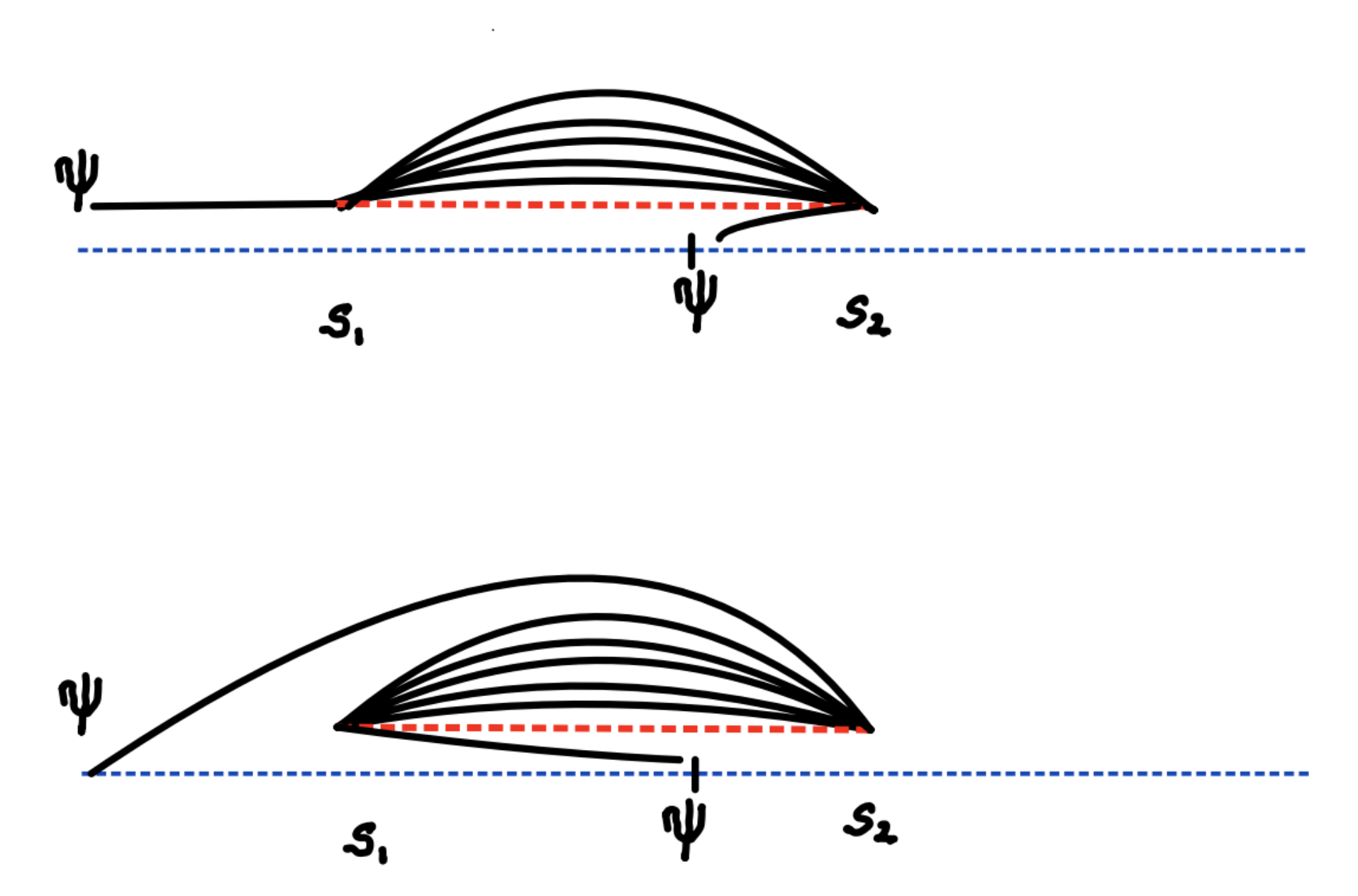}
\caption{Contractions for case 2.}
\label{case2}
\end{center}
\end{figure}

Case 2 shown in Figure~\ref{case2} also consists of two diagrams but this time they add. There is an overall minus sign due to the fact that the insertions are on opposite halves of the contour.
The time integrals for each diagram give a factor of $T^2$. 
Although we are studying the lowest order perturbation here,
let us introduce the symbols that will be used in the more
general context. One of the time coordinates to be integrated 
is the ``center of mass'' time of the vertices. 
We will call the result of its integration $T_{\rm c}$. 
(In fact, the integrand does not depend on this coordinate, 
so this is equal to the time separation $T$ of the external 
operators.) The other time coordinates (in the present case 
there is only one) represent relative times of the vertices. 
We will denote the contribution from each of them by the same 
symbol $T_{\rm r}$; for wee-irreducible diagrams 
(defined at the beginning of Section~4.3), they are of 
the same order $T_{\rm r}\sim 1/p\CJ$, as we will explain below. 

Taking into account the cancellation explained above, 
altogether we have 
$$-2T^2=-2 T_{\rm r}T_{\rm c}.$$ 
In addition, there is a multiplicative factor that depends of $\CJ,$ $p,$ and $N.$ We will call such factors the ``numerical coefficients." 
They
consist of several factors. 
First of all, for each dashed red line (Figure~\ref{vertex}) there is a 
$JJ$ propagator shown in Figure~\ref{propagators}, which gives 
$${\CJ^2\over 2}\frac{p!}{N^{p-1}}.$$
Next there is a combinatoric factor 
 which counts the number of ways the indices of the internal fermion lines can be chosen. These indices must not coincide with the ones for the external lines (which must be the same for the two external lines). Therefore there are $$\frac{(N-1)!}{(p-1)!(N-p)!}$$ ways of choosing the internal indices. For $N\gg 1$ and $N\gg p$ this tends to $$\frac{N^{p-1}}{(p-1)!}\,.$$

The overall numerical coefficient for the diagrams in Figures \ref{case1a}, \ref{case1b} and \ref{case2} are all the same and are given by,
\be  
\frac{\CJ^2p!}{N^{p-1}} \cdot \frac{N^{p-1}}{(p-1)!} = \CJ^2 p.
\ee
Combining this with the factor $-2T^2=-2 T_{\rm r}T_{\rm c}$ we find the order $\CJ^2$ contribution to $A(T)$  is,
\be 
\boxed{
A_2(T)= -2p\CJ^2 T^2 = -2p\CJ^2 T_{\rm r}T_{\rm c}.
}
 \label{C2}
 \ee
Note that the expression in \eqref{C2} is independent of $N$ and is therefore genus zero.
 
\begin{figure}[H]
\begin{center}
\includegraphics[scale=.8]{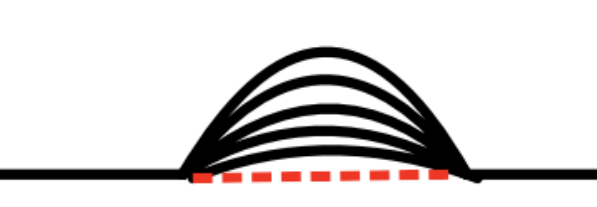}
\caption{The $A_2$ diagram, which has ``genus'' $h=0$.}
\label{top}
\end{center}
\end{figure}

 All of the diagrams for the three cases have a common ``topology" illustrated in Figure~\ref{top}.
\bn
We  refer to this  as the $A_2$ topology and the insertion of $A_2$ into a line in a diagram as a $A_2$ insertion. Summing $A_2$ insertions will play a major role in Section~\ref{S:decratingpd}.

We refer to the terms which scale as $1/N^{h}$ as having 
``genus'' $h$, with an understanding that this assignment 
is purely formal. With this definition of genus, the diagram 
$A_2$ in Figure~\ref{top} has $h=0$. 

\subsection*{Order $\CJ^4$}
 
\begin{figure}[H]
\begin{center}
\includegraphics[scale=.5]{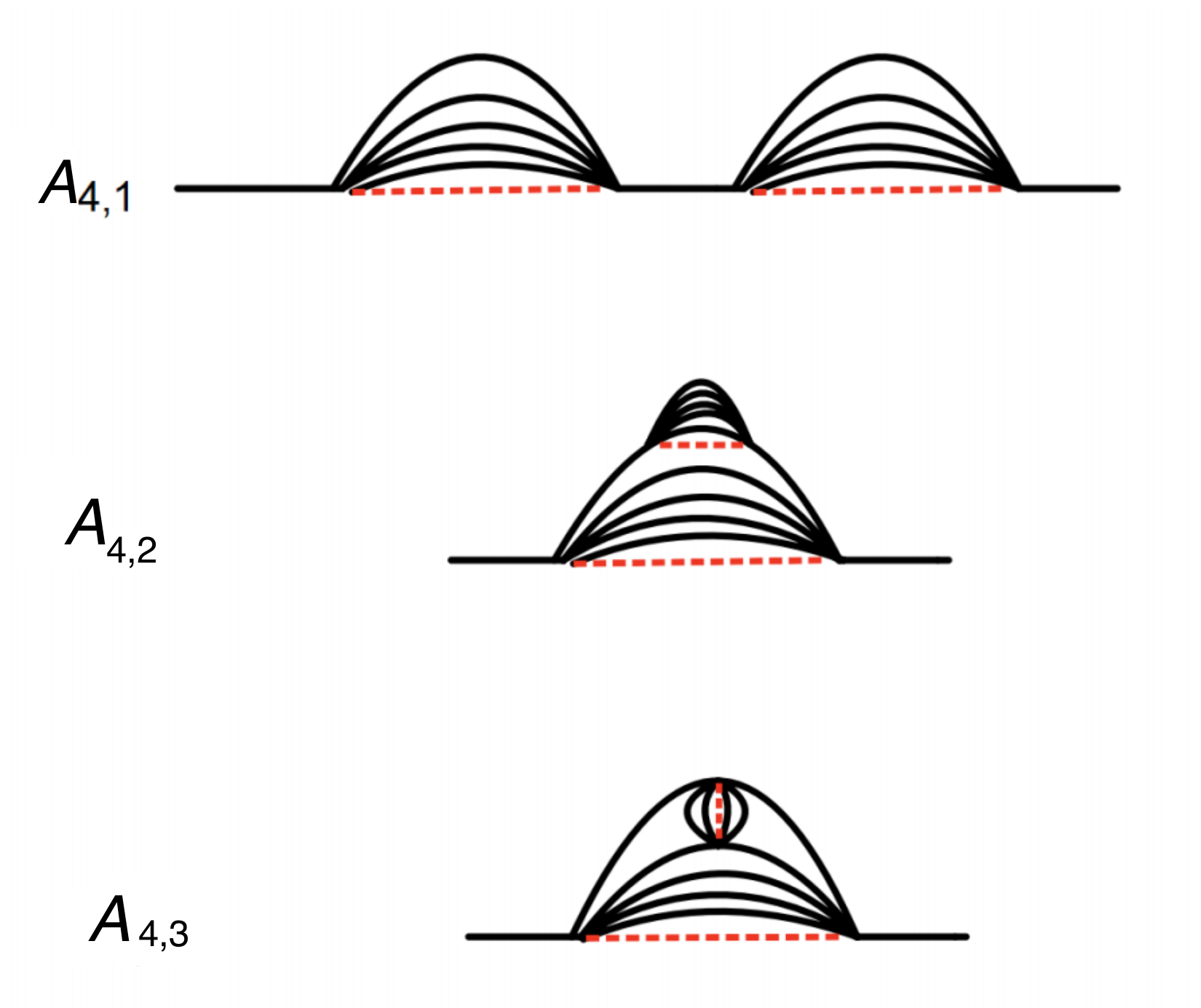}
\caption{Possible diagrams at order $\CJ^4$.}
\label{F4}
\end{center}
\end{figure}

Next let us consider the order $\CJ^4$ contribution to $A(T).$ There are several ``topologies'' to consider shown in Figure~\ref{F4}.

The diagram $A_{4,1}$ is simply a sequence of $A_2$-insertions, and is not one-particle irreducible (1PI). In our discussion, we will focus on the 1PI diagrams. In Appendix~A, we will explain that the sum over this kind of sequences of $A_2$-insertions exponentiates (for $\lambda=0$).

The diagram $A_{4,2}$ can be described as an $A_2$-insertion into an existing
 $A_2$ diagram. It has two dotted red propagators and two counting factors representing the number of ways of choosing $(p-1)$ fermionic indices in the interiors of the melons. Finally there is a factor of $(p-1)$ representing the fact that the melon can be inserted on any of $(p-1)$ fermion lines. Thus,
\be 
A_{4,2} \sim (p-1) \CJ^4 \lf \frac{p!}{N^{p-1}} \rg^2 \cdot \lf \frac{N^{p-1}}{(p-1)!}\rg^2  
T_{\rm r}^3 T_{\rm c}
\label{C42a}
\ee
The final factor of $T^4=T_{\rm r}^3 T_{\rm c}$ is required for dimensional consistency. 
It arises from the integration over the times of the four vertices. Alternatively it arises from expending the exponentials $e^{\pm iHT}.$

\begin{figure}[H]
\begin{center}
\includegraphics[scale=.5]{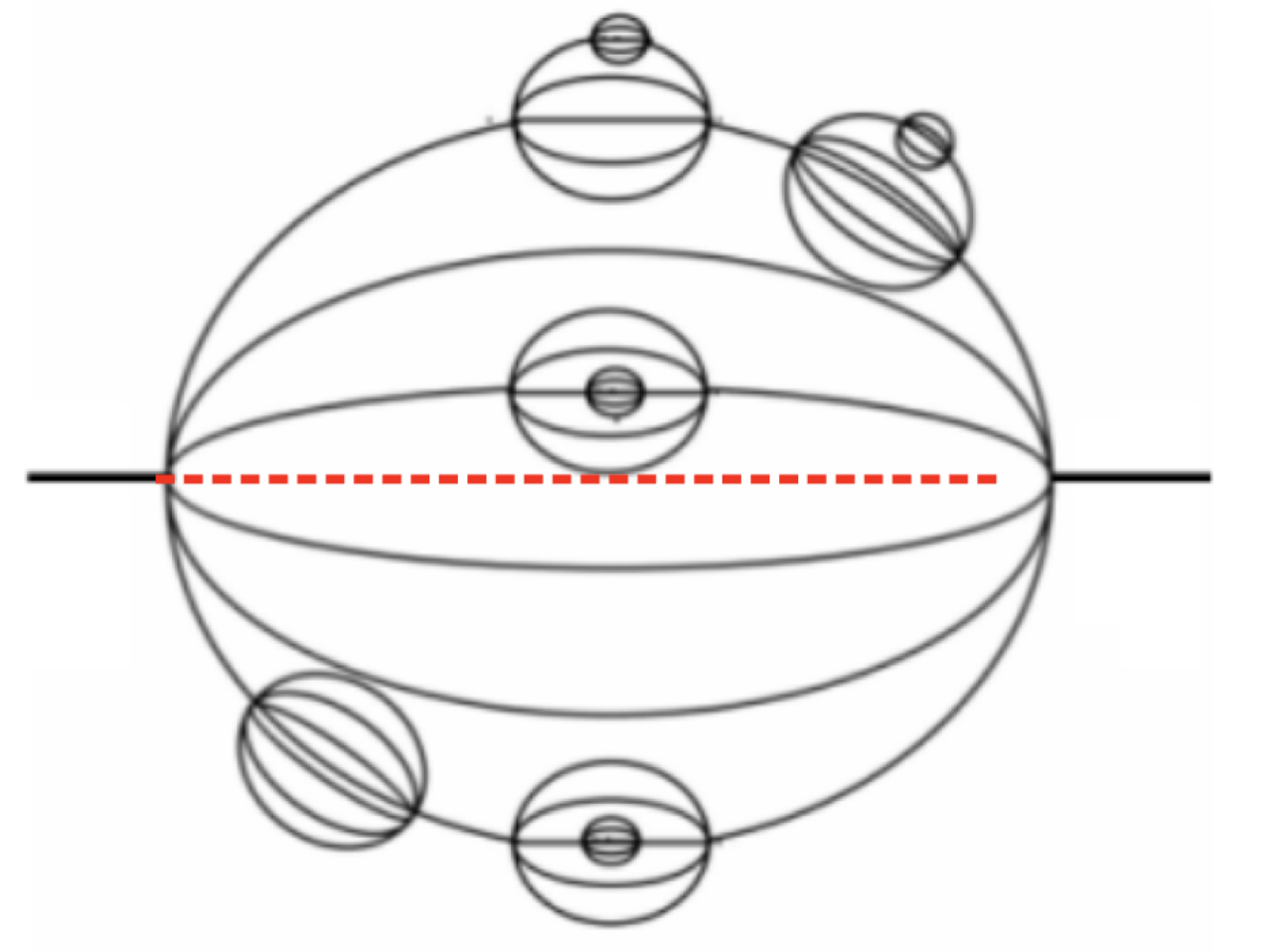}
\caption{A typical melon diagram.}
\label{cartoon}
\end{center}
\end{figure}

Combining the factors gives 
\be 
\boxed{
A_{4,2} \sim  \CJ^4(p-1) p^2 
T_{\rm r}^3 T_{\rm c}
\approx  \CJ^4 p^3 
T_{\rm r}^3 T_{\rm c}
\ \ \ \ \ \text{for} \ p \to \infty.
}
\label{C42b}
\ee
We see an example of a more general phenomenon: 
an $A_2$ insertion has no effect on the $N$ scaling of a diagram but increases the power of $p;$ in other words inserting $A_2$ does not change the genus.   When applied repeatedly to the basic primitive diagram (the free propagator in Figure~\ref{propagators}) they define the so-called melon diagrams and their sum supplements the  primitive diagram  to define the genus zero contribution.
It consists of Figure~\ref{propagators} and all melon insertions, melons within melons, etc. A typical example is shown in Figure~\ref{cartoon}.
The melonic diagrams for \dk \ can be summed by means of Schwinger-Dyson equations \cite{Maldacena:2016hyu}\cite{Roberts:2018mnp}. The result for general $p$ has not been explicitly given, but the limit for large $p$ is known, 
\be 
\CF^{(0)}(T)= \lim_{p\to \infty} \lf \f{1}{\cosh^2{p\CJ T}} \rg^{1/p} = e^{-2\CJ T}.
\label{cosh}
\ee
Thus we see that the $p\to\infty$ limit exists for the genus zero correlation
function.

Equation \eqref{cosh} has far-reaching implications. Perturbatively the only scale in \dk \ is $\CJ$ which enters into diagrams in a trivial way; each diagram has a power of $\CJ$ equal to the number of interaction vertices. But the non-perturbative effects of summing all melon diagrams in Figure~\ref{cartoon} is to {\it generate a new scale}, $\ell_{\rm string} = 1/p\CJ$. The emergent scale serves to cut off infrared divergences in a manner analogous to the way the formation of flux-tubes leads to confinement and cuts off IR divergences in QCD. It is for that reason that we refer to the emergent scale as the string scale. We will discuss this analogy further in Section~\ref{S: speclat}.

\subsection{Higher Genus}\label{S: highgen}

So far in all contributions to $A$ the explicit  powers of $N$ have canceled. In this respect $A_{4,3}$ is different. It contains: two dotted red propagators; four powers of time, $T_{\rm r}^3 T_{\rm c}$;
a factor $(p-1)(p-2)/2$ representing the choice of internal lines which are connected by the small melon: and finally counting factor for the indices,
$$
\frac{(N-2)^{p-2}}{(p-2)!}\cdot  \frac{(N-1)^{p-1}}{(p-1)!}.$$ 
Combining these factors we find (for large $p$),  
\be
A_{4,3}\sim  \frac{p^3(p-1)(p-2)}{N}\CJ^4 T_{\rm r}^3 T_{\rm c}
\approx \frac{p^5}{N}\CJ^4T_{\rm r}^3 T_{\rm c} .
\label{C43}
\ee
$A_{4,3} $  is an example of a $1/N$  (or genus one) term in the expansion \eqref{C(T)}. It is the first contribution to $\CF^{(1)}.$  From these examples one sees the pattern of \eqref{C(T)} emerging.

\begin{figure}[H]
\begin{center}
\includegraphics[scale=.5]{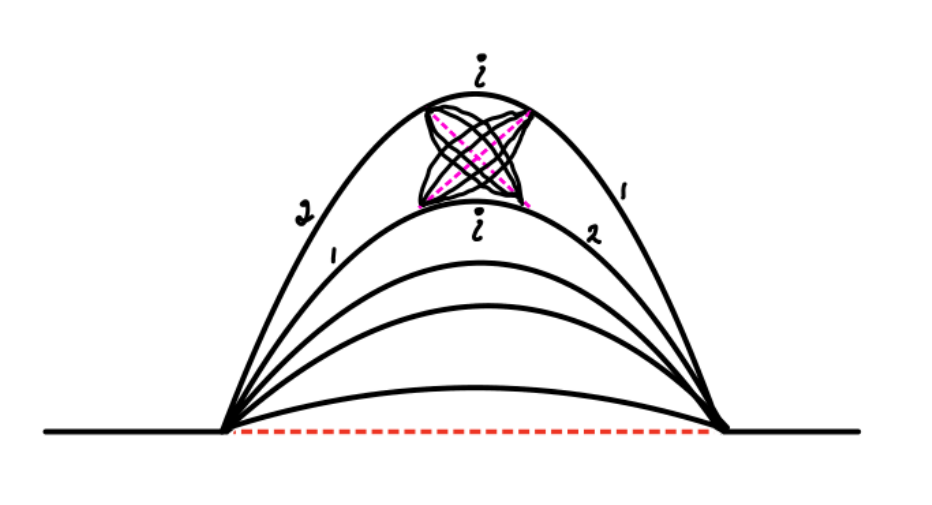}
\caption{``Crossed melon'' diagram.}
\label{crossed1}
\end{center}
\end{figure}

There is an infinite series of diagrams all of which scale the same way as 
eq.~\eqref{C43} that could be characterized as ``crossed melons.'' 
One of them is illustrated in Figure~\ref{crossed1}. This diagram
is obtained by adding a melon to diagram $A_{4,3}$. 
Off hand, it may seem that this diagram
is suppressed relative to $A_{4,3}$, since adding a melon which 
connects two different lines usually cost a factor of $1/N$ (as
is the case when making $A_{4,3}$ from $A_{2}$ diagram). 
However, one can see from the diagram in Figure~\ref{crossed1}
that the internal lines 
labeled $i$ can take on any value (other than 1 and 2), therefore summing over $i$ produces a factor of $(N-2)$. This compensates the factor of $1/N$, making this diagram of the same order as $A_{4,3}$. There are diagrams with multiple crossings generalizing this pattern. Summing up the series of these diagrams produces a numerical factor. (This type of diagrams are discussed also at the end of 
Appendix~A.3.)

\begin{figure}[H]
\begin{center}
\includegraphics[scale=.5]{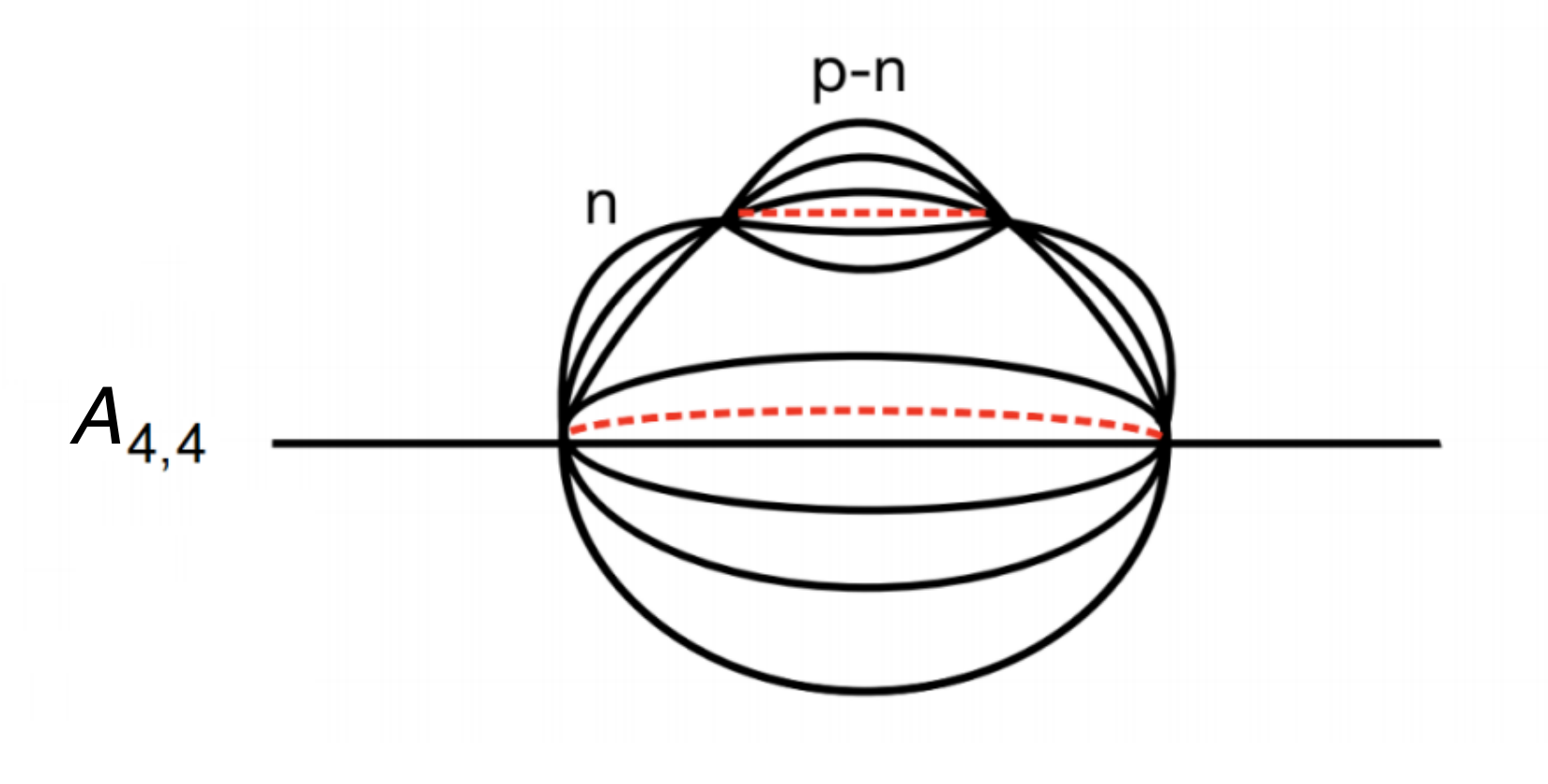}
\caption{An Example of genus $n-1$ diagram.}
\label{F44}
\end{center}
\end{figure}

\begin{figure}[H]
\begin{center}
\includegraphics[scale=.2]{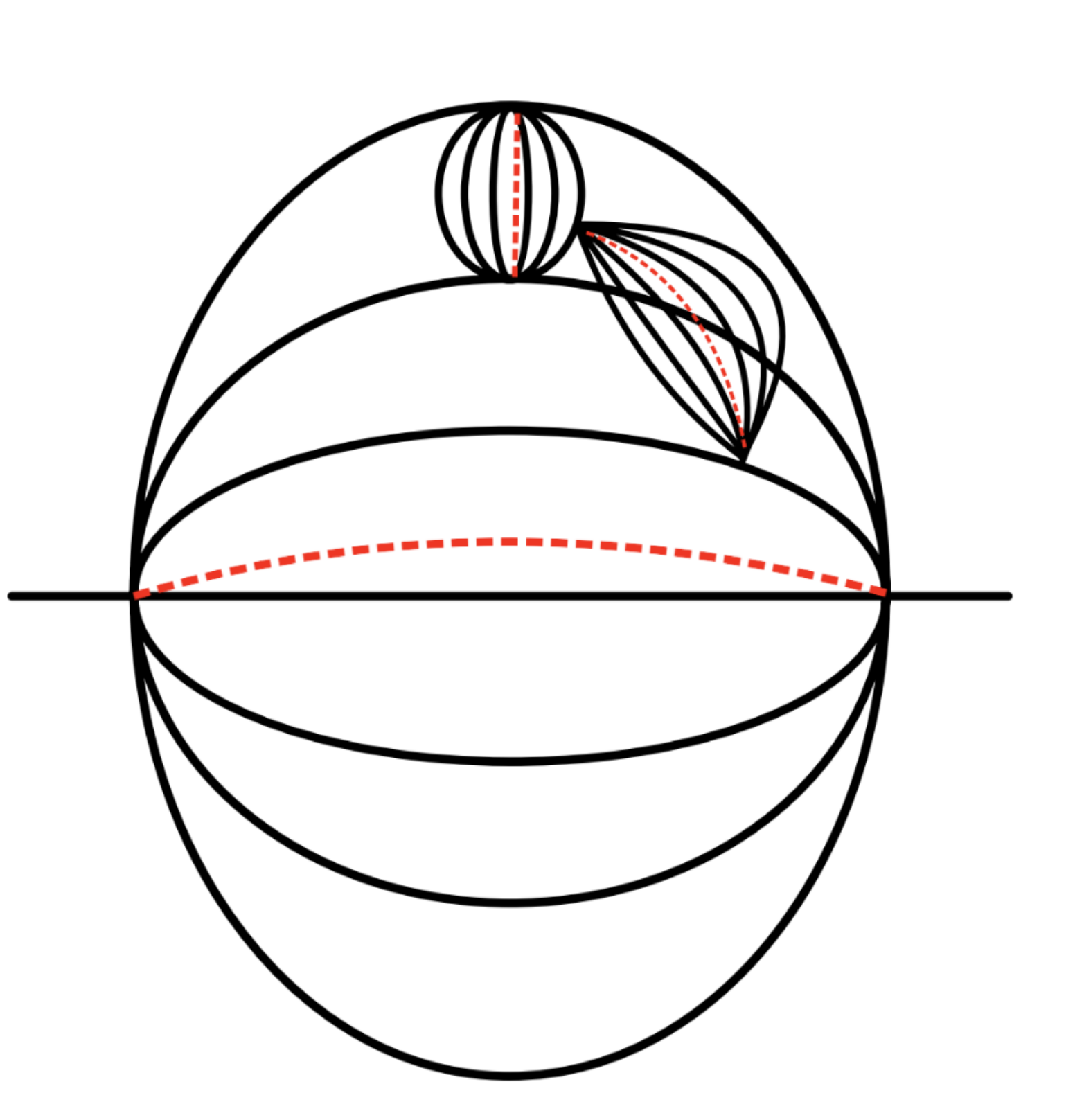}
\caption{A genus 2 diagram.}
\label{asym}
\end{center}
\end{figure}

Before leaving this section we will consider two more diagrams, Figures \ref{F44} and \ref{asym}.
We find the numerical coefficient for Figure~\ref{F44} is given by,
\be  
\boxed{
A_{4,4}=\frac{1}{n!} \ \CJ^4 \  \frac{p^{2n+1}}{N^{n-1}} \ 
T_{\rm r}^3 T_{\rm c}
}
\label{C44}
\ee
Formally it is a genus $n-1$ diagram.

Finally  Figure~\ref{asym}. We include it because it is the first diagram that is asymmetric, all the others being left-right symmetric. 
 The rules can be straightforwardly applied  and give
 \be 
\f{\CJ^6p^9}{N^2} T_{\rm r}^5 T_{\rm c},
\ee 
 evidently a genus two diagram.

\subsection{Note on $N$-scaling} \label{S: note}
\begin{figure}[H]
\begin{center}
\includegraphics[scale=.5]{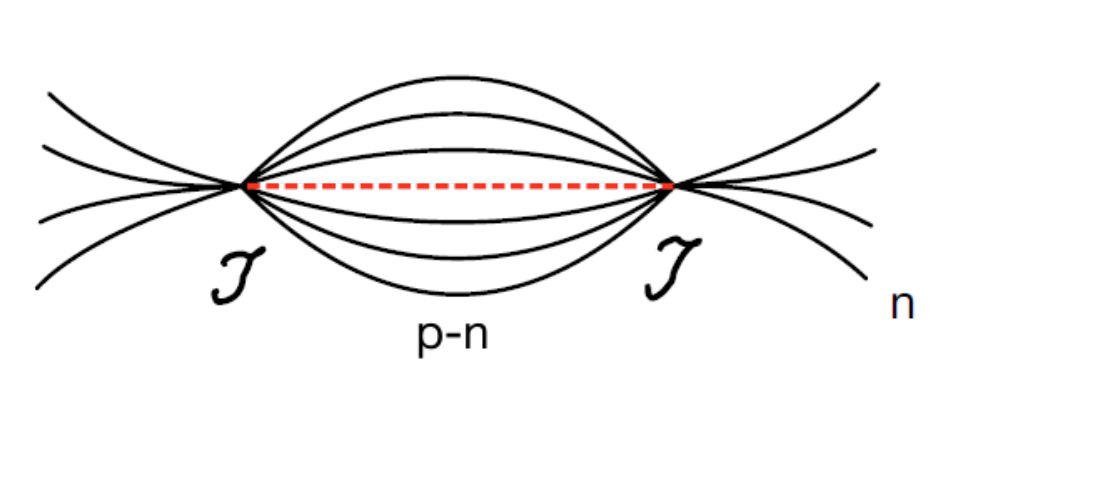}
\caption{A part of a diagram that has $n$ ``external'' lines from each end (which go elsewhere in a bigger diagram).}
\label{redline}
\end{center}
\end{figure}

The general rule for the scaling of a diagram with $N$ is the following: Begin by considering a dotted red propagator as shown in Figure~\ref{redline}. 
The red line is dressed with $(p-n)$
fermion lines, the remaining fermions that originate at the ends of the line go elsewhere in a bigger diagram.

The factor associated with such a sub-diagram consists of:
\begin{enumerate}
\item The  red propagator itself, 
\be  
 \frac{p!}{N^{p-1}} .
 \label{redprop}
\ee

\item The combinatoric factor for the indices of the internal lines,   
\be
\ {N-n \choose p-n} \approx \frac{N^{p-n}}{p!}  p^n,
\label{combp-n}
\ee
(where we assumed $n\ll p\ll N$ on the right hand side).

\item A factor of $T_{\rm r}$ from integrating over the relative time between the two vertices.
\end{enumerate}

\begin{figure}[H]
\begin{center}
\includegraphics[scale=.3]{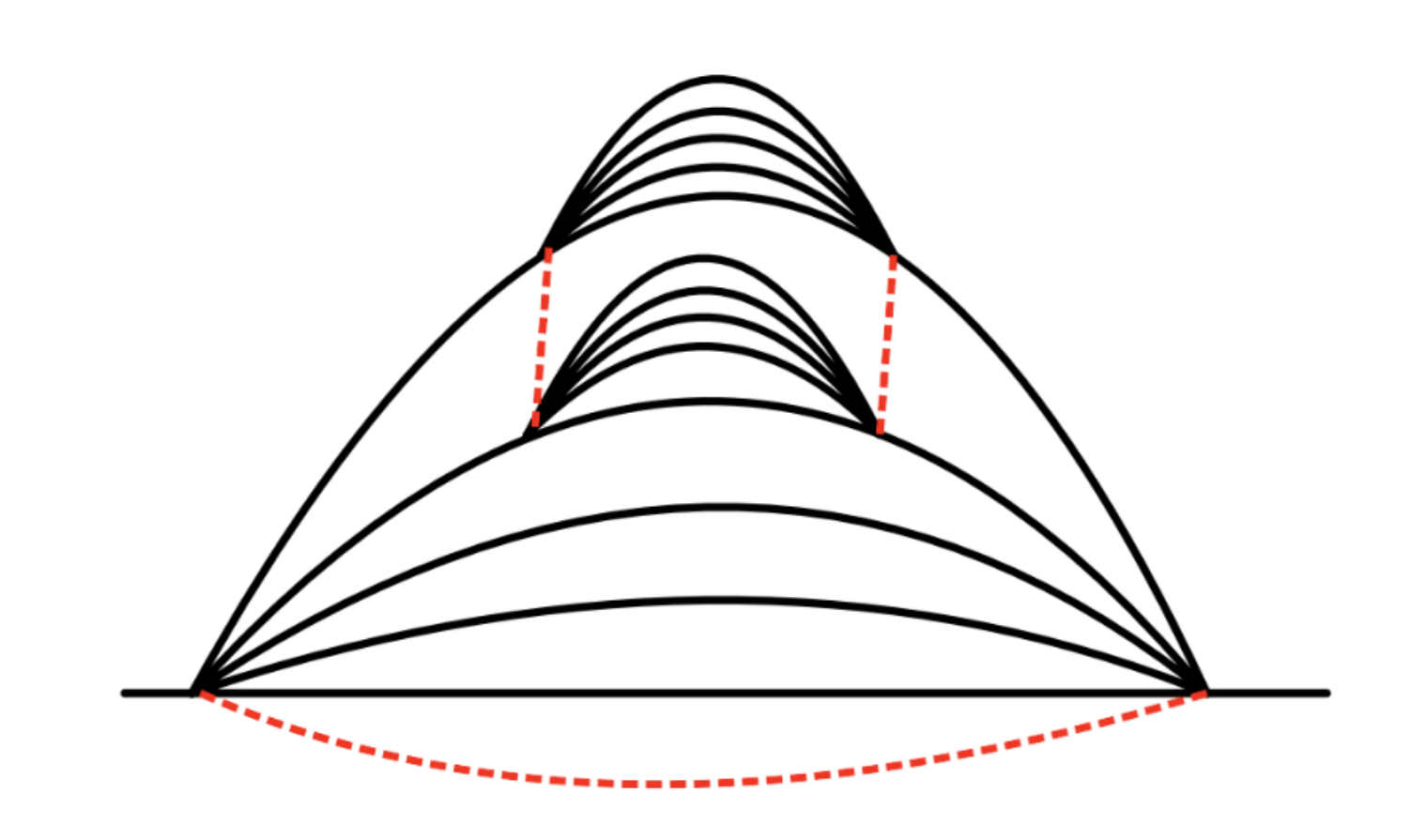}
\caption{A diagram which is infinitely suppressed when $p\to\infty$.}
\label{supres}
\end{center}
\end{figure}

\bn
All together we get,
\be 
 \frac{p^n}{N^{n-1}} T_{\rm r}.
 \label{Alltogether}
\ee
The larger the value of $n$ the more suppressed the diagram in $N,$ but at the same time the more enhanced it is in $p.$   Figure \ref{supres} is an extreme case in which the values of $n$ for two of  red propagators is $p.$ Since $p$ diverges in the double-scaled limit the diagram is infinitely suppressed.

One sees from these considerations that individual diagrams scale with negative powers of $N$ and positive powers of $p.$  The power of $p$ can be increased without  changing the genus by inserting an $A_2$ insertion in an internal line of the diagram. This pattern is the basis for the large $N$ expansion in equation \eqref{C(T)}.

We refer the reader to Appendix~A for 
the details of some of these calculations including precise numerical coefficients.

\section{Primitive Diagrams in DSSYK$_\infty$}\label{S:decratingpd}

In Section~\ref{tooftlim} we defined a primitive QCD ribbon diagram as one with the minimum power of $\alpha$ for a given genus. A ribbon diagram can represent several Feynman diagrams; thus we might speak of a primitive \it class \rm of Feynman diagrams.  We could define primitive \dk \ diagrams in a similar way: diagrams   which contain the lowest power of $p$ for a given genus. That is a possible definition but it is less convenient than a different inequivalent definition. 

Consider an $A_2$ insertion  into a line of a diagram. As we explained such an insertion always increases the power of $p,$ generally by two powers without  changing the genus. 
We will define primitive as meaning that a diagram has no $A_2$ insertions. Thus $A_{4,3}$ is the only primitive diagram in Figure~\ref{F4}. Figures \ref{F44} and  \ref{asym} are also primitive while ${A_{4,1}}$ and ${A_{4,2}}$ are not. For every primitive diagram there are an infinite number of ways of decorating it with $A_2$ insertions,  $A_2$ insertions within $A_2$ insertions......as in Figure~\ref{cartoon} without changing the genus.

 Every non-primitive diagram is a decoration of a unique primitive diagram which is obtained by removing all $A_2$ insertions until it is primitive. It follows that every primitive diagram defines an infinite sum of decorating diagrams with higher powers of $p.$ The value of the correlation function 
$A$ is given by a sum over decorated primitive diagrams.

This leads to a simple prescription; namely in every primitive diagram replace all the fermion propagators by dressed propagators which simply means replace the bare propagators $\epsilon(t)$ by,
\be 
\epsilon(t) \to \epsilon(t) e^{-2\CJ t}.
\ee

\subsection{A Rule of Thumb}

There is a simple rule that gives the correct scaling of a decorated primitive diagram.

We will consider what we call ``wee-irreducible'' diagrams. 
Wee cords have been defined as operators made of order one number of fermions, as opposed to ordinary cords which are made from order $p$ fermions~\cite{Rahman:2024iiu}.
Wee-irreducible diagrams refer to those that cannot be separated into pieces (one containing the initial vertex and the other containing the final vertex) by cutting order-one number of internal lines at an intermediate time. 
One example of wee-irredicuble diagram is the ``$A_2$ diagram'' (with the external lines removed) shown in Figure~\ref{top}. Its two ends are connected by $p-1$ internal lines and cannot be separated by cutting only order one lines.
%Wee-irreducible diagrams are the ones in which a red dotted line 
%($JJ$ propagator) connects the initial and the final vertex of 
%the diagram, as in the $A_2$ diagram. 

Every diagram is associated with the factor,
\be  
\CJ^{a} T_{\rm r}^{a-1} T_{\rm c}\f{p^b}{N^{h}}.
\label{abh}
\ee
where the factor of $\CJ^{a} T_{\rm r}^{a-1} T_{\rm c}$ 
comes from the integration over $a$ vertices.
For wee-irreducible diagrams, the integers $a,$  $b$ 
and $h$ satisfy\footnote{%
This relation does not apply to the free propagator.
It applies to the $A_2$ diagram  
and those that are obtained by adding melons to it
or deforming it, as explained in Appendix~B.} 
\be  
b = 2h +a -1.
\label{funeq}
\ee
This relation will be derived in Appendix~B.

In \eqref{abh}, 
the $(a-1)$ powers of $T_{\rm r}$ are bounded within  an overall melon. When the bare propagators are replaced by dressed propagators the integrals that gave rise to the factor $T_{\rm r}^{a-1}$ are "regulated" by the exponential decrease of the dressed propagators. The result is that each such factor is replaced by,
\be  
T_{\rm r} \to \f{1}{p\CJ}.
\label{replace}
\ee 
Thus correlation functions of single fermions whose 
primitive diagram is represented by \eqref{abh} 
is replaced by,
\be
\mbox{fermion correlator}\sim
\CJ^{a} T_{\rm r}^{a-1}T_{\rm c} \f{p^b}{N^{h}} \to 
\CJ T_{\rm c} \f{p^{b-a+1}}{N^h}.
\label{amost}
\ee
Now using \eqref{funeq}, \eqref{amost} becomes,
\be 
\mbox{fermion correlator}\sim
\CJ T_{\rm c} \lf \f{p^2}{N}\rg^h = \CJ T_{\rm c} \lf \f{\lambda}{2}\rg^h.
\label{JTlambda}
\ee

\subsection{String Worldsheet?}
Remarkably \eqref{JTlambda} not only has a fixed $\lambda$ limit that parallels the fixed $\g$ limit of gauge theory but by using the correspondence 
$$\lambda^h \leftrightarrow \g^{4h}  \leftrightarrow \gs^{2h},$$ 
it also matches \eqref{strhnt}, 
the hallmark of string theory. 
This is surprising to the present authors, since the only known theories which exhibit string-like behavior are those with matrix degrees of freedom. 

Does this mean that \dk \ is a string theory or has strings? This question is essentially the same as why the $1/N$ expansion closely resembles the 't~Hooft genus expansion although there does not seem to be any obvious relation between \dk \ diagrams and the topology of two-dimensional surfaces. 
%It seems
This makes us wonder if there is set of objects that lead to rules which parallel string theory but which are not themselves strings. Our best guess for what they are? \it Branched polymers\footnote{%
For a study on branched polymers that arise from melonic structures in tensor models, see~\cite{Gurau:2013cbh}}. \rm
\begin{figure}[H]
\begin{center}
\includegraphics[scale=.2]{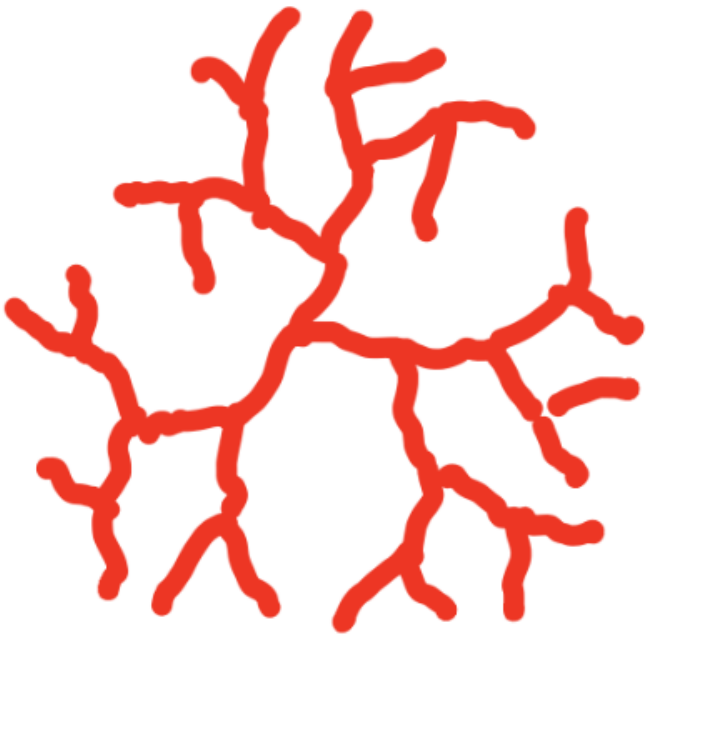}
\caption{Branched polymers.}
\label{BP}
\end{center}
\end{figure}

\section{QCD in Rindler Space and DDSYK$_{\infty}$}   \label{S:qcdr }

	In the limit of infinite de Sitter radius (of curvature)  the near horizon region of the static patch
 becomes flat Rindler space. It should be possible to formulate any quantum field theory in Rindler space but, unlike the case of the lightcone frame, there is not a great deal of research on QFT in Rindler coordinates. We will  fill some of the gap with intuitive observations and conclude with a speculation which at the moment we cannot prove, but which is central to the conjectured \dk-de Sitter duality.

In the next subsection we will focus on ordinary
 4-dimensional large $\N$ QCD in a flat background without gravity, but in Rindler coordinates. 

\subsection{The Phase Boundary and the Stretched 
Horizon}\label{S:pbsh }

The metric of Rindler space is,
\be 
ds^2 = -\rho^2 dt^2 +d\rho^2 +dx^idx^i
\label{Rmtric}
\ee
where $x^i$ are coordinates parameterizing the  $2$-dimensional plane of the horizon.  

The vacuum in Rindler space is described as a thermal state with dimensionless temperature $T_{\rm R} = \f{1}{2\pi}.$ The actual Unruh  temperature registered by a thermometer located at distance $\rho$ from the horizon is
\be 
T_{\rm U}(\rho) = \f{1}{2\pi \rho}.
\label{unruh}
\ee

\begin{figure}[H]
\begin{center}
\includegraphics[scale=.3]{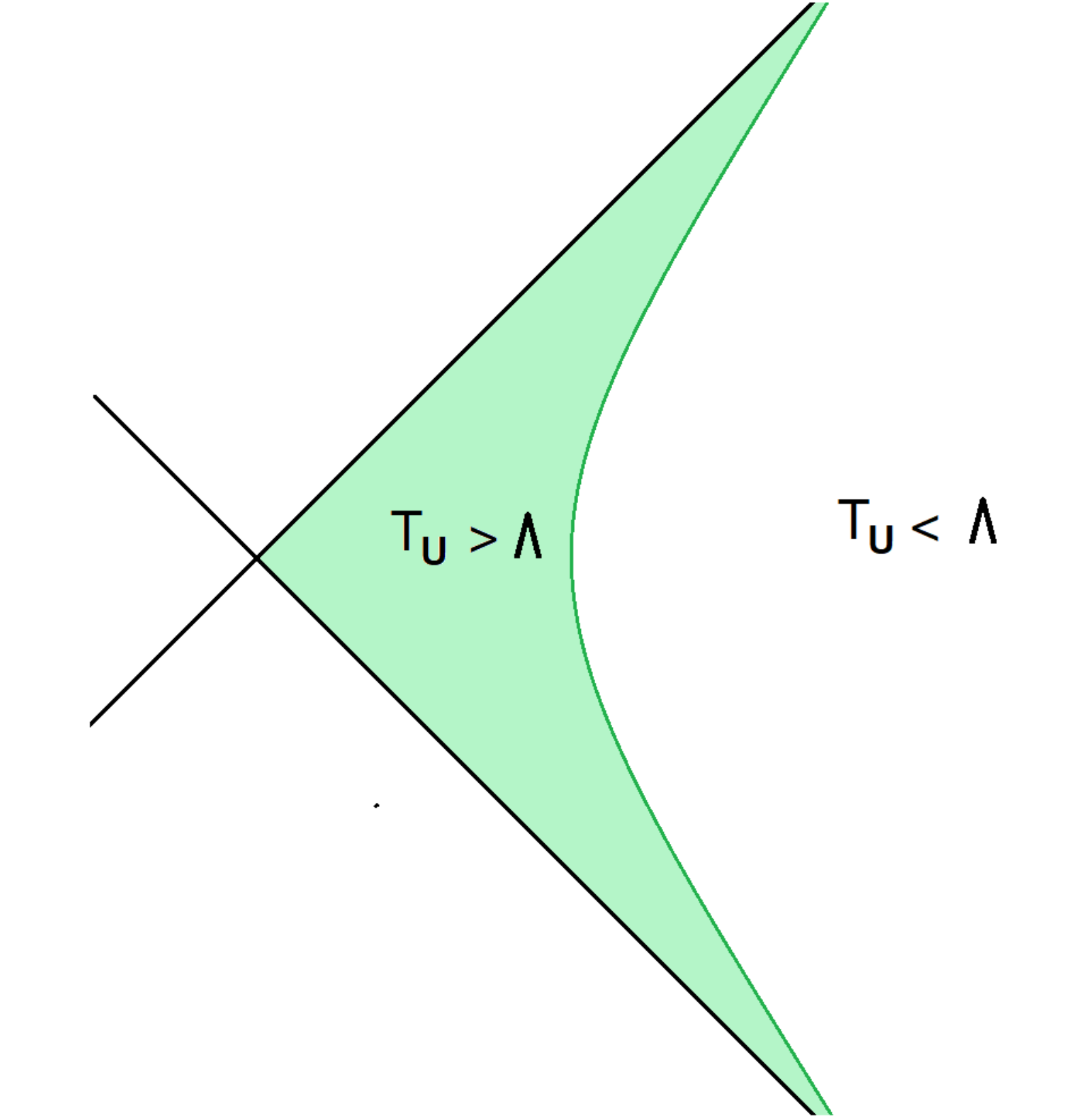}
\caption{Two regions in Rindler space.}
\label{ridler}
\end{center}
\end{figure}

Let us introduce a mathematical $(2+1)$-dimensional surface of fixed $\rho$ at the value of $\rho$ for which 
\be
T_{\rm U}(\rho) = \Lambda,
\label{Tu=Labda}
\ee
where $\Lambda$ is the usual QCD energy scale. To give it a precise definition $\Lambda$ can be taken to be the  temperature of the QCD confinement--de-confinement transition.  The surface (shown in Figure~\ref{ridler}) separates Rindler space into two regions: a hot plasma region where  QCD is in the deconfined phase (shown in green), and a cold region 
 where it is in the confined phase. This is shown in Figure~\ref{ridler}.
We may think of the surface as the phase boundary between the unconfined QCD-plasma phase and the confined phase. This phase boundary plays a role similar to the stretched horizon in \dkp 
\be
\text{QCD phase boundary}  \leftrightarrow \text{stretched horizon}
\label{morcorr}
\ee

In the hot  green deconfined region quarks and gluons propagate freely and the entropy per unit area\footnote{In a continuum field theory the entropy diverges due to UV divergences at $\rho =0.$ We can imagine a regulated version of QCD which renders the entropy finite. In that case the entropy per unit area will be of order $\N^2.$} is of order $\N^2.$ In the cold confined region only $SU(\N)$ singlets, i.e.,  mesons and glueballs can propagate%
\footnote{Baryons have mass proportional to $\N$ and in the $\N\to \infty$ disappear from the spectrum.}.
The number of species of hadrons with mass less than or order $\Lambda$ is finite and independent of $N.$ Therefore the entropy per unit area in the outer confined region is order one. If quarks are massless the entropy in the cold region would mainly be due to massless pions.

To a high approximation the phase boundary is the end of the world. The outer confined region is almost completely empty; the vast majority of degrees of freedom cannot escape the hot region. Only the light $SU(\N)$ singlets can escape. When a quark  from the hot region hits the phase boundary it is reflected back with a probability very close to $1.$ With a probability of order $1/\N^2$ it passes through the boundary dragging an antiquark with it, the two forming a meson. A similar thing can be said for gluons.
In other words to quote   \cite{susskind:confined} ``almost everything is confined" (to the green plasma region)%
\footnote{The word confined has two different meanings in this context. Non-singlet degrees of freedom are \it confined \rm to the deconfined (green) region. In the confined region quarks and gluons are \it confined \rm to form singlet hadrons.}.

 \subsection{A Speculation}\label{S: speclat} 

What we've said up to now is not especially speculative; it is based on direct comparison between calculations in QCD (ribbon graphs in QCD) and \dk perturbation theory. Some things were empirical observations gleaned from many diagrams, for example equation \eqref{funeq}. 
%They probably can  be proved with some additional work.
We'll now come to a more speculative parallel between QCD and \dk \ which is important to the interpretation of \dk \ as a holographic description of de Sitter space.

It was explained in \cite{susskind:confined}  that there are far too many degrees of freedom in any holographic description of the static patch for all, or even a tiny fraction, to propagate into the interior of the static patch. Almost all the degrees of freedom comprising the entropy must be confined to the  immediate  vicinity of the horizon. 

\begin{figure}[H]
\begin{center}
\includegraphics[scale=.5]{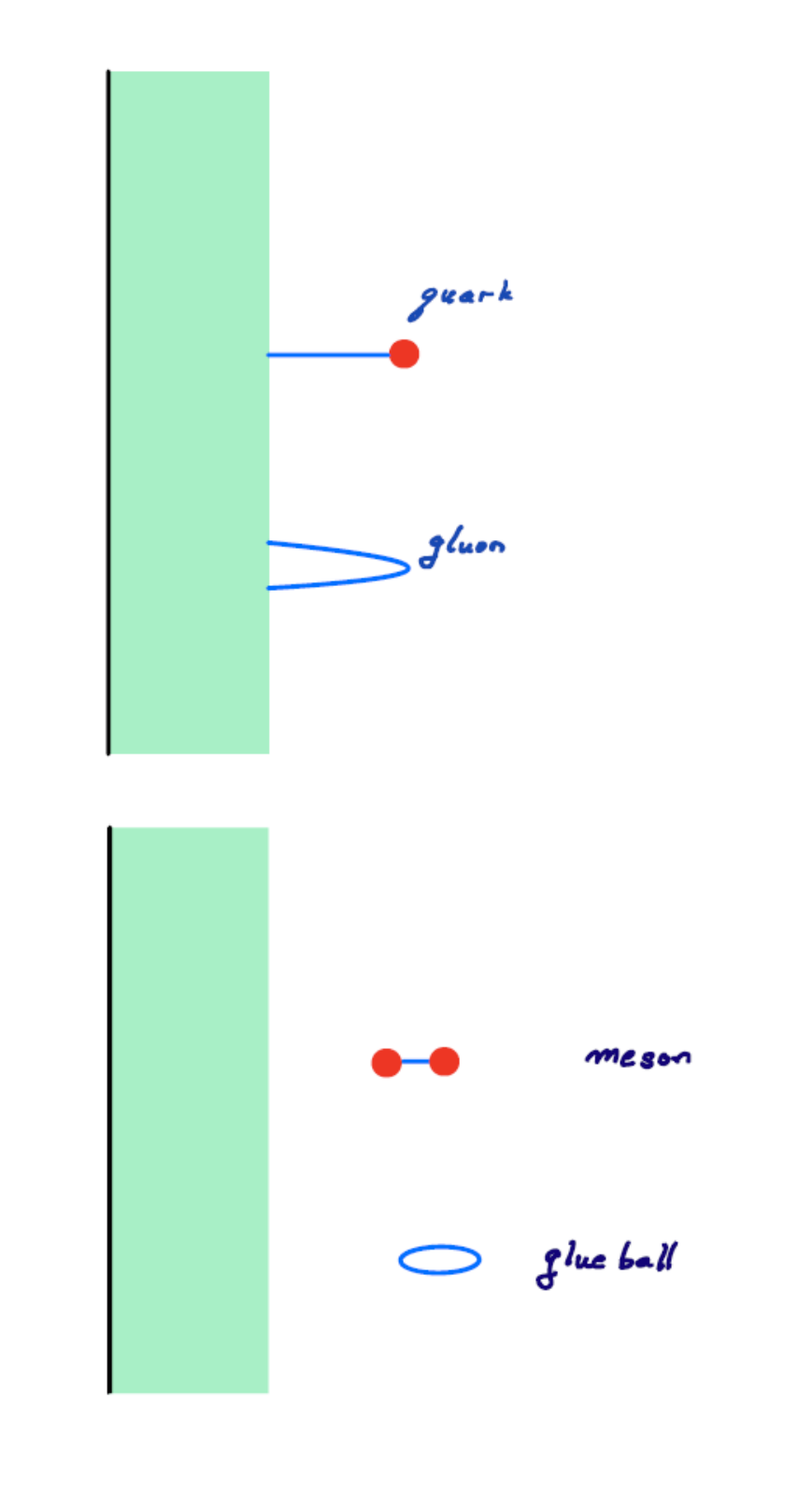}
\caption{Confinement of quarks and gluons to the stretched horizon, and free propagation of mesons and glueballs.}
\label{escape}
\end{center}
\end{figure}

This is completely understood in the
 example of QCD in Rindler space where the mechanism is ordinary confinement. In the hot plasma-like region (the stretched horizon) quarks and gluons are free to propagate independently but when they try to escape into the cold region they are held in place by QCD strings illustrated in the schematic Figure~\ref{escape}.
The objects which can escape the QCD-plasma are $SU(\N)$ singlets such as mesons and glueballs. The spectrum of singlets is very sparse and unlike the spectrum of quarks and gluons it  does not grow with increasing $\N.$
While these things may not be completely familiar they are not  speculative.

What is speculative is the application of these QCD ideas to \dk. It comes in several parts. The first is that the bulk dual of \dk \ includes a region outside
the stretched horizon that can be identified with the static patch of Jackiw-Teitelboim   de Sitter space\footnote{%
We are considering JT-de Sitter gravity with the dilaton background being 
proportional to a spacelike embedding coordinate. This is different 
from JT-de Sitter gravity studied in \cite{Maldacena:2019cbz} which has  
the dilaton background proportional to the timelike embedding 
coordinate. In the case of \cite{Maldacena:2019cbz}, the causal patch is 
not static due to the time dependence of the background.}.
The radius of curvature of the de Sitter space (in cosmic units) is 
\be 
\l_{\rm ds} = \CJ^{-1}
\label{ellds}
\ee
and the thickness of the stretched horizon is $\f{1}{\CJ p}.$ This follows from the cord two-point correlation functions (in cosmic units),   which for zero 
$\lambda$ have the form,
\be 
\la M_A(0)M_A(t) \ra = \lf  \f{1}{\cosh^2({p\CJ t})}  \rg^{\Delta_A}
\label{crdcrd}
\ee
where following \cite{Berkooz:2018jqr} we   define $M_A$ to be chord operators of fermionic weight $p\Delta_A.$  When continued to Euclidean signature \eqref{crdcrd} is periodic in imaginary cosmic time with period $\f{2\pi}{p\CJ},$ indicative of a physical temperature 
\be 
T_{\rm cord} = \f{p\CJ}{2\pi}.
\label{Tcord}  
\ee

The meaning of this is quite simple; in Rindler space, which is a good approximation to the near-horizon region of the static patch, the local temperature is given by \eqref{unruh}. Setting the Unruh temperature to $T_{\rm cord}$ \eqref{Tcord}  gives 
\be 
\rho_{\rm sh}=\f{\l_{\rm ds}}{p} 
\ee
for the location of the stretched horizon.
 Another way to put it is that the thickness of the stretched horizon is string scale,
\be 
\rho_{\rm sh}=\f{\l_{\rm ds}}{p} = \ell_{\rm string}
\ee
In the limit $p\to \infty$, $\rho_{\rm sh}$ goes to zero in cosmic units, while it stays finite in string units. This is a manifestation of ``sub-cosmic locality."

Our speculation, previously given in \cite{susskind:confined},  is that the stretched horizon at $\rho_{\rm sh}=\f{\l_{\rm ds}}{p} $      is
the phase boundary between confined and unconfined phases, and the ``charges" that are confined are the generators of the $O{(N)}$ symmetry of \dk. In other words:

\bn 
  \it  Only  $O(N)$ singlets can propagate into the bulk of the static patch.   In the limit of large $N$ the  $O(N)$ singlets are a tiny fraction of all the operators that can be made from the fermionic degrees of freedom.\rm
  
  \bn
  To a very good approximation ``almost everything is confined"---confined, that is,   to the stretched horizon. \rm
   Some evidence for this was given in \cite{susskind:confined}  but it remains very much a conjecture.

\section{Summary}\label{S: summary}
It had been our impression that the structure of the QCD 
large $\N$ expansion---its classification by genus, and  the relation between the 't~Hooft  and the fixed $\g$  limits---was something special to theories with matrix degrees of freedom and single trace actions: Yang Mills theory being  of this form. It seems  very surprising  that so similar a pattern should show  up in \dk  which has nothing to do with matrices.

Both theories can be expressed in terms of  a sum over primitive diagrams multiplying  a power series expansion, either  in $\alpha$ or $p.$ The primitive diagrams manifestly have fixed $\g$ or fixed $\lambda$ limits.  Whether these limits exist for the full series (genus by genus) turns on the existence of finite asymptotic limits of the functions defined by  power series. For Yang Mills theory the limit is only assured for theories with holographic duals and flat space limits.  Other than that we know very little.
Remarkably for \dk \ the functions $\CF^{(h)}(p)$ are known to converge as $p\to \infty,$ the limits  being obtained by replacing the bare fermionic propagators $\epsilon(t)$ by the corrected propagators $\epsilon(t)e^{-2\CJ t}$  in primitive graphs.

The parallel between large $\N$ gauge theory and \dk \ is an empirical fact justified by comparing their perturbative expansions. If there is a deeper reason   it is at present  a mystery that we need to unravel. We might express it this way: There appears to be a connection between the Riemann surfaces that encode the structure of QCD diagrams, and the graph structures that occur in  \dk \ perturbation theory. The relationship is often surprising; for example the correspondence between the 't~Hooft coupling $\alpha$ and the locality parameter $p.$  Off hand these seem to have nothing to do with each other. Our guess is that there is some mathematical framework that the \dk \ graphs fall into that parallels the topology of  two-dimensional surfaces.

The close parallel with QCD suggests that \dk \ may exhibit a similar form of confinement to what we discussed in Section~\ref{S: speclat} in which only 
$O(N)$ or $SU(N)$ singlets can propagate into the bulk of the static patch. This is a speculation but it is an important one for the conjectured duality between \dk \ and JT de Sitter space. We hope to come back to this in the future.

\section*{Acknowledgements}

We would like to thank Douglas Stanford for very helpful discussions on the use of Schwinger-Keldysh formalism in \dk. We also thank Henry Lin, Adel Rahman, Steve Shenker for discussions. Y.S.\ would like to thank Tomotaka Kitamura and Shoichiro Miyashita for helpful discussions. This work has been done while Y.S.\ was visiting Stanford Institute for Theoretical Physics (SITP) on sabbatical leave from Takushoku University under the ``Long-term Overseas Research'' program. He is grateful to Takushoku University for support and SITP for hospitality. The work of Y.S.\ is also supported in part by MEXT KAKENHI Grant Number 21H05187.

\appendix

\section{SYK Correlators from the Schwinger-Keldysh Formalism}

In this Appendix, we will perform explicit calculations of 
the SYK correlators based on the Schwinger-Keldysh formalism.
The analysis in the main text has been focused on showing 
the presence of the fixed $\lambda$ limit in SYK. 
Here, we will supplement it by describing all the steps of
calculations.
As in the main text, we consider the SYK model with real 
(Majorana) fermions at infinite temperature, and 
compute two-point functions of single fermions. We will use
the cosmic units for the time coordinate throughout the appendix.

As mentioned in the main text, the general strategy for the
calculation is to first draw the ``primitive diagrams'' 
appropriate for the order in $N$ of interest. These 
diagrams represent naive perturbations in powers of 
the strength of the couplings $\CJ$. 
Then, by replacing the free-fermion 
propagators by the dressed propagators (exact propagator 
for $\lambda=0$, given by the ``melon diagrams''), 
we obtain the result.

The following two procedures will be important computationally,
as well as conceptually. One is the summation over the ``up'' and 
``down'' paths, meaning the segment of the forward and backward 
time evolution, 
as defined below. Only certain configurations
of the interaction vertices survive the summation. This will be
explained using diagrams (e.g.,\ Figures~\ref{fig:line} and 
\ref{fig:line2}) with line segments between vertices
representing the pattern of contractions of fermions, 
intended to keep track of the signs of each contribution.
The other is the integration over the relative time of a
melon structure (which is made from two vertices). 
In the $p\to \infty$ limit, this integral is dominated by 
the region of almost zero relative time. Thus, 
a melon represents an almost instantaneous interaction
in this limit.  

In \ref{sec:preliminaries}, 
we will explain the formalism and set the notations.
In \ref{sec:lambda0}, we compute the  
two-point function at 
$\lambda=0$ (i.e.\ the leading term for $N\to\infty$ with fixed $p$). 
The result is well known; the main purpose of this analysis is to confirm
the consistency of the formalism, and to set the stage for the 
subsequent analysis. In A.~3, we compute the first $\lambda$
correction to the two-point function.

\subsection{Preliminaries}
\label{sec:preliminaries}

\subsubsection*{Closed-time contour}
The two-point function of fermions at time $t=0$ and at $t=T$
is written in the Hamiltonian formalism as
\begin{equation}
\langle{\rm tr}\left[\psi_{i}(T)\psi_{j}(0)\right]\rangle_{J}
=\langle{\rm tr}\left[e^{iHT}\psi_{i}e^{-iHT}\psi_{j}\right]\rangle_{J}
\label{eq:correldef}
\end{equation}
where ${\rm tr}$ means the normalized trace ${\rm tr}1=1$. We are taking
an ensemble average over parameters collectively referred as $J$; 
the expectation value after averaging is denoted by
$\langle \cdots \rangle_{J}$. We will not use the
$\langle \cdots \rangle$ symbol for the quantum mechanical expectation 
value to avoid confusion (except for a very few cases where noted).

\begin{figure}[H]
  \centering
  \includegraphics[width=10cm]{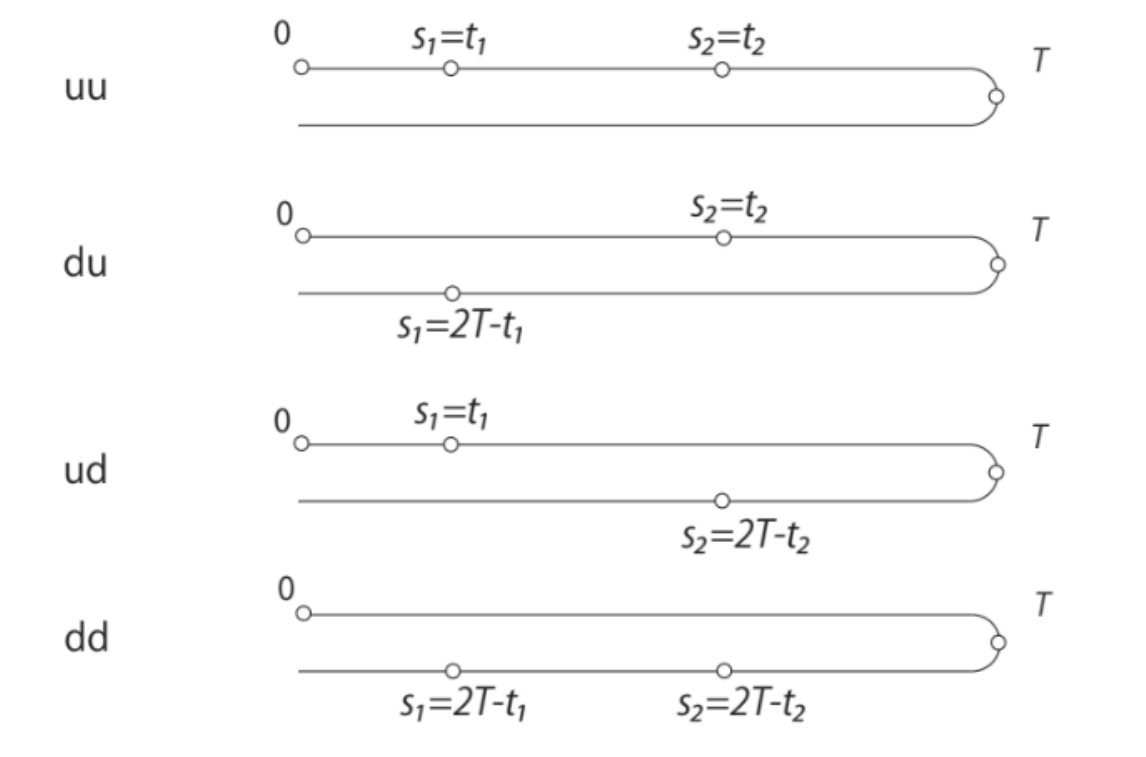}
  \caption{The parameter $s$ on the closed-time contour for the forward
(up) and backward (down) paths.}
  \label{fig:contour}
\end{figure}

The operation in \eqref{eq:correldef} can be interpreted as preparing a
state at time $t=0$, acting $\psi_{i}$ on it, 
evolving it to $t=T$, acting $\psi_{j}$ on it, then evolving
backwards in time to $t=0$ and take expectation value with respect to
the state we started with, and sum over all states. 
So, \eqref{eq:correldef} can be rewritten in the Lagrangian formulation as
\begin{equation}
\langle{\rm tr}\left[ \psi_{i}(T)\psi_{j}(0) \right]\rangle_{J}=
\left\langle \int{\cal D}\psi\, \psi_{i}(T)\psi_{j}(0) e^{i\int_{0}^{2T} {\cal L}(s)ds}\right\rangle_{J}
\label{eq:psipsigeneral}
\end{equation}
where the Lagrangian is integrated over the closed-time contour
parametrized by $s$. The part from $s=0$ to $s=T$
 (which will be called the ``up'' path) represents the forward time 
evolution, and the part from $s=T$ to $s=2T$ (the ``down'' path) represents the backward time evolution from $t=T$ to $t=0$. See Figure~\ref{fig:contour}. The parameter $s$ corresponding to a given physical time $t$ ($0\le t\le T$) for the up and down paths are 
\begin{equation}
s_{\rm u}(t)=t,\qquad s_{\rm d}(t)=2T-t,
\label{eq:sud}
\end{equation}
respectively. (The path integral on the r.h.s.\ of \eqref{eq:psipsigeneral} is understood to be divided by the partition function $\int{\cal D}\psi\, e^{i\int_{0}^{2T} {\cal L}(s)ds}$. With this understanding, we will omit this factor 
and consider only the connected correlators.)

Higher-point functions can be represented similarly using a closed-time contour which goes from the earliest time to the latest time then go back to the earliest time.

\subsubsection*{Lagrangian}
The Lagrangian for the real SYK is
\begin{align}
{\cal L}(s)&={\cal L}_{0}(s)+{\cal L}_{I}(s)
\nonumber\\
{\cal L}_{0}(s)&=
-{i\over 2}\sum_{i}\psi_{i}\partial_s\psi_{i}\\
{\cal L}_{I}(s)&=
-(\pm) i^{p/2}\sum_{i_1<\ldots <i_p} J^{i_1,\ldots,i_p}\psi_{i_1}\cdots\psi_{i_p}
\end{align}
We will treat the kinetic term ${\cal L}_{0}$ as the free part, and calculate
the correlators in an expansion in powers of ${\cal L}_{I}$. 
The $+$($-$) sign in ${\cal L}_{I}$ is for the up (down) path; Hamiltonian (or the interaction term) for the down path should have the opposite sign relative to the up path to represent the backward time evolution. 

The couplings 
$J^{i_1,\ldots,i_p}$ are random variables with the variance
\begin{equation}
\langle J^{i_1,\ldots,i_p} J^{i_1,\ldots,i_p}\rangle_{J}
={{\cal J}^2\over 2}{p! \over N^{p-1}},
\label{eq:J2}
\end{equation}
where the indices are not summed over. The $J^{i_1,\ldots,i_p}$'s with different indices are independent random variables.

The fermions obey the
equal-time anticommutation relation,
\begin{equation}
\{\psi_i,\psi_j\}=2\delta_{ij}.
\end{equation}
 
\subsubsection*{Perturbative calculation}

The free propagator for the fermion satisfies
\begin{equation}
{1\over 2}{d\over ds}\langle \psi_i(s)\psi_j(s')\rangle_{\rm free}
=\delta_{ij}\delta(s-s'),
\end{equation}
and is given by
\begin{equation}
\langle \psi_i(s)\psi_j(s')\rangle_{\rm free}
=\delta_{ij}\epsilon(s-s'),
%\label{eq:freeG}
\end{equation}
(Here, the symbol $\langle \cdots \rangle$ means the
quantum mechanical expectation value, and not 
ensemble average. We will use the symbol $G_0(s,s')$ in 
what follows.)
The free propagator apart from $\delta_{ij}$ 
will be denoted by $G_0(s,s')$,
\begin{align}
\langle \psi_i(s)\psi_j(s')\rangle_{\rm free}&=\delta_{ij}G_0(s,s'),
\nonumber\\
\qquad G_0(s,s')&=\epsilon(s-s').
\label{eq:freeG}
\end{align}

\subsection{Single-Fermion Two-Point Function for $\lambda=0$}
\label{sec:lambda0}

Let us compute the single-fermion two-point function 
in the double scaling limit with $\lambda=0$. We keep only  
the terms with the highest power of $N$ (i.e., the
``genus'' zero diagrams), then take the 
$p\to\infty$ limit in the calculation. As we will see
(and as explained in the main text), there is a finite limit.

At $\lambda=0$, 
the one-particle irreducible (1PI) 
self-energy $\Sigma$ is given by the ``melon diagram,''
in which $(p-1)$ lines connect two vertices representing the 
Hamiltonian. This is the diagram 
in Figure~\ref{fig:basic} with the external lines removed.
Each line represents the dressed single-fermion propagator, 
which is given in turn by the sum of diagrams in which arbitrary number of
self-energies are connected in series by the free propagators. 
In the melon diagram, each line is independent, so this self-energy is 
the product of $(p-1)$ dressed fermion propagators. 

\subsubsection*{Perturbation at order $\CJ^2$}

\begin{figure}[H]
  \centering
  \includegraphics[width=10cm]{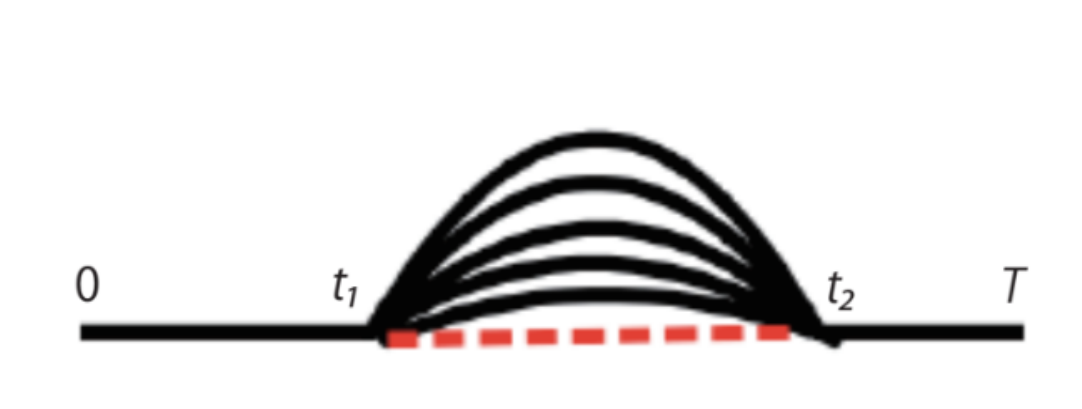}
  \caption{The $A_{2}$ diagram.}
  \label{fig:basic}
\end{figure}

Let us first consider the order $\CJ^2$ contribution, represented
by the simplest melon diagram, called the ``$A_2$ diagram.'' 
It is the diagram shown in Figure~\ref{fig:basic} with each line interpreted
as the free propagator. 

The two-point function at order $\CJ^2$ is calculated by
bringing down two interaction Lagrangian form the
exponential, 
\begingroup\makeatletter\def\f@size{14}\check@mathfonts
\def\maketag@@@#1{\hbox{\m@th\large\normalfont#1}}%
\begin{align}
&{i^2\over 2}
\left\langle\int{\cal D}\psi\, \psi_i(T)\psi_j(0) \left(\int_{0}^{2T} 
{\cal L}_{I}(s_1) ds_1\right)
\left(\int_{0}^{2T} {\cal L}_{I}(s_2) ds_2\right)
\, e^{i\int_{0}^{2T} {\cal L}_{0}(s)ds}\right\rangle_{J}
\nonumber\\
&=
{i^2\over 2}\cdot {\cal J}^2{p!\over 2 N^{p-1}}\sum_{i_1<\ldots <i_p}
\int{\cal D}\psi\, \psi_i(T)
\left( \int_{0}^{2T}(\pm) \psi_{i_1}(s_1)\cdots\psi_{i_p}(s_1)ds_1
\right)
\nonumber\\
&\qquad 
\times \left(
\int_{0}^{2T}(\pm) \psi_{i_p}(s_2)\cdots\psi_{i_1}(s_2)ds_2
\right)
\, \psi_j(0)\, e^{i\int_{0}^{2T} {\cal L}_{0}(s)ds}\nonumber\\
&\hspace*{-1cm}=
{i^2\over 2}\cdot{\cal J}^2{p!\over 2N^{p-1}}
\sum_{{\rm u},{\rm d}}\sum_{{\rm u},{\rm d}}\sum_{i_1<\ldots <i_p}
\int{\cal D}\psi\, \psi_i(T)
\left( \int_{0}^{T}(\pm) \psi_{i_1}(s_{{\rm u,d}}(t_1))\cdots
\psi_{i_p}(s_{{\rm u,d}}(t_1))dt_1
\right)\nonumber\\
&\qquad 
\times \left(
\int_{0}^{T}(\pm) \psi_{i_p}(s_{{\rm u,d}}(t_2))\cdots\psi_{i_1}(s_{{\rm u,d}}(t_2))dt_2
\right)
\, \psi_j(0)\, e^{i\int_{0}^{2T} {\cal L}_{0}(s)ds}
\label{eq:orderJ2}
\end{align}
\endgroup
The factor $i^2/2$ comes from expanding the exponential 
to the second order. In the second equality, we took the ensemble average over 
$J$ and got the factor given in \eqref{eq:J2}. 
$(\pm)$ are the signs of the interaction term 
for up $(+)$ and down $(-)$ path, mentioned above.
(We have reversed the order of fermions
using $i^{p/2}\psi_{i_1}\cdots\psi_{i_p}
=(-i)^{p/2}\psi_{i_p}\cdots\psi_{i_1}$ in the
second ${\cal L}_I$ factor, and used $(i)^{p/2}(-i)^{p/2}=1$.) 
In the last expression, we parametrized the position on 
the closed-time contour by the physical time $t$ with 
the symbol u or d, related to the parameter $s$ by \eqref{eq:sud}.
We will contract the fermions and the couplings $J$ 
in \eqref{eq:orderJ2} in the way shown in Figure~\ref{fig:basic}.

\subsubsection*{Summation over the u and d paths}
\label{sec:closedtime}

The locations of the interaction points on the full 
contour are depicted in Figure~\ref{fig:contour} for the 
four combinations of u and d.
In the following, we will first sum over u and d for a fixed $t$. 
An advantage of this method is that the contribution from 
particular configurations 
of the vertices vanish upon summation. 
(Of course, this is equivalent to directly using the parameter 
$s$ ranging from 0 to $2T$, as we did in the main text.) 

For definiteness, let us assume $t_1<t_2$. 

\begin{figure}[H]
  \centering
  \includegraphics[width=10cm]{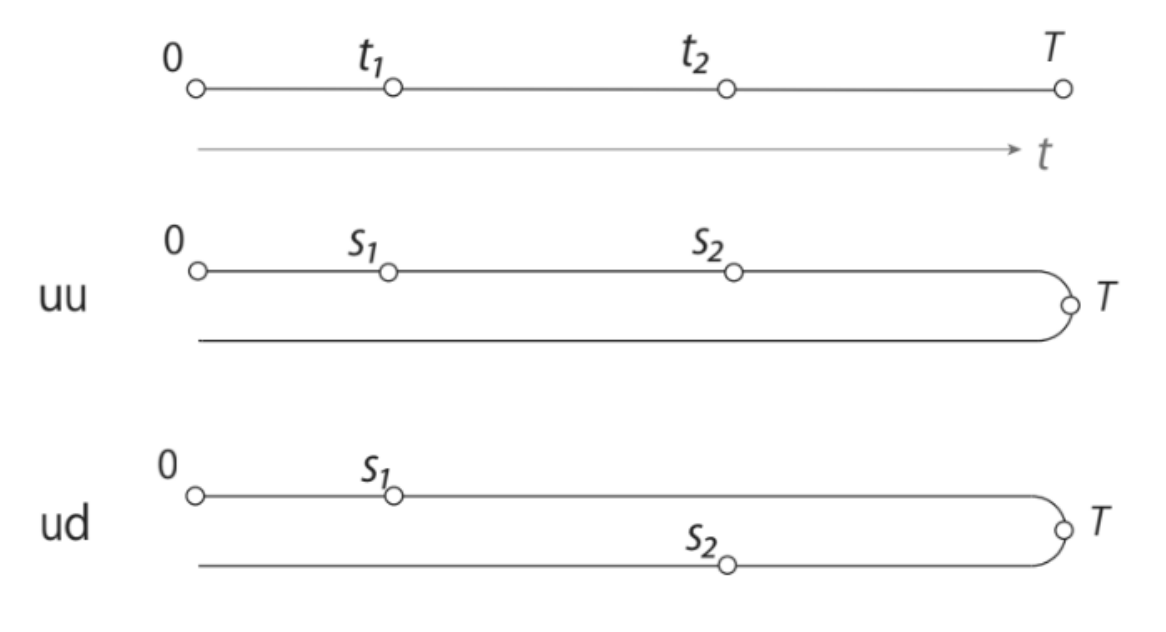}
  \caption{Top panel shows the pattern of contractions respecting the order of physical time. In the middle panel we have $s_2<T$, but in the bottom panel
we have $T<s_2$. We have $0<s_1$ and $s_1<s_2$ in both panels.}
  \label{fig:line}
\end{figure}

Let us first consider
the case in which the fermions are contracted in a way 
that respects their ordering in physical time $t$: We contract  
$\psi(0)$ with one of the fermions from the vertex at $t_1$,
and $\psi(T)$ with one from the vertex at $t_2$. 
In this case, the pattern of
the contraction is as shown in the top panel of Figure~\ref{fig:line}.
The horizontal line represents the physical time $t$, and a 
line segment denotes a free fermion propagator. Although we really 
have $(p-1)$ propagators between the vertices at $t_1$ and $t_2$, we 
represent them as a single line. This is because we are 
interested in only the signs, and the sign of the product of
$(p-1)$ propagators is the same as that of 
one propagator, since $(p-1)$ is odd.

When the vertices are both on the u path, their time ordering on 
the closed-time contour is of course
the same as in physical time (the middle panel of 
Figure~\ref{fig:line}). Now, if we move the vertex at $t_2$ from u
to d, their ordering w.r.t.\ the parameter $s$ is as shown in 
the bottom panel of Figure~\ref{fig:line}. The order of $s_2$ and
$T$ is flipped, but other orders are unchanged. This introduces
one $-$ sign. Also, there is a relative $-$ sign for u and d
in the coefficient of $J$ due to the opposite direction of the 
time evolution. In all, the sign
does not change when we change u to d for the vertex at $t_2$.
The same is true for the vertex at $t_1$. Thus, all 
four combinations of u and d have the same sign, giving rise 
to the factor of 4 as a result of the summation.

\begin{figure}[H]
  \centering
  \includegraphics[width=10cm]{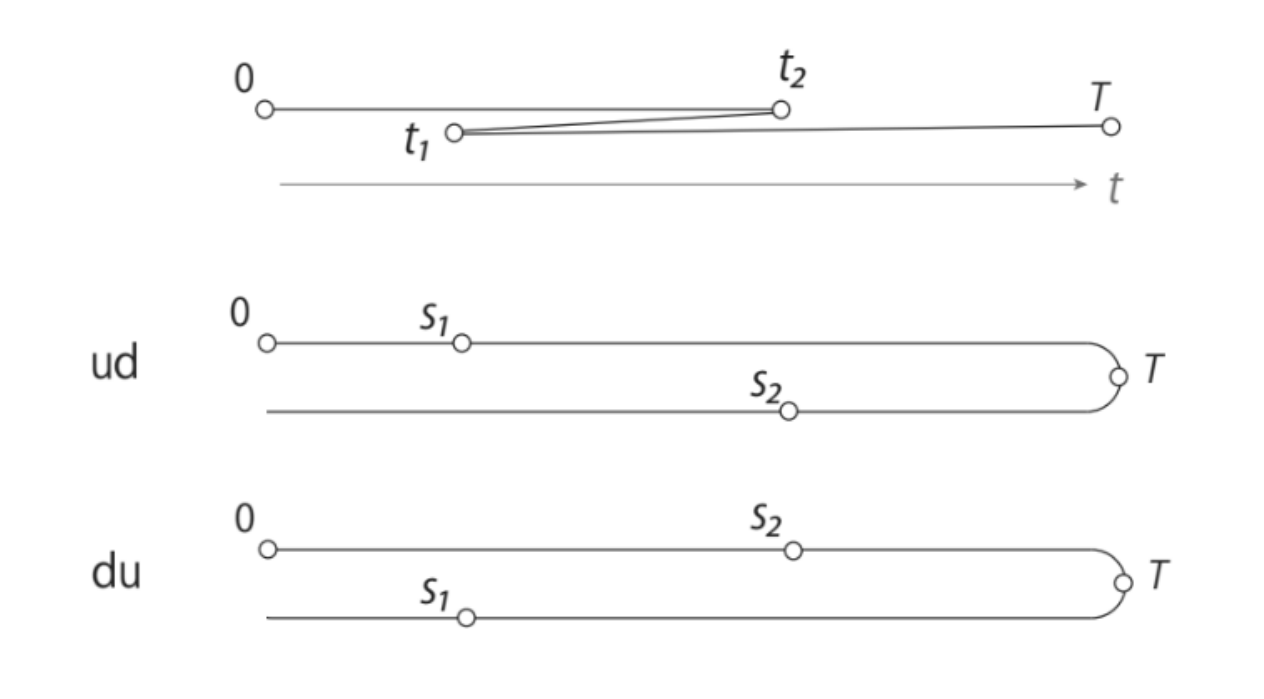}
  \caption{Top panel shows a pattern of the contractions which 
does not respect the order in
physical time. In the middle panel, 
we have $s_1<s_2$ and $0<s_2$, which is the
same as in the uu case (i.e. the same as the ordering in $t$, which is
$t_1<t_2$ and $0<t_2$). In the bottom panel, 
we have $s_2<s_1$ and $T<s_1$, so two orders are flipped compared 
to the uu case.}
  \label{fig:line2}
\end{figure}

Next, consider the case in which the fermions are contracted
in a way not respecting the ordering of $t$:  
We contract $\psi(0)$ with one of the fermions 
from the vertex at $t_2$, and $\psi(T)$ with one 
from the vertex at $t_1$ (with $t_1<t_2$). 
The pattern of the contractions is as shown in the top panel of 
Figure~\ref{fig:line2}. Suppose both of the vertices at $t_1$ and $t_2$ 
are u, and we change the vertex at $t_2$ from u to d. 
As we can see from the middle panel of Figure~\ref{fig:line2},
the ordering on the closed-time contour does not change, 
since $s_2$ is still ahead of $s_1$ and 0. So, we only have one
$-$ factor from the coefficient of $J$, and (d,u)
has the opposite sign relative to (u,u). Now, if we start from 
(u,u) and change the vertex at $t_1$ from u to d, there are two 
changes of the order: $s_1$ and $T$,  
and $s_1$ and $s_2$. So, together with the $-$ factor from the 
coefficient of $J$, we again have $-$ sign due to this change. 
We saw the sign change starting from (u,u), but the same thing
occurs starting from arbitrary configurations.
Thus, we conclude that the contributions from the contractions that
do not respect the order of physical time vanish as a result of 
the summation over u and d. 

The case of $t_1>t_2$ is exactly the same as above, except that the roles of
$t_1$ and $t_2$ are interchanged: To get a non-zero result, $\psi(T)$ should be contracted with one of the fermions at $t_1$, and $\psi(0)$ should be 
contracted with one of the fermions $t_2$.

The general rule for whether a particular configuration 
in the line-segment diagram is vanishing or not is as follows:
\begin{itemize}
\item
If there are odd number of lines from a vertex that goes to the future
(meaning that there are also odd number to the past, since the total
number from a vertex is even), the contributions from u and d for 
that vertex are of the same sign, and add up.
\item
If the above condition is satisfied for all the vertices, the
diagram survives the u and d summation, and we get a factor 
of $2^{n_{\rm v}}$ where $n_{\rm v}$ is the number of vertices, 
but if the condition is not satisfied at any one of the vertices, 
cancellation occurs, and the diagram vanishes.
\end{itemize}

\subsubsection*{Primitive diagram}

After contracting the fermions and the couplings $J$, and 
summing over u and d for the vertices at $t_1$ and $t_2$
in \eqref{eq:orderJ2}, the
single-fermion two-point function at order $\CJ^2$ 
becomes
\begin{align}
&\langle{\rm tr} \left[ \psi_i(T)\psi_j(0)\right]\rangle_J 
=
\delta_{ij}{i^2\over 2}{\cal J}^2{p!\over 2N^{p-1}}
{N^{p-1}\over (p-1)!}\nonumber\\
&\qquad\quad \times 4\cdot 2\int_{0}^{T}dt_2 \int_{0}^{t_2}dt_1\,
G_{0}(T,t_2)\left[G_{0}(t_1,t_2)\right]^{p-1}G_{0}(t_1,0)\nonumber\\
&\qquad=\delta_{ij}{\cal J}^2 p\cdot (-2)\int_{0}^{T}dt_2 \int_{0}^{t_2}dt_1
\left[G_{0}(t_1,t_2)\right]^{p-1}.
\label{eq:correlator}
\end{align}
where $G_{0}(t,t')$ is the free propagator 
\eqref{eq:freeG}.
In the first expression, the factor ${N^{p-1}\over (p-1)!}$ is
the large-$N$ approximation (assuming $N\gg p$)
of ${N-1 \choose p-1}$ from the index sums for the
$(p-1)$ internal lines. (Together with the factor ${p!\over N^{p-1}}$
from the contraction of $J$'s, this becomes $p$. This is the factor
we get when we decorate a line with a melon, i.e.,
insert a melon which starts and ends on a same line, as 
mentioned in the main text.) The factor $4$ 
comes from the summation over the u and d contributions 
explained above. The above integrand is the
expression for $t_1<t_2$; the factor $2$ just before the
integration symbol is due to the fact that we have  
the same contribution when $t_1>t_2$. 
For the type of contractions that we consider here, there is 
no crossing of the contraction lines, 
so there is no extra minus sign due to the interchange of
fermion ordering\footnote{%
In the main text, we summed over the contributions from
the contraction that respect the ordering in $s$ and 
the one that does not, before summing over u and d. 
We saw that when the two vertices have (u, u) or (d,d) 
they cancel, but when they have (u,d) or (d,u) they add up;
this is due to the sign from the crossing of the 
contraction lines.}.
In the last expression, we have used the fact that 
the free propagators connected to the external fermions are
$G_{0}(T,t_2)=G_{0}(t_1,0)=1$.

%In the following, we will replace the part 
%$\left[G_{0}(t_1,t_2)\right]^{p-1}$ which represents
%the product of $(p-1)$ internal lines by the known expression
%for the melon diagram (which is exact for $\lambda=0$). 

\begin{figure}[H]
  \centering
  \includegraphics[width=12cm]{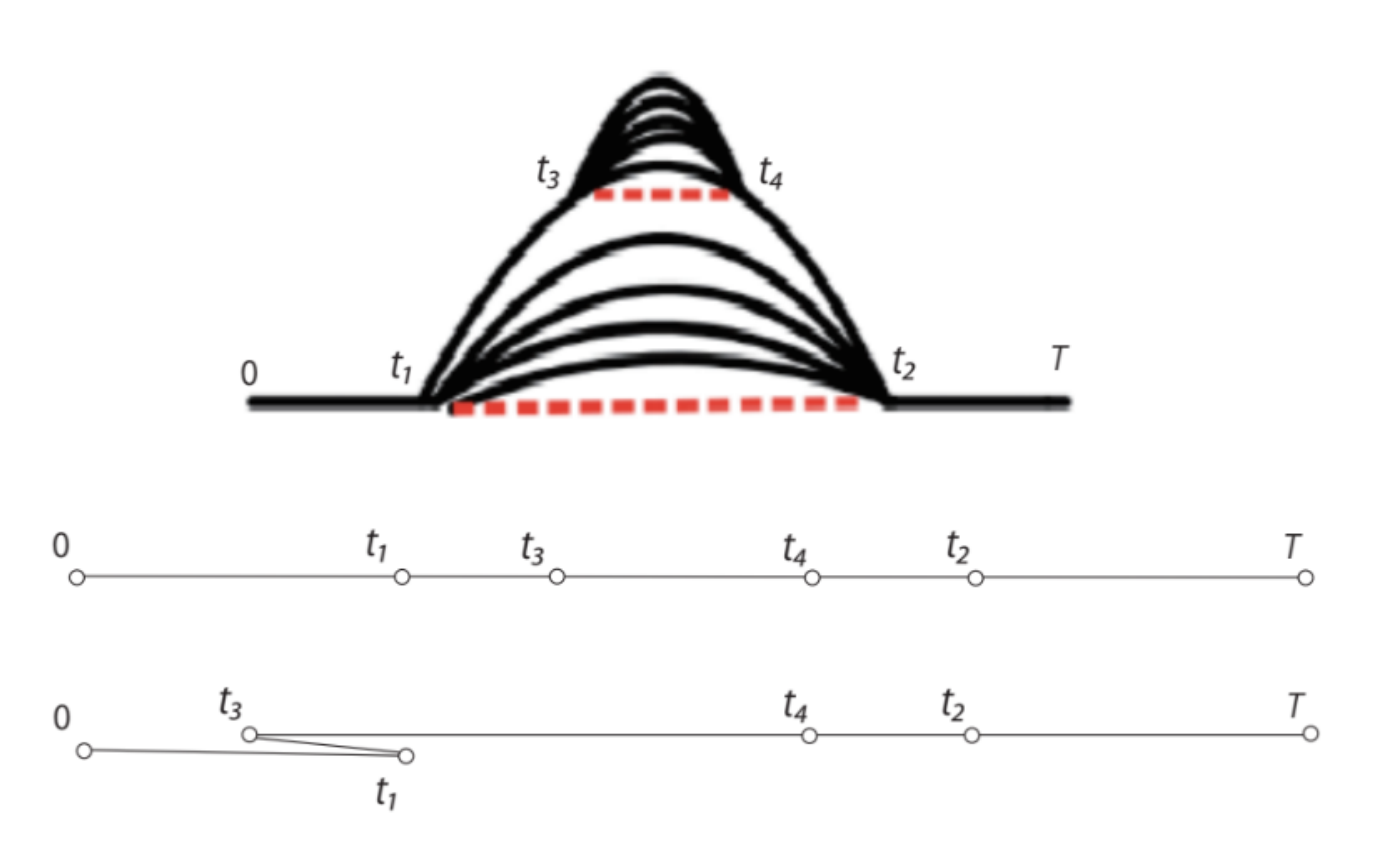}
  \caption{Top panel: The $\CJ^4$ diagram with $\lambda=0$ (melon within
melon). Middle panel: pattern of contractions that survives
the u and d summations. Bottom panel: the one that vanishes.}
  \label{fig:line3}
\end{figure}

Although we do not directly use it in the following analysis, 
we shall make a comment
about the u and d summations for 
higher order diagrams for $\lambda=0$,
such as the one shown 
in the top panel of Figure~\ref{fig:line3}, in which 
a small melon decorates one of the internal lines 
of the outer melon. From the summation of u and d
for the vertices, we find that the time of the
vertices, $t_3$, $t_4$, of the small melon should
be between the time vertices, $t_1$, $t_2$, 
of the outer melon. We can see 
from the middle and bottom panels of Figure~\ref{fig:line3}
that if $t_1<t_3<t_4<t_2$, one line is going
to the future from each vertex, so the u and d 
contributions add up. For other contractions, such as
the one shown in the bottom panel of Figure~\ref{fig:line3},
we have even number of lines to the future at some vertex,
and the diagram vanishes.

\subsubsection*{Exact self-energy at $\lambda=0$}

We now would like to obtain the two-point function for $\lambda=0$ 
to the full orders in $\CJ$. 

If we perform the summation of infinite number of the melon
diagrams (which do not contribute factors of $1/N$)
the free propagator
$G_{0}(t_1,t_2)$ that appeared in $\left[G_{0}(t_1,t_2)\right]^{p-1}$ 
is replaced with the dressed propagator $G(t_1,t_2)$. The one-particle 
diagram $\left[G_{0}(t_1,t_2)\right]^{p-1}$ is the self-energy, whose 
explicit form in the $p\to\infty$ limit 
(and at infinite temperature) is given in the position space  
by~\cite{Maldacena:2016hyu, Roberts:2018mnp} 
\begin{equation}
\left[ G(t_1,t_2)\right]^{p-1}
=\Sigma(t_1,t_2)={1\over \cosh^2(p{\cal J}(t_2-t_1))}.
\label{eq:Sigma}
\end{equation}

We could get the dressed single-fermion propagator $G(t_1,t_2)$ by 
taking the $1/(p-1)$-th power 
of \eqref{eq:Sigma}, since each line is independent 
in the melon diagram.
Here we will calculate it as a sequence of 
self-energy, to check the consistency 
of our calculational method by reproducing the known 
result, and to set the stage for computing the 
$\lambda$ corrections.

\subsubsection*{Correlator with one self-energy insertion}

We first compute the two-point function with one self-energy inserted,
which is given by replacing $\left[G_0(t_1,t_2)\right]^{p-1}
\to \left[G(t_1,t_2)\right]^{p-1}$, 
\begin{align}
\langle{\rm tr} \left[ \psi_i(T)\psi_j(0)\right]\rangle_J 
&=\delta_{ij}{\cal J}^2 p\cdot (-2)\int_{0}^{T}dt_2 \int_{0}^{t_2}dt_1
\left[G(t_1,t_2)\right]^{p-1}.
\end{align}

We change the integration variables from $t_1$ and $t_2$ to
the ``center of mass'' time $t_{\rm c}$ and the
time separation $\tilde{t}$,
\begin{equation}
t_{\rm c}={1\over 2}(t_1+t_2),\quad \tilde{t}=t_2-t_1.
\end{equation}
The absolute value of the Jacobian for this transformation
is unity, and the integral is transformed to 
\begin{equation}
\int_{0}^{T}dt_2 \int_{0}^{t_2}dt_1\to
\int_{0}^{T/2}dt_{\rm c} \int_{0}^{t_{\rm c}}d\tilde{t} 
+\int_{T/2}^{T}dt_{\rm c} 
\int_{0}^{T-t_{\rm c}}d\tilde{t}.
\label{eq:integrange}
\end{equation}
%The range of integration for $t_{\rm c}$ and $\tilde{t}$
%is depicted in Figure~\ref{fig:integregion}
%\begin{figure}[H]
%  \centering
%  \includegraphics[width=8cm]{integregion.pdf}
%  \caption{The range of integration: the darker (lighter) shaded 
%region corresponds to the first (second) term of \eqref{eq:integrange}.}
%  \label{fig:integregion}
%\end{figure}

Now, we notice the important fact that 
the self-energy \eqref{eq:Sigma} has a support only at the time 
separation of the order $|t_2-t_1|\lesssim 1/(p{\cal J})$,
which approaches zero in the limit of large $p$. This means 
that the integration over the relative time $\tilde{t}=t_2-t_1$ can be 
approximated by the integral with an infinite range,
\begin{align}
&\int_{0}^{t_{\rm c}}d\tilde{t} {1\over \cosh^2(p{\cal J}\tilde{t})}
\sim
\int_{0}^{T-t_{\rm c}}d\tilde{t}
{1\over \cosh^2(p{\cal J}\tilde{t})}\nonumber\\
&\quad \sim
\int_{0}^{\infty}d\tilde{t}
{1\over \cosh^2(p{\cal J}\tilde{t})}={1\over p{\cal J}}.
\end{align}
(For finite $p$, the finiteness of the range of support of 
self-energy \eqref{eq:Sigma} introduces an 
important scale-dependence in the problem. We will not
consider it in the present paper, and defer it 
to future study.)

By using these, the correlator \eqref{eq:correlator} becomes
\begin{align}
\langle{\rm tr} \left[ \psi_i(T)\psi_j(0)\right]\rangle_J 
&=\delta_{ij}{\cal J}^2 p\cdot (-2)\int_{0}^{T}dt_c {1\over p{\cal J}}
\nonumber\\
&=\delta_{ij}(-2){\cal J}T.
\label{eq:oneselfenergy}
\end{align}
The ratio of this with the free (order $\CJ^0$) correlator,
which is $\delta_{ij}$, 
is supposed to be the expansion of the correlation function
to the first order in time $T$.
The above equation suggests that the decay rate defined by
$e^{-\gamma T}\sim 1-\gamma T$ 
is given by 
\begin{equation}
\gamma=2{\cal J}. 
\end{equation}
This is consistent with the propagator obtained by taking the 
$1/(p-1)$-th power of \eqref{eq:Sigma},
\begin{equation}
\left[{1\over \cosh^2(p{\cal J}T)}\right]^{1/(p-1)}
\sim \left[{1\over e^{2p{\cal J}T}}\right]^{1/p}=e^{-2{\cal J}T}.
\end{equation}

\subsubsection*{Self-energy insertions in series}
Let us briefly explain that the contributions from the
sequence of the self-energy insertions,
connected by free propagators in series,
exponentiates to $e^{-2{\cal J}T}$. 

The term which have $n$ self-energy insertions is obtained from 
the order $J^{2n}$ term in the perturbative expansion. 
From the coefficient of the action and from the Taylor series,
we have the factor
\begin{equation}
{i^{2n}\over (2n)!}={(-1)^n \over (2n)!}.
\end{equation}

Then, we contract these $J$'s in $n$ pairs. The number of
such combinations is
\begin{equation}
{(2n)!\over 2^{n} n!}.
\end{equation}
Within each pair, we contract the $(p-1)$ fermions. The factor 
for each pair (melon) is obtained as in the single self-energy case
studied above. By combining
the factors from the $J$ contractions and the summation over the 
indices of the $(p-1)$ fermions, and raising the result to the $n$-th
power (for $n$ melons), we obtain
\begin{equation}
\left({\cal J}^2{p!\over 2N^{p-1}}{N^{p-1}\over (p-1)!}\right)^n
=\left({p\over 2}{\cal J}^2\right)^n.
\end{equation}

So far we had free propagators in mind, but here we
replace them with the dressed propagator, as we did above.
By summing over the u and d path, and  
integrating over the relative time of the two 
vertices of a melon, 
we get the factor $4\cdot 2/(p{\cal J})$ for each melon.
For $n$ melons, we have
\begin{equation}
\left({4\cdot 2\over p{\cal J}}\right)^n.
\end{equation}
The range of integrations for the ``center of mass'' times of 
the melons can be taken to be from 0 to $T$, for all $n$ of them.
Although the contractions should respect the ordering of
physical time to give a non-zero answer, the permutation of
the melons effectively makes the range to be the full time
interval. Thus, this gives the factor $T^n$.

Combining the above factors and summing over $n$, 
we obtain the two-point functions with 
arbitrary number of self-energy insertions, 
\begin{align}
\langle{\rm tr} \left[ \psi_i(T)\psi_j(0)\right]\rangle_J 
&=\delta_{ij}\sum_{n=0}^{\infty}
{(-1)^n \over (2n)!}{(2n)!\over 2^{n} n!}
\left({p\over 2}{\cal J}^2\right)^n \left({4\cdot 2\over p{\cal J}}\right)^n
T^n\nonumber\\
&=\delta_{ij}\sum_{n=0}^{\infty}{1\over n!}(-2{\cal J}T)^n
=\delta_{ij}e^{-2{\cal J}T}.
\end{align}

\subsection{Single-Fermion Two-Point Function at Order $\lambda$}
\label{sec:lambda1}
\begin{figure}[H]
  \centering
  \includegraphics[width=10cm]{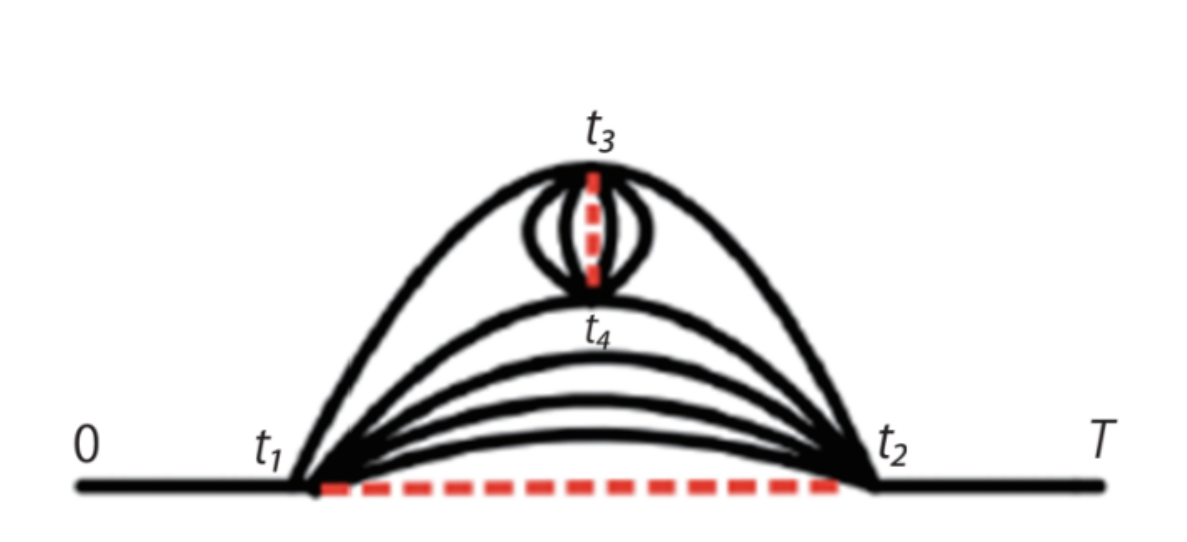}
  \caption{The order $\lambda$ diagram.}
  \label{fig:lambda}
\end{figure}

Let us now compute the first $\lambda$ correction to the 
single-fermion two-point function. We will compute the 
contribution from the 
diagram shown in Figure~\ref{fig:lambda} (which was
called the ``$A_{4,3}$ diagram'' in the main text,
shown in the bottom panel of Figure~\ref{F4}), in which a 
small melon connects two different lines of the outer melon. 
It can connect 
any two lines in the outer melon, so we will later multiply 
the result by a factor of $p(p-1)/2$
for the choice of two lines from $p$ lines.  

We have to note that this diagram does not give the full 
answer at order $\lambda$. In fact, there are infinite 
series of diagrams, which have the structure of 
``crossed melons,'' that contribute at 
the same order, as mentioned in Section~4.4 in the main text. 
As the end of this appendix, we will 
have a brief discussion on those diagrams.

In the following, the time coordinates 
of the vertices for the outer melon will be called $t_1$ and
$t_2$, and the ones for the small melon will be called
$t_3$ and $t_4$. (Different labeling represents a different
configuration. We will multiply the results by a 
factor which accounts for the choice of labeling.) 
The parameters for the closed-time contour
are called $s_1, \cdots, s_4$, correspondingly. 

\subsubsection*{Primitive diagram}
What we really would like to compute is the diagram in
Figure~\ref{fig:lambda} with the lines being 
the dressed propagators, but for the moment, let us 
suppose these lines are the free propagators. Then,
this diagram is of order $\CJ^4$, and is given by
bringing down four factors of interaction terms
from the exponential,
\begingroup\makeatletter\def\f@size{14}\check@mathfonts
\def\maketag@@@#1{\hbox{\m@th\large\normalfont#1}}%
\begin{equation}
{i^4\over 4!}
\left\langle\int{\cal D}\psi\, \psi_i(T)\psi_j(0) \left(\int_{0}^{2T} 
{\cal L}_{I}(s_1) ds_1\right)\cdots 
\left(\int_{0}^{2T} {\cal L}_{I}(s_4) ds_4\right)
\, e^{i\int_{0}^{2T} {\cal L}_{0}(s)ds}\right\rangle_{J},
\end{equation}
\endgroup
and contracting the fermions and the $J$'s in the way shown
in Figure~\ref{fig:lambda}. 

By contracting (ensemble averaging over) $J$'s, the above 
expression becomes
\begingroup\makeatletter\def\f@size{14}\check@mathfonts
\def\maketag@@@#1{\hbox{\m@th\large\normalfont#1}}%
\begin{align}
&{i^4\over 4!}\left({\cal J}^2 {p!\over 2N^{p-1}}\right)^2 
{4!\over 2^2\cdot 2!}\sum_{i_1<\ldots <i_p}\sum_{j_1<\ldots <j_p}
\int{\cal D}\psi\, \psi_i(T) 
\left((\pm)\int_{0}^{2T}\psi_{i_1}\cdots \psi_{i_p}(s_2) ds_2\right)\nonumber\\
&\qquad \times\left((\pm)\int_{0}^{2T}\psi_{j_1}\cdots \psi_{j_p}(s_4) ds_4\right) 
\left((\pm)\int_{0}^{2T}\psi_{j_p}\cdots \psi_{j_1}(s_3) ds_3\right)\nonumber\\
&\qquad \times\left((\pm)\int_{0}^{2T}\psi_{i_p}\cdots \psi_{i_1}(s_1) ds_1\right)\psi_j(0) \, e^{i\int_{0}^{2T} {\cal L}_{0}(s)ds},
\label{eq:contractJ}
\end{align}
\endgroup
where the factor $\left({\cal J}^2 {p!\over 2N^{p-1}}\right)^2$
comes from two contractions of $J$'s. In \eqref{eq:contractJ},
the contractions of $J$'s are taken for the pairs ($s_1$, $s_2$)
and ($s_3$, $s_4$). The factor 
${4!\over 2^2\cdot 2!}$
accounts for the different pairings, which give the same results.
The $(\pm)$ sign means we take the $+$ sign for up path 
and $-$ for down path.

Now, we contract the fermions with the free propagator,
and get
\begingroup\makeatletter\def\f@size{14}\check@mathfonts
\def\maketag@@@#1{\hbox{\m@th\large\normalfont#1}}%
\begin{align}
&\langle{\rm tr}\left[ \psi_i(T)\psi_j(0)\right]\rangle_{J}=
\delta_{ij}{i^4\over 4!}\left({\cal J}^2 {p!\over 2N^{p-1}}\right)^2 
 {4!\over 2^2\cdot 2!} {N^{p-1}\over (p-1)!}
{N^{p-2}\over (p-2)!} {p(p-1)\over 2}\nonumber\\
&\quad
\times 4\int_{0}^{2T}(\pm)ds_1 
\cdots\int_{0}^{2T}(\pm)ds_4 \,
G_0(T,s_2)G_0(s_2,s_3)G_0(s_2,s_4)
\nonumber\\
&\quad\times\left( G_0(s_2,s_1)\right)^{p-3}
\left( G_0(s_3,s_4)\right)^{p-2}
G_0(s_3,s_1)G_0(s_4,s_1)G_0(s_1,0),
\label{eq:contractPsi}
\end{align}
\endgroup
where the factors ${N^{p-1}\over (p-1)!}
\cdot{N^{p-2}\over (p-2)!}$ comes from the summation
of the indices in the internal lines; the first 
from the outer melon, and the second from the inner melon. 
(The latter factor ${N^{p-2}\over (p-2)!}$ times 
one factor of ${p!\over N^{p-1}}$ from a $J$
contraction gives $p^2/N\sim \lambda$. This 
is the factor mentioned in the main text,
associated to a melon which connects 
two different lines.) The factor ${p(p-1)\over 2}$ is
the number of the choices of the lines in the outer melon
between which 
the small melon is located. In the above expression, 
the fermion $\psi_i(T)$ is contracted with a fermion
from the vertex at $s_2$, but we could contract $\psi_i(T)$ with
a fermion at other vertices; the factor 
of 4 just before the integration symbol 
accounts for this. There is no extra sign from 
the commutations of fermions to bring them next to 
each other before contraction; 
the contraction lines do not cross when we 
contract the fermions in \eqref{eq:contractJ}
to get \eqref{eq:contractPsi}.

\subsubsection*{Summation over the u and d paths}
We now take a summation over u and d, and write an
expression in terms of the physical time $t$ instead
of the parameter $s$. 

\begin{figure}[H]
  \centering
  \includegraphics[width=10cm]{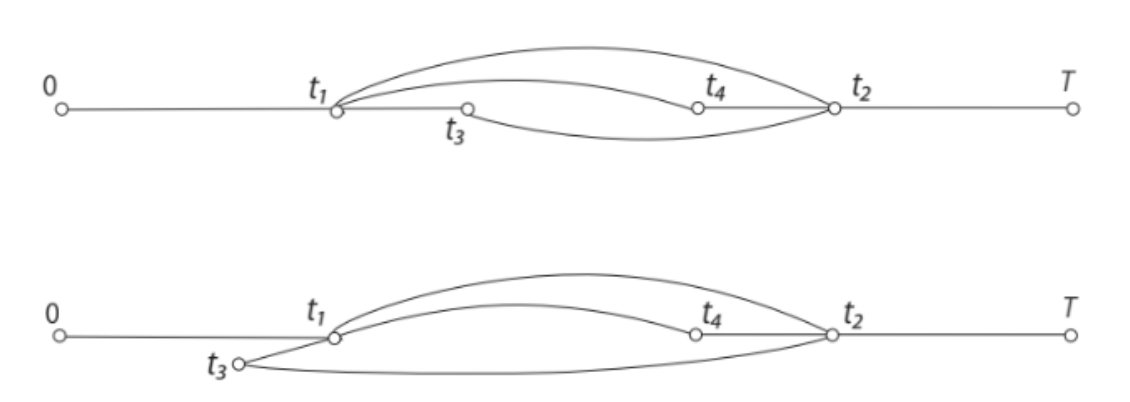}
  \caption{Pattern of contractions that survives the summation over 
u and d (top panel) and the one that vanishes (bottom panel).}
  \label{fig:line4}
\end{figure}

Let us consider the case in which
the physical time coordinates for the inner vertices
$t_3$ and $t_4$ are both between the time coordinates 
for the outer vertices, i.e.,
$t_1<t_3<t_2$ and $t_1<t_4<t_2$. 
Then the pattern of contraction is as
shown in the top panel of Figure~\ref{fig:line4}.
In this case, for every vertex, there are odd number of
lines going to the future, and odd number 
to the past. ($t_3$ can be either larger or smaller than $t_4$.)
Thus, the u and d at each vertex contribute
with the same sign as we explained in Appendix~\ref{sec:closedtime}, 
so we have a factor of $2^4$ from 
the summation at four vertices.  

If the vertices of the inner melon are not inside those
of the outer melon, cancellation occurs. The bottom 
panel of Figure~\ref{fig:line4} represents
a case in which $t_3$ is outside the vertices of the
outer melon, $t_3<t_1<t_2$ (though $t_4$ is inside,
$t_1<t_4<t_2$). We can see that there are even
(two) lines going to the future from $t_3$ or
$t_1$, so this configuration vanishes as a result of 
summation over u and d at these vertices. Similarly, 
one can check that if both $t_3$ and $t_4$ are outside 
the vertices of the outer melon, cancellation occurs. 
Also, we need $t_1>t_2$, as in the case of $\lambda=0$. 

\subsubsection*{Vertex integrations using dressed propagators}
The correlator now becomes
\begingroup\makeatletter\def\f@size{14}\check@mathfonts
\def\maketag@@@#1{\hbox{\m@th\large\normalfont#1}}%
\begin{align}
&\langle{\rm tr}\left[ \psi_i(T)\psi_j(0)\right]\rangle_{J}=
\delta_{ij}i^4{\cal J}^4{p^5\over N}\int_{0}^{T}dt_2 
\int_{0}^{t_2}dt_1 \int_{t_1}^{t_2}dt_3 
\int_{t_1}^{t_2}dt_4\, G_0(T,t_2)G_0(t_2,t_3)\nonumber\\
&\times
G_0(t_2,t_4)\left( G_0(t_1,t_2)\right)^{p-3}\left( G_0(t_3,t_4)\right)^{p-2}
G_0(t_3,t_1)G_0(t_4,t_1)G_0(t_1,0),
\label{eq:aftersum}
\end{align}
\endgroup
where we have used the fact that the factors
on the first line in \eqref{eq:contractPsi} times
the factor $2^4$ explained above (from the 
summation over u and d) equals $i^4{\cal J}^4{p^5\over N}$
(in the large $p$ limit in which we can replace
$p-1\to p$). 

The correlator at order $\lambda$ can be obtained by 
replacing the free propagators by the dressed propagators, 
$G_0(t,t')\to G(t,t')$, in \eqref{eq:aftersum}. Using
the dressed propagators, we will perform the integration
over the vertex times.

As in the last subsection, we define 
the ``center of mass'' time and the relative time
for the pair of vertices in each melon,
\begin{align}
t^{(1)}_{\rm c}&={1\over 2}(t_1+t_2),\quad \tilde{t}{}^{(1)}=t_2-t_1,
\nonumber\\
t^{(2)}_{\rm c}&={1\over 2}(t_3+t_4),\quad \tilde{t}{}^{(2)}=t_4-t_3.
\label{eq:t1t2}
\end{align}

Then, the correlator can be written in the form
(using special properties of the $p\to\infty$ 
limit as explained below),
\begingroup\makeatletter\def\f@size{14}\check@mathfonts
\def\maketag@@@#1{\hbox{\m@th\large\normalfont#1}}%
\begin{align}
&\langle{\rm tr}\left[ \psi_i(T)\psi_j(0)\right]\rangle_{J}=
\delta_{ij}i^4{\cal J}^4{p^5\over N}
\int_{0}^{T}dt^{(1)}_c \int_{0}^{\infty} d\tilde{t}^{(1)} 
\int_{t_1}^{t_2}dt^{(2)}_c \int_{-\infty}^{\infty} d\tilde{t}^{(2)} 
%\,
\nonumber\\
&\times G(T,t_1)\left( G(t_1,t^{(2)}_c)\right)^2
\left( G(\tilde{t}^{(1)})\right)^{p-3}
\left( G(\tilde{t}^{(2)})\right)^{p-2}
\left( G(t^{(2)}_c,t_2) \right)^2 G(t_2,0).
\label{eq:newvariables}
\end{align}
\endgroup
The integrals are supposed to be done ``inside first,'' meaning
in the order, $\tilde{t}^{(2)}$, $t_c^{(2)}$, 
$\tilde{t}^{(1)}$, $t_c^{(1)}$. The integration ranges for the
relative times, $\tilde{t}^{(1)}$ and $\tilde{t}^{(2)}$, really 
depend on the center of mass times.
(For $\tilde{t}^{(1)}$, it is given by the
range for $\tilde{t}$ in \eqref{eq:integrange}.) But as explained
in the last subsection, the integrand, 
$\left(G(\tilde{t}^{(1)})\right)^{p-3}$
or $\left( G(\tilde{t}^{(2)})\right)^{p-2}$, is sharply
peaked around relative time zero when $p\to\infty$, so the
integration range can be effectively taken to infinity.
The integration for $\tilde{t}^{(1)}$ is over positive values,
since $\psi_i(T)$ is contracted with a fermion at $t_2$ and
we must have $t_2>t_1$. 
(We have already included a factor which accounts for 
the case in which $\psi_i(T)$ is contracted with a 
fermion at $t_1$, which gives a contribution for 
$t_1>t_2$.) The integration for $\tilde{t}^{(2)}$ 
is for both signs, since $t_3>t_4$ and $t_3<t_4$ both survives 
the summation over u and d, as explained above. The argument of
$( G(t_1,t^{(2)}_c))^2$ and 
$( G(t^{(2)}_c, t_2))^2$ really depend on 
the relative time $\tilde{t}^{(2)}$ of $t_3$ and $t_4$, 
but since this function is multiplied by a function sharply
peaked around $t_3=t_4$ for $p\to\infty$ for the reason
that we have 
just mentioned, we can ignore this dependence. 
The integration range for the ``center of mass time,''
$t_c^{(2)}$, of the inner melon is restricted to 
the range inside the outer melon,
$t_1\le t_c^{(2)}\le t_2$; if it is outside
this range (as in
the bottom panel of Fig~\ref{fig:line4}), the 
contributions from the u and d paths cancel each other,
as explained above.

Now, let us perform the integrations. First the
integration over $\tilde{t}^{(2)}$ gives the factor,
\begin{align}
\int_{-\infty}^{\infty} d\tilde{t}^{(2)}
\left( G(\tilde{t}^{(2)})\right)^{p-2}
=\int_{-\infty}^{\infty} d\tilde{t}^{(2)}
{1\over \cosh^2(p{\cal J}\tilde{t}^{(2)})}
={2\over p{\cal J}}.
\label{eq:integtildet2}
\end{align}
Next the integration over $t_c^{(2)}$ is
\begingroup\makeatletter\def\f@size{14}\check@mathfonts
\def\maketag@@@#1{\hbox{\m@th\large\normalfont#1}}%
\begin{align}
&\int_{t_1}^{t_2} dt_c^{(2)}
\left( G(t_1,t^{(2)}_c)\right)^2
\left( G(t^{(2)}_c,t_2) \right)^2\nonumber\\
&\quad=\int_{t_1}^{t_2} dt_c^{(2)}
\left({1\over \cosh^2(p{\cal J}(t_2-t^{(2)}_c)}\right)^{2\over p-1}
\left({1\over \cosh^2(p{\cal J}(t_1-t^{(2)}_c)}\right)^{2\over p-1}\nonumber\\
&\quad \sim \int_{t_1}^{t_2} dt_c^{(2)}
\left(e^{-2p{\cal J}|t_2-t^{(2)}_c|}e^{-2p{\cal J}|t^{(2)}_c-t_1|}
\right)^{2\over p}=\int_{t_1}^{t_2} dt_c^{(2)}e^{-4{\cal J}(t_2-t_1)}
\nonumber\\
&\quad 
=(t_2-t_1) e^{-4{\cal J}(t_2-t_1)},
\label{eq:integt2c}
\end{align}
\endgroup
where we see that the integrand in fact does not depend on $t_c^{(2)}$,
and the answer is proportional to the integration range
$(t_2-t_1)$. Then, the integral for $\tilde{t}^{(1)}=t_2-t_1$, by using 
\eqref{eq:integt2c} in the integrand, becomes
\begingroup\makeatletter\def\f@size{14}\check@mathfonts
\def\maketag@@@#1{\hbox{\m@th\large\normalfont#1}}%
\begin{align}
\int_{0}^{\infty} d\tilde{t}^{(1)} 
\left( G(\tilde{t}^{(1)})\right)^{p-3}
\cdot \tilde{t}^{(1)} e^{-4{\cal J}\tilde{t}^{(1)}}
\sim \int_{0}^{\infty} d\tilde{t}^{(1)} 
\tilde{t}^{(1)} e^{-2p{\cal J}\tilde{t}^{(1)}}
={1\over 4(p{\cal J})^2}
\end{align}
\endgroup
By combining these factors, the correlator becomes
\begingroup\makeatletter\def\f@size{14}\check@mathfonts
\def\maketag@@@#1{\hbox{\m@th\large\normalfont#1}}%
\begin{align}
\langle{\rm tr}\left[ \psi_i(T)\psi_j(0)\right]\rangle_{J}
&=\delta_{ij}i^4{\cal J}^4{p^5\over N}\cdot {2\over p{\cal J}}\cdot{1\over 4(p{\cal J})^2}
\int_{0}^{T}dt^{(1)}_c 
G(T,t^{(1)}_c)G(t^{(1)}_c, 0)\nonumber\\
&=\delta_{ij}{{\cal J}\over 2}{p^2\over N}\int_{0}^{T}dt^{(1)}_c 
G(T,t^{(1)}_c)G(t^{(1)}_c, 0)
\label{eq:final}
\end{align}
\endgroup
where we have replaced $t_1$ or $t_2$ in the propagator by 
their center of mass time $t^{(1)}_c$, since only the
region around zero relative time contributes,
as in the case for $t^{(2)}_c$ explained above. 
The integral in \eqref{eq:final}
just gives $T$ if we use the free propagator. Thus,
\eqref{eq:final} can be regarded as a part of 
the correlation function 
at the first order in $T$, which gives a 
correction to the decay rate defined by
the behavior of the two-point function,
$\delta_{ij}e^{-\gamma T}\sim 
\delta_{ij}(1 -\gamma T)$.
The correction is proportional to 
$\lambda={2p^2\over N}$, 
as dictated by the dimensional analysis in the main text. 

Thus, the decay rate at $\lambda=0$,
$$\gamma|_{\lambda^{0}}=2\CJ,$$
gets a correction at order $\lambda^{1}$ 
from the diagram in Figure~\ref{fig:lambda},
\begin{equation}
\gamma|_{\lambda^{1},0}
%\rm Fig.~\ref{fig:lambda}}
=-{\lambda\over 4}\CJ.
\end{equation}
We have put a symbol 0 after
the subscript $\lambda^{1}$,
to indicate that the diagrams described below
has not been taken into account and this is not
the full answer at order $\lambda^{1}$. 

\subsubsection*{Crossed melon diagrams at order $\lambda$}

\begin{figure}[H]
\begin{center}
\includegraphics[scale=.4]{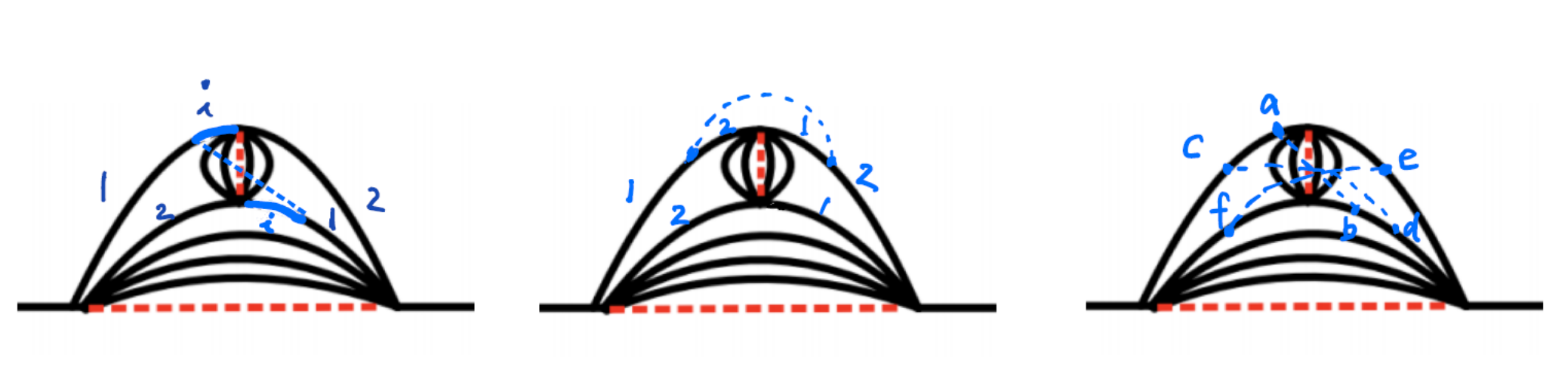}
\caption{Adding melons to the order $\lambda$ diagram in Figure~\ref{fig:lambda}. A blue dotted line collectively denotes the $JJ$ propagator and the $(p-2)$ fermion lines connecting the same end points. The number 1 and 2 indicate particular values of the indices; the symbol $i$ indicates an index not fixed by the index 1 and 2, which should be summed over. 
Left panel: an example of a crossed melon, which is of order $\lambda$. Middle panel: a diagram of order $\lambda^2$. Right panel: diagram with three crossed melons, connecting (a and b), (c and d ) and (e and f), which is of order $\lambda$.}
\label{crossed}
\end{center}
\end{figure}

As mentioned in Section~4.4, 
there are infinite number of diagrams that 
contribute at the same order as above. 
They are the ones in which melons ``cross'' the
small melon in Figure~\ref{fig:lambda}. The simplest
example is given in the left panel of Figure~\ref{crossed}
(and Figure~\ref{crossed1} in the main text).
At first sight, it may seem that adding a melon connecting two different 
lines gives a suppression factor $1/N$. But it does not when
the melon connects two lines with the same index 
(in the left panel of Figure~\ref{crossed}, 
two lines with the same index ``1''). 
In this case, the segments which are
marked in blue have an index (denoted by $i$) that is not
fixed by the indices ``1'' and ``2'' and should be
summed over. This gives a factor of $N$, and the   
counting of the indices in this case becomes effectively 
the same as in the case of a melon connecting the same line. 

For comparison, we show a diagram which is suppressed by 
$1/N$ (or $\lambda$) relative to the one in the left panel.
In the diagram shown in the middle panel, a melon connects 
two lines which have different indices (``1'' and ``2'').
In this case, the indices in the short segments are fixed 
(as indicated by ``1'' and ``2'' in the figure), 
so there is no extra factor of $N$. 

There could be an arbitrary number of the crossed melons, 
and also melons that is crossing in the ``other direction.'' 
The right panel of Figure~\ref{crossed} shows such an example. 
When we have more than one crossed melons, 
the times of the melons should be ``nested''; i.e., in the
above example, if $t_{c}\le t_{a}$, then we need 
$t_{b}\le t_{d}$, in order to have a contribution that 
survives the summation over u and d, explained earlier
in this appendix. 
The integration over the vertices should be done from 
the inside to the outside. We may be able to regard this 
structure as a vertex renormalization.

It is possible to compute the contributions from 
the crossed melons, at least in the limit of 
$p\to \infty$, in which the integration
over the relative time of the vertices of a melon 
effectively makes the interaction instantaneous. 
We defer this analysis to future study.

\section{Derivation of the Scaling of Each Diagram}

In this appendix, we derive the scaling of each 
diagram in SYK, which was introduced in Section~5.1 and
played a major role in the argument for the existence
of fixed $\lambda$ limit. 

To each diagram, we assign the factor
\be  
\CJ^{a} T_{\rm r}^{a-1} T_{\rm c}\f{p^b}{N^{h}},
\label{abh_a}
\ee
where 
$T_{\rm r}$ and $T_{\rm c}$ denote the 
relative time and the ``center of mass'' time
of the vertices. 
We will explain that for wee-irreducible diagrams, 
the integers $a,$  $b$ and $h$ satisfy the
relation
\be  
b = 2h +a -1.
\label{funeq_a}
\ee
In the present case\footnote{%
For the case of real (Majorana) fermions 
and order-one external lines, wee-irreducible
diagrams have this property.}, 
wee-irreducible diagrams are the ones in which the
JJ propagator directly connects the initial and final 
vertices.

By a diagram, we really mean a 
set of diagrams in the following sense: by the $A_{4,2}$ diagram 
we mean all the diagrams in which one melon ($A_2$ insertion) 
is inserted in any one of the internal lines of one outer melon 
(not only in the top line, as shown in Figure~\ref{F4}); 
By the $A_{4,3}$ diagram, we mean all the diagrams in which one melon
connects any combination of two internal lines of one outer melon 
(not only the top two lines, as shown in Figure~\ref{F4}). 

\subsection{Insertion of Melons}
One can understand the relation \eqref{funeq_a} by imagining 
a process of making a new diagram by adding a melon to
an existing diagram. There are two types of such an 
addition. Let us see how the factors change
by this procedure for each of them.

\begin{enumerate}
\item 
One can add a melon which connects two points on the
same line, which means replacing an internal line of
an existing diagram by the ``$A_2$'' structure in 
Figure~\ref{top}. 
(For example, we can make  $A_{4,2}$
in Figure~\ref{F4} from $A_2$ in Figure~\ref{top}
this way.) In this case, the factor between the 
new and the old diagram is
$$ p^2 (\CJ T_{\rm r})^2.$$
One power of $p$ comes 
from the product of the $JJ$ propagator and the 
combinatoric factor; another power of $p$ comes 
from the choice of which line the melon is attached;
$(\CJ T_{\rm r})^2$ comes from the integration over 
two extra vertices.

\item
One can also add a melon which connects two different 
lines\footnote{%
By two different lines, we really mean two lines that have
different indices. If the two lines have the same
index, the index summation for the internal lines of
the melon between them becomes equivalent to 
the case of connecting the same line, and we do not 
get the $1/N$ suppression. The ``crossed melon'' 
diagrams mentioned at the end of Appendix~A.3 are
such examples.}
 (e.g., to make $A_{4,3}$
in Figure~\ref{F4} from $A_2$ in Figure~\ref{top}).
In this case, we get the factor
$$ {p^4\over N} (\CJ T_{\rm r})^2.$$
The power ${p^2\over N}$ comes from the product of the 
$JJ$ propagator and the combinatoric
factor; another power of $p^2$ comes from the choice of 
which two lines are connected; $(\CJ T_{\rm r})^2$ comes 
from the integration over two extra vertices, as above.
\end{enumerate}

Here we note that the $A_2$ diagram in Figure~\ref{top},
which has $a=2$, $b=1$, $h=0$, satisfies the 
the relation \eqref{funeq_a}.
Consider making new diagrams by applying the above
procedure repeatedly starting from the $A_2$ diagram.

In the first case, $a$ increases by 2, and $b$ also increases 
by 2, so \eqref{funeq_a} is kept invariant. In the second
case, $a$ increases by 2, $b$ increases by 4, $h$ increases
by 1, so also \eqref{funeq_a} is kept invariant.
Therefore, any diagram that can be produced by the
above procedure from the $A_2$ diagram satisfies \eqref{funeq_a}.

\subsection{Reconnecting the Lines}
The diagram in Figure~\ref{asym} cannot be generated in the
above way, but one can see that it also satisfies \eqref{funeq_a} from the 
following considerations. 

This diagram can be obtained from the $A_{4,2}$ diagram in 
Figure~\ref{F4} by cutting open a line in the small melon,
and also cutting open a line in the outer melon, and reconnecting
them (by gluing open ends in the former to the open ends of the 
latter). The index of the former has to match the one
of the latter if we glue them, while they have been unconstrained 
if we did not do this cutting and gluing. So we lose one power 
of $N$ in this operation. Since we have $p^2$ choices of which
two lines to cut, we have the factor $p^2/N$ (i.e.,
$\lambda$) in total. In this case, $h$ increase by 1, 
$b$ increases by 2, and $a$ does not change, so the diagram
that results from this operation satisfies the 
relation~\eqref{funeq_a} if the original diagram satisfies it.
To get the diagram in Figure~\ref{asym} from the $A_2$ 
diagram in Figure~\ref{top}, we repeat this $n-1$ times. 

In fact, the $A_{4,3}$ diagram (which has $h=1$) in 
Figure~\ref{F4}, which was discussed in the last 
subsection, can also be obtained from 
$A_{4,2}$ diagram by a similar procedure:
When we glue two pairs of the open ends, we take the
combination different from the above. 

Diagrams with an arbitrary $h$ can be constructed from 
an $h=0$ diagram with certain number of $A_{2}$, by 
applying the above procedure. The constructions 
given here and the previous subsection guarantee
that those diagrams satisfy the relation~\eqref{funeq_a}.


\begin{thebibliography}{99}

\bibitem{susskind:confined}
L.~Susskind,
"De Sitter Space has no Chords. Almost Everything is Confined,"
[arxiv:2303.00792 [hep-th]].

\bibitem{Cotler:2016fpe}
J.~S.~Cotler, G.~Gur-Ari, M.~Hanada, J.~Polchinski, P.~Saad, S.~H.~Shenker, D.~Stanford, A.~Streicher and M.~Tezuka,
``Black Holes and Random Matrices,''
JHEP \textbf{05}, 118 (2017)
[erratum: JHEP \textbf{09}, 002 (2018)]
[arXiv:1611.04650 [hep-th]].

\bibitem{Berkooz:2018jqr}
M.~Berkooz, M.~Isachenkov, V.~Narovlansky and G.~Torrents,
``Towards a full solution of the large N double-scaled SYK model,''
JHEP \textbf{03}, 079 (2019)
[arXiv:1811.02584 [hep-th]].

\bibitem{Susskind:2022bia}
L.~Susskind,
``De Sitter Space, Double-Scaled SYK, and the Separation of Scales in the Semiclassical Limit,''
[arXiv:2209.09999 [hep-th]].

\bibitem{Susskind:2021esx}
L.~Susskind,
``Entanglement and Chaos in De Sitter Holography: An SYK Example,''
[arXiv:2109.14104 [hep-th]].

\bibitem{Susskind:2022dfz}
L.~Susskind,
``Scrambling in Double-Scaled SYK and De Sitter Space,''
[arXiv:2205.00315 [hep-th]].


\bibitem{Narovlansky:2023lfz}
V.~Narovlansky and H.~Verlinde,
``Double-scaled SYK and de Sitter Holography,''
[arXiv:2310.16994 [hep-th]].


\bibitem{Verlinde:2024znh}
H.~Verlinde,
``Double-scaled SYK, Chords and de Sitter Gravity,''
[arXiv:2402.00635 [hep-th]].

\bibitem{Rahman:2023pgt}
A.~A.~Rahman and L.~Susskind,
``Comments on a Paper by Narovlansky and Verlinde,''
[arXiv:2312.04097 [hep-th]].

\bibitem{Rahman:2024vyg}
A.~A.~Rahman and L.~Susskind,
``Infinite Temperature is Not So Infinite: The Many Temperatures of de Sitter Space,''
[arXiv:2401.08555 [hep-th]].

\bibitem{Rahman:2024iiu}
A.~A.~Rahman and L.~Susskind,
``$p$-Chords, Wee-Chords, and de Sitter Space,''
[arXiv:2407.12988 [hep-th]].

\bibitem{Lin:2022nss}
H.~Lin and L.~Susskind,
``Infinite Temperature's Not So Hot,''
[arXiv:2206.01083 [hep-th]].

\bibitem{Lin:2023trc}
H.~W.~Lin and D.~Stanford,
``A symmetry algebra in double-scaled SYK,''
SciPost Phys. \textbf{15}, no.6, 234 (2023)
[arXiv:2307.15725 [hep-th]].

\bibitem{Miyashita:2025rpt}
S.~Miyashita, Y.~Sekino and L.~Susskind,
``DSSYK at Infinite Temperature: The Flat-Space Limit and the 't Hooft Model,''
[arXiv:2506.18054 [hep-th]].

\bibitem{Polchinski:1999ry}
J.~Polchinski,
``S matrices from AdS space-time,''
[arXiv:hep-th/9901076 [hep-th]].

\bibitem{Susskind:1998vk}
L.~Susskind,
``Holography in the flat space limit,''
AIP Conf. Proc. \textbf{493}, no.1, 98-112 (1999)
[arXiv:hep-th/9901079 [hep-th]].


\bibitem{Polchinski:1999yd}
J.~Polchinski, L.~Susskind and N.~Toumbas,
``Negative energy, superluminosity and holography,''
Phys. Rev. D \textbf{60}, 084006 (1999)
[arXiv:hep-th/9903228 [hep-th]].

\bibitem{Banks:1996vh}
T.~Banks, W.~Fischler, S.~H.~Shenker and L.~Susskind,
``M theory as a matrix model: A conjecture,''
Phys. Rev. D \textbf{55}, 5112-5128 (1997)
[arXiv:hep-th/9610043 [hep-th]].

\bibitem{Maldacena:2016hyu}
J.~Maldacena and D.~Stanford,
``Remarks on the Sachdev-Ye-Kitaev model,''
Phys. Rev. D \textbf{94}, no.10, 106002 (2016)
[arXiv:1604.07818 [hep-th]].

\bibitem{Roberts:2018mnp}
D.~A.~Roberts, D.~Stanford and A.~Streicher,
``Operator growth in the SYK model,''
JHEP \textbf{06}, 122 (2018)
[arXiv:1802.02633 [hep-th]].

\bibitem{Gurau:2013cbh}
R.~Gurau and J.~P.~Ryan,
``Melons are branched polymers,''
Annales Henri Poincare \textbf{15}, no.11, 2085-2131 (2014)
[arXiv:1302.4386 [math-ph]].

\bibitem{Maldacena:2019cbz}
J.~Maldacena, G.~J.~Turiaci and Z.~Yang,
``Two dimensional Nearly de Sitter gravity,''
JHEP \textbf{01}, 139 (2021)
[arXiv:1904.01911 [hep-th]].


\end{thebibliography}
\end{document}